\documentclass[acmsmall,screen]{acmart}

\usepackage{amsmath, bm}
\usepackage{multirow, multicol}
\usepackage{makecell, float, diagbox}
\usepackage{algorithmicx,algorithm}
\usepackage[noend]{algpseudocode}
\usepackage{xcolor}
\usepackage{CJKutf8}
\usepackage{tabularx, array}
\usepackage{longtable}
\usepackage{colortbl}

\usepackage{forest}
\usepackage{enumitem}

\AtBeginDocument{%
  }

\setcopyright{acmlicensed}
\copyrightyear{2018}
\acmYear{2018}
\acmDOI{XXXXXXX.XXXXXXX}
\acmConference[Conference acronym 'XX]{Make sure to enter the correct
  conference title from your rights confirmation email}{June 03--05,
  2018}{Woodstock, NY}
\acmISBN{978-1-4503-XXXX-X/2018/06}

\begin{document}

\title{A Survey on Multimodal Retrieval-Augmented Generation}


\author{Lang Mei}
\affiliation{
    \institution{Huawei Cloud BU}
    \city{Beijing}
    \country{China}
}
\email{meilang1@huawei.com}

\author{Siyu Mo}
\affiliation{
    \institution{Huawei Cloud BU}
    \city{Beijing}
    \country{China}
}
\email{mosiyu@huawei.com}

\author{Zhihan Yang}
\affiliation{
    \institution{Huawei Cloud BU}
    \city{Beijing}
    \country{China}
}
\email{yangzhihan4@huawei.com}

\author{Chong Chen}
\authornote{Chong Chen is the corresponding author.}
\affiliation{
    \institution{Huawei Cloud BU}
    \city{Beijing}
    \country{China}
}
\email{chenchong55@huawei.com}

\renewcommand{\shortauthors}{Trovato et al.}

\begin{abstract}
Multimodal Retrieval-Augmented Generation (MRAG) represents a significant advancement in enhancing the capabilities of large language models (LLMs) by integrating multimodal data, such as text, images, and videos, into the retrieval and generation processes. Traditional Retrieval-Augmented Generation (RAG) systems, which primarily rely on textual data, have shown promise in reducing hallucinations and improving response accuracy by dynamically incorporating external knowledge. However, these systems are limited by their reliance on text-only modalities, which restricts their ability to leverage the rich, contextual information available in multimodal data. MRAG addresses this limitation by extending the RAG framework to include multimodal retrieval and generation, thereby enabling more comprehensive and contextually relevant responses.
In MRAG, the retrieval step involves locating and integrating relevant knowledge from diverse modalities, while the generation step utilizes multimodal large language models (MLLMs) to produce answers that incorporate information from multiple data types. This approach not only enhances the quality of question-answering systems but also significantly reduces the incidence of hallucinations by grounding responses in factual, multimodal knowledge. Recent research has demonstrated that MRAG outperforms traditional text-modal RAG, particularly in scenarios where visual and textual information are both critical for understanding and responding to queries.
This survey systematically reviews the current state of MRAG research, focusing on four key aspects: essential components and technologies, datasets, evaluation methods and metrics, and existing limitations. By analyzing these dimensions, we aim to provide a comprehensive understanding of how MRAG can be effectively constructed and improved. Additionally, we highlight current challenges and propose future research directions, encouraging further exploration into this promising paradigm. Our work underscores the potential of MRAG to revolutionize multimodal information retrieval and generation, offering a forward-looking perspective on its development and applications.
\end{abstract}



\received{20 February 2007}
\received[revised]{12 March 2009}
\received[accepted]{5 June 2009}

\maketitle



\section{Introduction}
Large language models (LLMs), especially the Transformer-based variants, have achieved extraordinary success in many language-related tasks.
Through pre-training on extensive, high-quality instruction datasets, LLMs can learn a wide range of language patterns, structures, and factual knowledge. 
These pre-trained LLMs can generate human-like text with high degrees of fluency and coherence, and attain strong performance on question-answering tasks, which demonstrates their ability to understand and respond to a wide range of queries.
However, despite their impressive capabilities, LLMs still face significant limitations. One of the primary challenges lies in their performance within specific domains or knowledge-intensive tasks. While these models are often trained on diverse and extensive datasets, such datasets may not cover the depth of knowledge required for highly specialized fields or real-time information updates. This can be particularly problematic in areas like medicine, law, finance, and other technical fields where precision and up-to-date knowledge are to be prioritized.
When handling queries that extend beyond the scope of their training knowledge or require the most current information, LLMs may generate responses that are speculative or based on patterns they have learned, rather than on verified facts. This can result in misleading, incorrect, or even entirely fabricated answers, a phenomenon known as "hallucination".
Minimizing the incidence of hallucinations is important for enhancing the reliability of LLMs in providing accurate and context-relevant information across different domains.

Recently, Retrieval-Augmented Generation (RAG) has emerged as an effective solution to mitigate hallucinations, by enhancing the generation capabilities of large language models (LLMs) through the retrieval of relevant external knowledge.
Existing RAG systems typically operate through a two-step process: retrieval and generation.
In the retrieval step, the goal is to quickly locate relevant knowledge that is semantically similar to the query from a large-scale document collection. Since the relevant knowledge is often scattered across various parts of documents, each document is pre-processed into multiple chunks. Additional chunks may be created through manual or automated methods. This process, known as document chunkerization, ensures that fine-grained knowledge can be retrieved more efficiently.
In the generation step, the retrieved document chunks are combined with the query to form an augmented input. This augmented input provides the LLM with context that includes external knowledge.
Furthermore, RAG allows LLMs to dynamically integrate the latest information during the inference stage. This capability ensures that the model's responses are not only based on static, pre-trained knowledge but are continuously updated with current and relevant data.
By retrieving and referencing external knowledge, RAG grounds the generated responses in factual information, thereby significantly reducing the occurrence of hallucinations.
However, previous research on RAG systems has primarily focused on knowledge bases built from plain text and LLMs pre-trained on plain text, ignoring other rich sources of knowledge available for query responses in the real world, such as videos and images, referred to as "multimodal data".

Multimodal data refers to data that comes from multiple sources or formats. This can include text, images, audio, video, and other types of data. In real-world scenarios, humans naturally interact with multimodal data, such as browsing web pages that combine text, images, and videos in mixed layouts.
By analyzing images or videos alongside text, the user can better understand the context of the content, and thus improve the satisfaction with the quality of the answers. For example, if a passenger inquires about how to store luggage while flying, it will be clearer that the system provides relevant graphic guides or instructional videos.
However, transferring the capabilities of LLMs to the domain of multimodal text and images remains an active area of research, as plain-text LLMs are typically trained only on textual corpora and lack perceptual abilities for visual signals. 
How to effectively incorporate multimodal data is important to enhance the capability of LLMs.
In recent years, the development of multimodal generative models has showcased additional application possibilities. Apart from textual generative models, multimodal generative models have been increasingly applied in fields such as human-computer interaction, robot control, image search, and speech generation. 
Similarly, based on multimodal generative models and multimodal data, how to effectively process Multimodal Retrieval-Augmented Generation (\textbf{MRAG}) is an issue that needs to be explored.

Recently, some research have demonstrated that MRAG with multimodal data outperforms traditional text-modal RAG. 
By enhancing the generation capabilities of multimodal large language models (MLLMs) through the retrieval of external multimodal knowledge, MRAG system can further enhance question answering capabilities and quality, thereby further reducing hallucination issues.
The main differences between text-modal RAG and MRAG lie in retrieval and generation.
In the retrieval step, the former only needs to consider retrieving relevant textual knowledge from a large document collection, while the latter needs to consider how to retrieve and integrate the relevant knowledge under different modalities, as well as the relationships between knowledge in different modalities.
In the generation step, the former only needs to consider the input text query and relevant textual knowledge, and output a text answer based on the LLM. The latter, however, needs to consider how to utilize the input query from different modalities and multimodal retrieval knowledge, and output an answer that includes information from different modalities based on the MLLM.

Considering the immense potential of MRAG in this field, this survey aims to systematically review and analyze the current state and main challenges of MRAG. We discuss existing research from several key perspectives: 
1) \emph{What important components and technologies are involved in MRAG?}
2) \emph{What types of datasets can be used for the evaluation of MRAG?}
3) \emph{What methods and metrics are used to evaluate MRAG?}
4) \emph{What limitations exist in the different aspects of MRAG?}
We explore the main challenges faced by MRAG, and hope to provide clearer guidelines for their future development.
In summary, the main contributions of this paper are as follows:
\begin{itemize}[leftmargin=1em]
\item \textbf{Comprehensive and Timely Survey:} We conducted an extensive survey on the emerging paradigm of multimodal Retrieval-Augmented Generation, systematically reviewing the current state of research and development in this field.
\item \textbf{Systematic Analysis from Four Key Perspectives:} Our survey is organized around four key aspects: essential components and technologies, datasets, evaluation methods and metrics, and limitations. This structured approach allows for a detailed understanding of how MRAG can be efficiently constructed, its reliability issues, and how it can be further improved.
\item \textbf{Current Challenges and Future Research Directions:} We discuss the existing challenges of MRAG, highlight potential research opportunities and directions, and provide a forward-looking perspective on the future development of this paradigm, encouraging researchers to delve deeper into this exciting field.
\end{itemize}

We have provided an overall introduction to this survey paper. 
The section \ref{sec:OverviewofMRAG} presents a comprehensive overview of multimodal retrieval-augmented generation, covering multiple developmental stages. 
The section \ref{sec:ComponentsTechnologiesMRAG} delves into the technical details of multimodal retrieval-augmented generation, focusing on key components such as multimodal retrieval, multimodal generation, etc. 
In section \ref{sec:datasetMRAG}, we discuss how to comprehensively evaluate multimodal retrieval-augmented generation systems using datasets, including specialized assessments for different competency areas. 
The section \ref{sec:EvaluationMetricsMRAG} introduces relevant metrics for evaluating multimodal retrieval-augmented generation systems. 
In section \ref{sec:ChallengesMRAG}, we outline the current technical challenges associated with multimodal retrieval-augmented generation. 
In section \ref{sec:FutureWork}, based on previous investigation of MRAG, we summarize future work in this field, and provide some suggestions.
Finally, we give the conclusion of the paper in section \ref{sec:Conclusion}.





\section{Overview of MRAG}
\label{sec:OverviewofMRAG}
Multimodal Retrieval-Augmented Generation (\textbf{MRAG}) represents a significant evolution of the traditional Retrieval-Augmented Generation (\textbf{RAG}) framework, building upon its foundational structure while extending its capabilities to process diverse data modalities. While RAG is limited to processing plain text, MRAG integrates multimodal data, including images, audio, video, and text, enabling it to address more complex and diverse real-world applications where information spans multiple modalities.

In the early stages of MRAG development, researchers converted multimodal data into unified textual representations. This approach allowed for a seamless transition from RAG to MRAG by leveraging existing text-based retrieval and generation mechanisms. Although this strategy simplified multimodal data integration and improved the end-to-end user experience, it introduced significant limitations. For instance, the conversion process often resulted in the loss of modality-specific information, such as visual details in images or tonal nuances in audio, restricting the system's ability to fully exploit the potential of multimodal inputs. Subsequent research has focused on addressing these limitations by developing more advanced methods to optimize MRAG systems. These advancements have substantially enhanced MRAG's performance and versatility, achieving state-of-the-art results across various multimodal tasks. This paper categorizes the evolution of MRAG into three distinct stages:

\subsection{MRAG1.0}
The initial stage of the MRAG framework, commonly termed "pseudo-MRAG", emerged as a straightforward extension of the highly successful RAG paradigm. This stage was rapidly adopted due to its adherence to RAG's core principles, with modifications to support multimodal data. As illustrated in Figure \ref{MRAG1.0}, the MRAG1.0 architecture consists of three key components: Document Parsing and Indexing, Retrieval, and Generation.
\begin{itemize}[leftmargin=1em, listparindent=\parindent]
\item \textbf{Document Parsing and Indexing:} This component is responsible for processing multimodal documents in formats such as Word, Excel, PDF, and HTML. It extracts textual content using Optical Character Recognition (OCR) or format-specific parsing techniques. A document layout detection model is then utilized to segment the document into structured elements, including titles, paragraphs, images, videos, tables, and footers. For textual content, a chunking strategy is applied to segment or group semantically coherent passages. For multimodal data, specialized models are used to generate captions describing images, videos, and other non-textual elements. These chunks and captions are encoded into vector representations using an embedding model and stored in a vector database. The choice of embedding model is crucial, as it significantly impacts the performance and effectiveness of downstream retrieval tasks.
\item \textbf{Retrieval:} This component processes user queries by encoding them into vector representations using the same embedding model applied during indexing. The query vectors are then utilized to retrieve the top-$k$ most relevant chunks and captions from the vector database, typically employing cosine similarity as the relevance metric. Duplicate or overlapping information from chunks and captions is merged to create a consolidated set of external knowledge, which is subsequently integrated into the prompt for the generation phase. This ensures the system retrieves contextually relevant information to deliver accurate and informed responses.
\begin{figure}[htbp]
\centering
\includegraphics[width=0.99\textwidth]{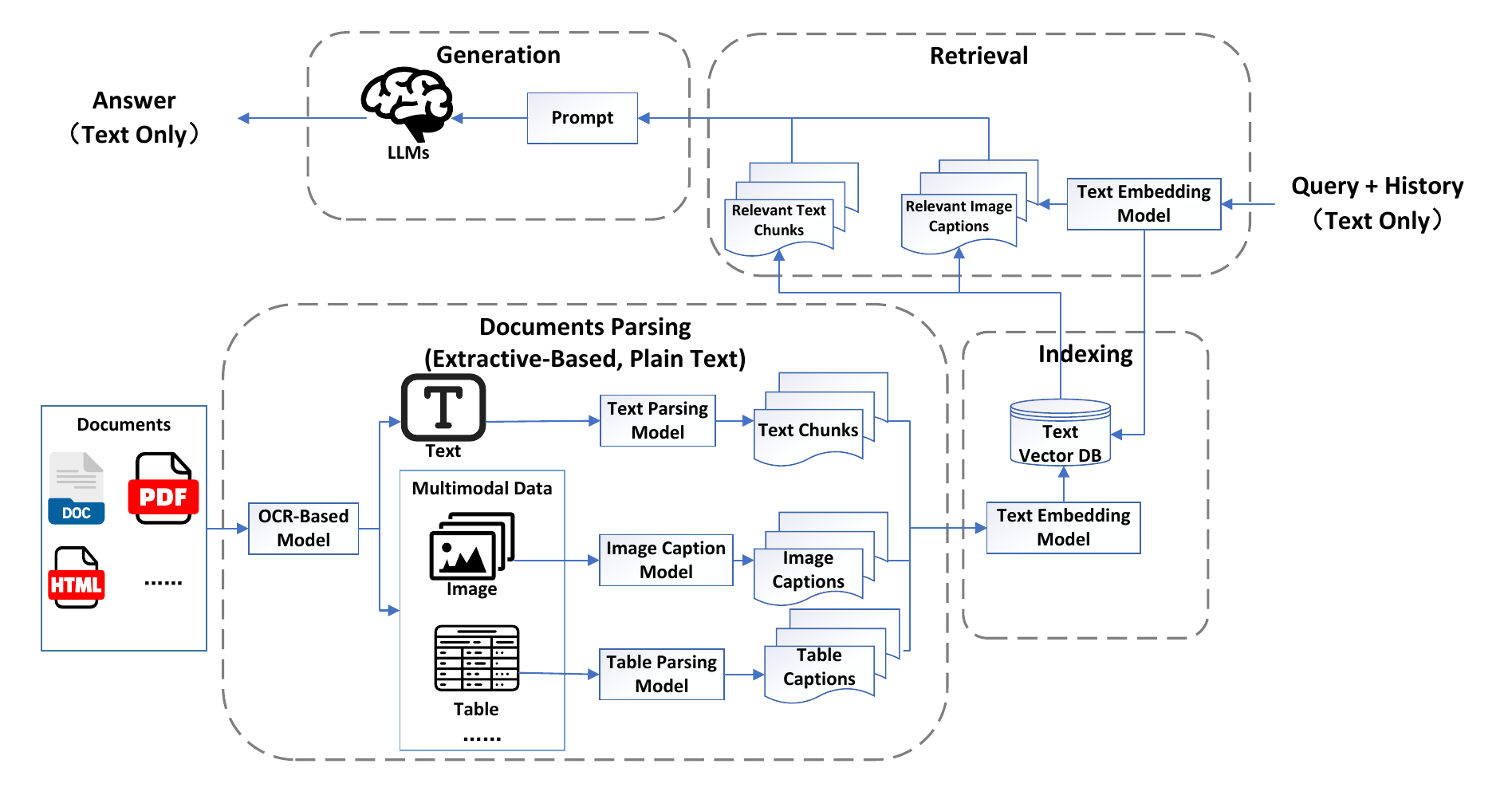}
\caption{The architecture of MRAG1.0, often termed "pseudo-MRAG", closely resembles traditional RAG, consisting of three modules: Document Parsing and Indexing, Retrieval, and Generation. While the overall process remains largely unchanged, the key distinction lies in the Document Parsing stage. In this stage, specialized models are employed to convert diverse modal data into modality-specific captions. These captions are then stored alongside textual data for utilization in subsequent stages.} 
\label{MRAG1.0}
\end{figure}
\item \textbf{Generation:} In the Generation phase, the MRAG system synthesizes the user's query and retrieved documents into a coherent prompt. A large language model (LLM) generates a response by integrating its parametric knowledge with the retrieved external information. This approach enhances response accuracy and timeliness, particularly in domain-specific contexts, while reducing the risk of hallucinations common in LLM outputs. In multi-turn dialogues, the system incorporates conversational history into the prompt, enabling contextually aware and seamless interactions.
\end{itemize}
Despite its initial success, MRAG1.0 exhibited several notable limitations that constrained its effectiveness:
\begin{itemize}[leftmargin=1em, listparindent=\parindent]
\item \textbf{Cumbersome Document Parsing:} Converting multimodal data into textual captions introduced substantial complexity to the system. This necessitated distinct models for processing different data modalities, increasing both computational overhead and system intricacy. Additionally, the conversion process frequently often to multimodal information loss. For instance, image captions typically provided only coarse-grained descriptions, failing to capture fine-grained details essential for accurate retrieval and generation.
\item \textbf{Bottleneck of Retrieval:} While text vector retrieval technology is well-established, MRAG1.0 encountered challenges in achieving high recall accuracy. Similar to traditional RAG, the chunking strategy for text segmentation often fragmented keywords, making some content irretrievable. Additionally, transforming multimodal data into text, while enabling non-textual data retrieval, introduced additional information loss. These issues collectively created a bottleneck, limiting the system's ability to retrieve comprehensive and accurate information.
\begin{figure}[htbp]
	\centering
	\includegraphics[width=0.99\textwidth]{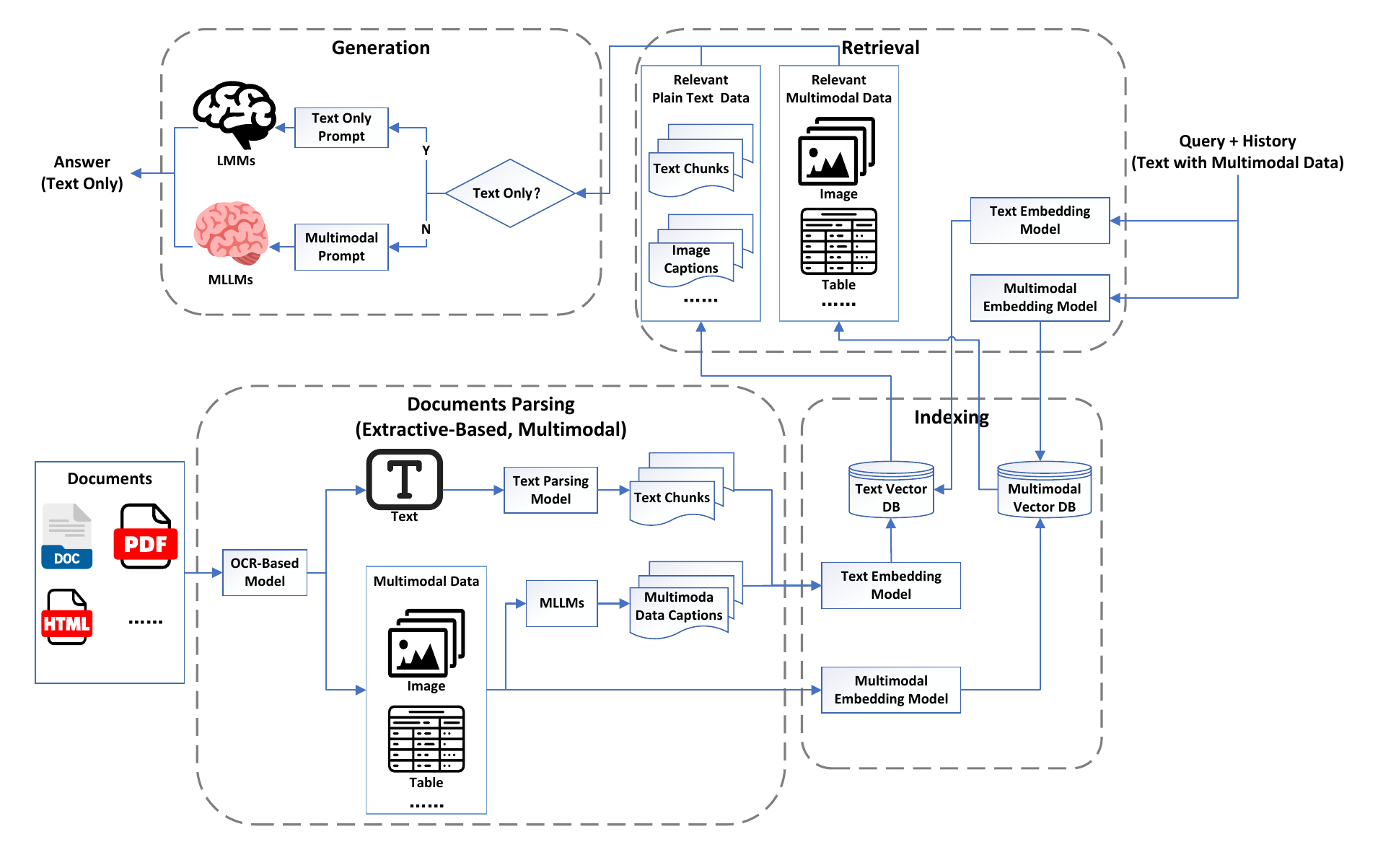}
	\caption{The architecture of MRAG2.0 retains multimodal data through document parsing and indexing, while introducing multimodal retrieval and MLLMs for answer generation, truly entering the multimodal era.} 
	\label{MRAG2.0}
\end{figure}
\item \textbf{Challenges in Generation:} Unlike traditional RAG, MRAG1.0 required processing not only text chunks but also image captions and other multimodal data. Effectively organizing these diverse elements into coherent prompts while minimizing redundancy and preserving relevant information posed a significant challenge. Additionally, the "Garbage In, Garbage Out" (GIGO) principle highlighted the sensitivity of LLMs to input quality. Information loss during parsing and retrieval increased the risk of incorporating irrelevant data, compromising the robustness and reliability of the generated responses.
\end{itemize}
The limitations of MRAG1.0 created a performance ceiling, highlighting the need for more advanced technological solutions. The system's reliance on text-based representations for multimodal data, along with inherent challenges in retrieval and generation, revealed critical gaps in multimodal understanding, retrieval efficiency, and generation robustness. Subsequent iterations of MRAG must address these issues by adopting more sophisticated models, enhancing information retention during parsing, and improving the integration of multimodal data into retrieval and generation processes.

\subsection{MRAG2.0}
With the rapid evolution of multimodal technologies, MRAG has transitioned into a "true multimodal" era, termed MRAG2.0. Unlike its predecessor MRAG1.0, MRAG2.0 not only supports user queries with multimodal inputs but also preserves the original multimodal data within the knowledge base. By leveraging the capabilities of MLLMs, the generation module can now process multimodal data directly, minimizing information loss during data conversion. As illustrated in Figure \ref{MRAG2.0}, the MRAG2.0 architecture incorporates several key optimizations:
\begin{itemize}[leftmargin=1em, listparindent=\parindent]
\item \textbf{MLLMs Captions:} The representational capabilities of MLLMs have significantly advanced, especially in captioning tasks. MRAG2.0 leverages a single, unified MLLM—or multiple MLLMs—to extract captions from multimodal documents. This approach replaces the conventional paradigm of using separate models for different modalities, simplifying the document parsing module and reducing its complexity.
\item \textbf{Multimodal Retrieval:} MRAG2.0 enhances its retrieval module to support multimodal user inputs by preserving original multimodal data and enabling cross-modal retrieval. This allows text-based queries to directly retrieve relevant multimodal data, combining caption-based recall with cross-modal search capabilities. The dual retrieval approach enriches data sources for downstream tasks while minimizing data loss, improving accuracy and robustness for downstream tasks.
\item \textbf{Multimodal Generation:} To fully leverage original multimodal data, the generation module in MRAG2.0 has been enhanced by integrating MLLMs, enabling the synthesis of user queries and retrieval results into a coherent prompt. When retrieval results are accurate and the input comprises original multimodal data, the generation module mitigates information loss typically associated with modality conversion. This enhancement has significantly improved the accuracy of question-answering (QA) tasks, especially in scenarios involving interrelated multimodal data.
\end{itemize}
Despite these advancements, MRAG2.0 encounters several emerging challenges:
1) Integrating multimodal data inputs may reduce the accuracy of traditional textual query descriptions. Furthermore, current multimodal retrieval capabilities remain inferior to text-based retrieval, potentially limiting the overall accuracy of the retrieval module.
2) The diversity of data formats presents new challenges for the generation module. Efficiently organizing these diverse data forms and clearly defining inputs for generation are critical areas requiring further exploration and prioritization.

\subsection{MRAG3.0}
\begin{figure}[htbp]
	\centering
	\includegraphics[width=0.99\textwidth]{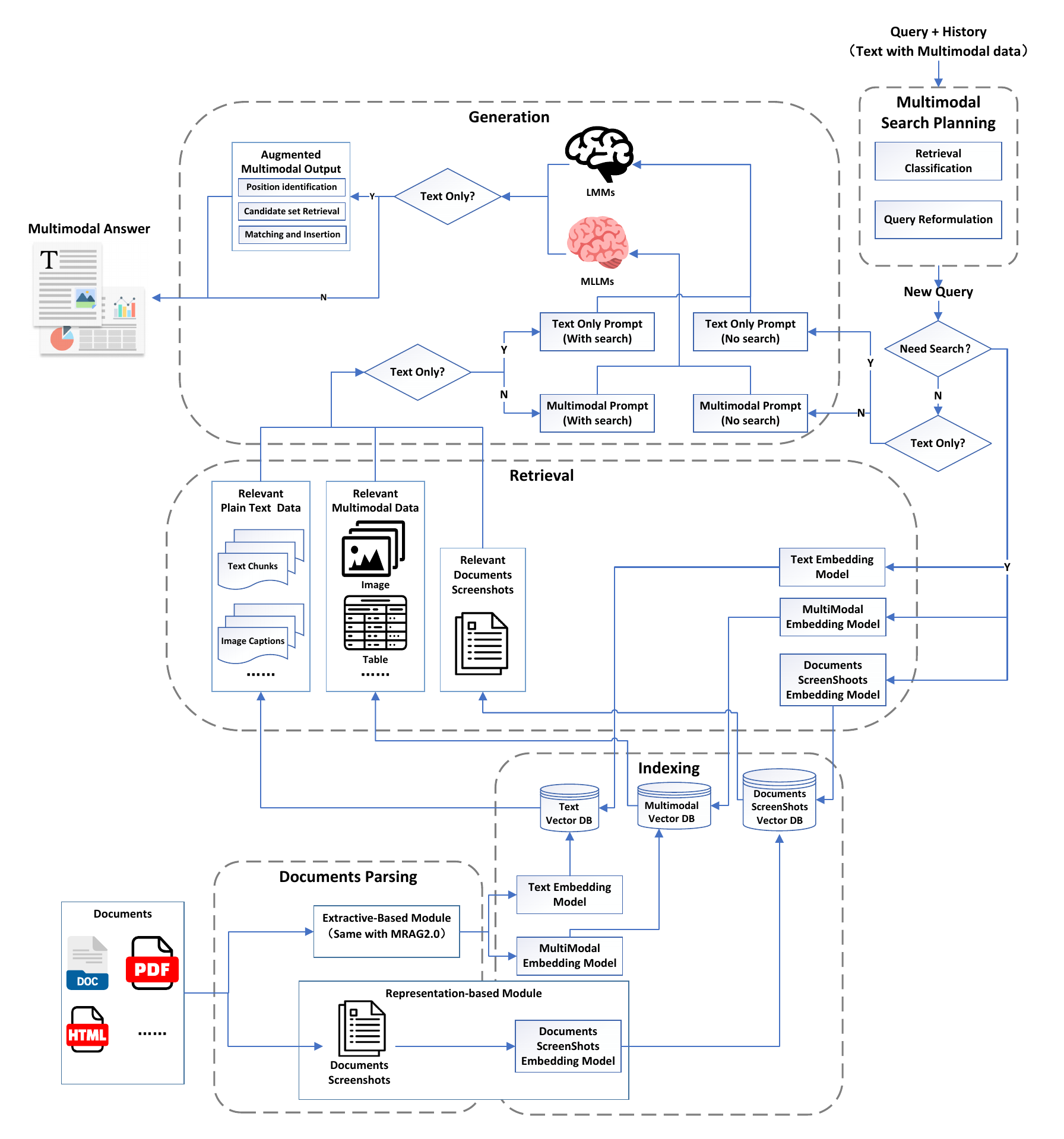}
	\caption{MRAG3.0 architecture integrates document screenshots during the document parsing and indexing stages to minimize information loss. At the input stage, it incorporates a Multimodal Search Planning module, unifying Visual Question Answering (VQA) and Retrieval-Augmented Generation (RAG) tasks while refining user query precision. At the output stage, the Multimodal Retrieval-Augmented Composition module enhances answer generation by transforming plain text into multimodal formats, thereby enriching information delivery.} 
	\label{definition}
\end{figure}

 As illustrated in Figure \ref{definition}, the MRAG3.0 system represents a significant evolution from its predecessors, introducing structural and functional innovations that enhance its capabilities across multiple dimensions. This new paradigm shift is characterized by three key advancements:
1) Enhanced Document Parsing: A novel approach retains document page screenshots during parsing, minimizing information loss in database storage.
2) True End-to-End Multimodality: While earlier versions emphasized multimodal capabilities in knowledge base construction and system input, MRAG3.0 introduces multimodal output capabilities, completing the end-to-end multimodal framework.
3) Scenario Expansion: Moving beyond traditional focus on understanding capabilities—primarily applied in VQA (Visual Question Answering) scenarios reliant on knowledge bases, the new paradigm integrates understanding and generation capabilities through module adjustments and additions. This unification significantly broadens the system's applicability.
In the following sections, we will detail the scenarios supported by MRAG3.0 and the specific module modifications enabling these advanced capabilities.

\subsubsection{Scenario for MRAG}
\begin{figure}[htbp]
	\centering
	\includegraphics[width=0.9\textwidth]{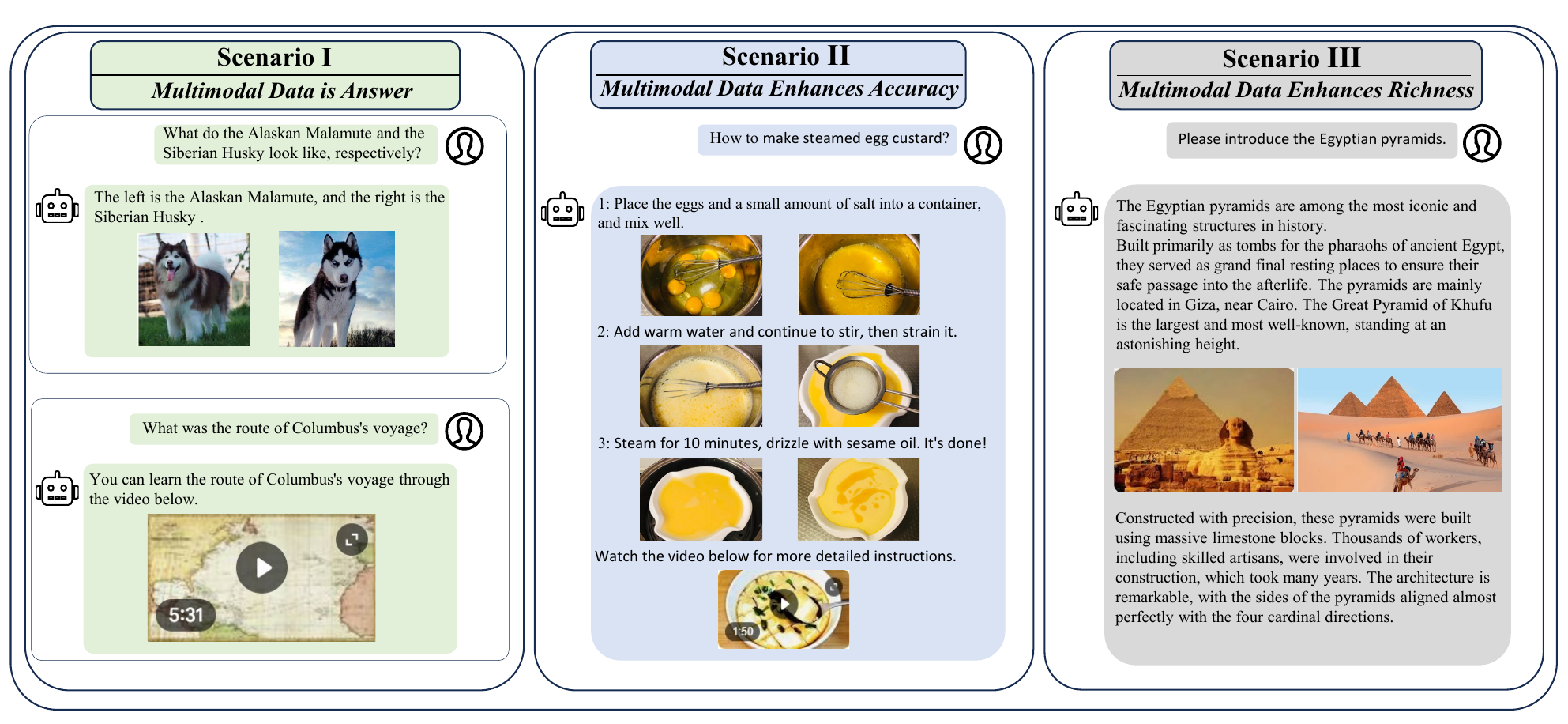}
	\caption{Multimodal output in QA scenarios can be categorized into three distinct types. In sub-scenario I, the user's query can be fully addressed using only images or videos, without requiring supplementary textual information. Sub-scenario II involves a step-by-step explanation that combines text and images to ensure clarity and precision; omitting the images may lead to user confusion at specific steps. In sub-scenario III, supplementary images enrich the information conveyed in the answer, but their removal does not compromise the answer's accuracy.} 
	\label{fig:mm output}
\end{figure}

\begin{itemize}[leftmargin=1em, listparindent=\parindent]
\item\textbf{Retrieval-Augmented Scenario:} This scenario addresses cases where LLMs or MLLMs alone cannot adequately answer user queries. MRAG3.0 retrieves relevant content from external knowledge bases to provide accurate answers, leveraging its enhanced retrieval capabilities.
\item\textbf{VQA Scenario:} This scenario serves as a critical test for evaluating the fundamental capabilities of MLLMs, which generate responses directly from user inputs containing text and multimodal queries without retrieval. The new MRAG paradigm introduces a search planning module, enabling dynamic routing and retrieval to minimize unnecessary searches and the inclusion of irrelevant information.
\begin{figure}[htbp]
	\centering
	\includegraphics[width=0.7\textwidth]{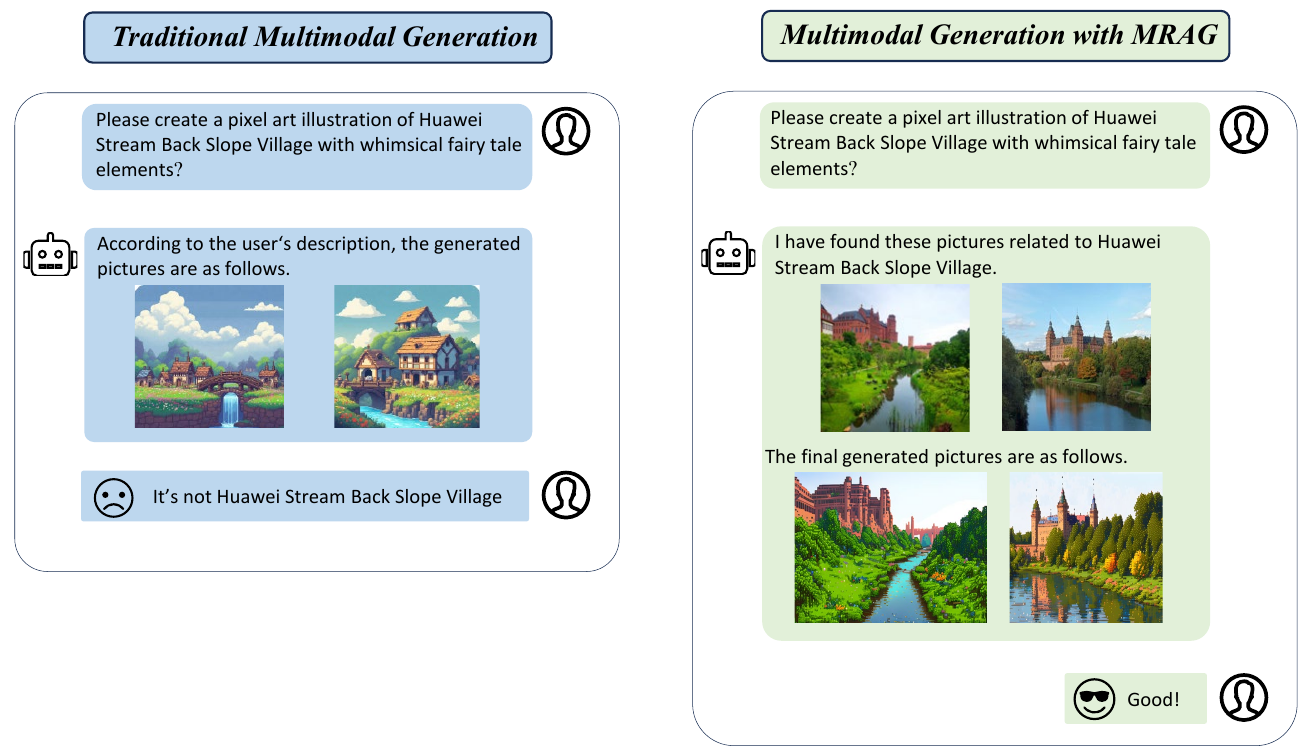}
	\caption{The user aims to generate images depicting "Huawei Stream Back Slope Village." Due to the location's obscurity and the model's limited knowledge, it may produce inaccurate representations, such as images of houses by a stream. By integrating retrieval-augmented capabilities, the model can access relevant information beforehand, enabling the generation of precise and contextually accurate images.} 
	\label{fig:mm-generation}
\end{figure}
\item\textbf{Multimodal Generation Scenario:} This primarily pertains to multimodal generation tasks, such as text-to-image or text-to-video generation. While the original MRAG framework primarily addressed understanding tasks, the new MRAG paradigm extends its capabilities by modifying multiple generation modules, unifying the solutions for both understanding and generation tasks within a single framework. Following integration, the generation scenarios are further enhanced by Retrieval-Augmentation (RA), which significantly improves the overall performance of generation tasks (see Figure \ref{fig:mm-generation}).
\item\textbf{Fusion Multimodal Output Scenario:} This scenario is distinct from those previously mentioned but represents a significant aspect of the new paradigm, warranting separate discussion. In traditional settings, the final output is typically a plain text response. However, the new paradigm enhances the generation module to produce outputs that integrate multiple modalities within a single response (e.g., combining text, images, or videos). This can be further categorized into three sub-scenarios (see Figure \ref{fig:mm output}).
    \begin{itemize}[listparindent=\parindent]
    \item \textbf{Multimodal Data is Answer:} The query can be answered directly through multimodal data without any text, as the adage "a picture is worth a thousand words" suggests.
    \item \textbf{Multimodal Data Enhances Accuracy:} The integration of multimodal data enhances the accuracy of responses, particularly in instructional contexts such as "How to register for a Gmail account.". By generating answers that interweave text and image, users can more effectively comprehend and follow the required operations.
    \item \textbf{Multimodal Data Enhances Richness:} While multimodal data is not essential, its inclusion can significantly enhance user experience. For instance, when responding to a query such as "Please introduce the Eiffel Tower.", supplementing the textual explanation with relevant images or a brief video can offer users a more engaging and visually enriched experience.
    \end{itemize}
\end{itemize}

\subsubsection{Modified Modules}
\begin{itemize}[leftmargin=1em, listparindent=\parindent]
\item \textbf{Documents Parsing and Indexing:} To minimize information loss and enhance the accuracy of document retrieval, the document parsing and indexing module has been upgraded with innovative technologies. This new approach preserves document screenshots during parsing, addressing the information loss issues inherent in previous methods. By utilizing fine-tuned MLLMs, the system vectorizes and indexes these document screenshots, enabling efficient retrieval of relevant document screenshots based on user queries. This optimization not only improves the reliability of the knowledge base but also paves the way for advanced multimodal retrieval capabilities.
\item \textbf{Generation:} Previously, the generation module relied exclusively on large models with understanding capabilities. The new paradigm integrates large models with generation capabilities, unifying reasoning and generation scenarios at the system architecture level. Additionally, by incorporating a multimodal output enhancement submodule, it facilitates a shift from text-based answers to mixed multimodal answers. The implementation methods can be categorized into two types:
    \begin{itemize}[listparindent=\parindent]
    \item \textbf{Native MLLM-Based Output:} In this task, the generation of multimodal data is entirely model-driven, eliminating the need for external data sources to supplement the model responses. The most straightforward approach involves using a unified MLLM to produce the desired multimodal output in a single step, ensuring seamless integration of diverse data types, such as text, images, or audio, within a cohesive framework. 
    \item \textbf{Augmented Multimodal Output:} This method utilizes pre-existing multimodal data to enhance textual responses. After generating the text, the system executes three sequential subtasks to create the final multimodal output:
    1) Position Identification: The system determines optimal insertion points within the text where multimodal elements (e.g., images, videos, graphs) can be integrated to complement or clarify the content. This step ensures that the multimodal data aligns contextually with the text.
    2) Candidate Set Retrieval: Relevant multimodal data is retrieved from external sources, such as the web or a knowledge base, by querying and filtering potential candidates that best match the text's context and intent.
    3) Matching and Insertion: The system selects the most appropriate multimodal element from the retrieved candidate set based on relevance, quality, and coherence. The chosen data is then seamlessly integrated into the identified positions, producing a cohesive and enriched multimodal answer.
    \end{itemize}
\end{itemize}

\subsubsection{New Modules}
\begin{itemize}[leftmargin=1em, listparindent=\parindent]
\item \textbf{Multimodal Search Planning:} This module tackles key decision-making challenges in MRAG systems by focusing on two core tasks: retrieval classification and query reformulation. Given a multimodal query $\mathcal{Q} = (q, v)$, where $q$ is the textual component and $v$ is the visual component, the module is designed to optimize information acquisition. Specifically, retrieval classification involves determining the relevance and category of the multimodal query to guide the search toward the most appropriate data sources. Query reformulation, on the other hand, refines the query by integrating textual and visual cues to improve retrieval accuracy and comprehensiveness. By combining these tasks, the module strengthens the system's ability to handle complex multimodal inputs, ensuring more effective and contextually relevant information retrieval.
    \begin{itemize}[listparindent=\parindent]
    \item \textbf{Retrieval Classification:} This task determines the optimal retrieval strategy $a^*$ from the action space $\mathcal{A} = \{a_{none}, a_{text}, a_{image}\}$ based on the current query and optionally the retrieved historical documents. The decision process is formulated as:
    \begin{equation}
    a^* = \underset{a \in \mathcal{A}}{argmax} \mathcal{F}_{RC}\left( a \mid \mathcal{Q}, \mathcal{D} \right) 
    \end{equation}
    where the retrieval control module $\mathcal{F}_{RC}$ evaluates the utility of retrieval actions by considering query characteristics, the MLLM's inherent capabilities, and, when available, the retrieved documents $\mathcal{D}$ from previous iterations. For example, in multi-hop scenarios, after retrieving visual information in the initial round, the module may leverage accumulated knowledge to determine subsequent actions, such as text-based retrieval or direct generation. Existing MRAG frameworks typically follow a rigid pipeline with predetermined retrieval actions, which poses significant limitations. Recent studies \cite{hu2024mragbench} have shown that compulsive image-to-image retrieval can be counterproductive, as retrieved images may introduce misleading information, degrading MLLM performance. This highlights the necessity of dynamic retrieval strategy selection.
    \item \textbf{Query Reformulation:} In scenarios where external information is required ($a^* \neq a_{none}$) for queries, the task of query reformulation involves generating an enhanced query ${\mathcal{Q}}^*$ by integrating visual information and, when applicable, retrieved documents from previous iterations. This process can be formulated as:
    \begin{equation}
    {\mathcal{Q}}^* = \mathcal{F}_{QR}\left(\mathcal{Q}, \mathcal{D}\right)
    \end{equation}
    where $\mathcal{F}_{QR}$ denotes the query enhancement function, which utilizes visual cues and, if available, historical retrieval results to refine the query's precision. This task is particularly critical in real-world human interactions, where queries often rely heavily on visual context and frequently employ anaphoric references. The inherent challenges of visual incompleteness and textual ambiguity pose significant obstacles to retrieving relevant information through straightforward search mechanisms.
    For complex queries that necessitate multi-hop reasoning, the enhanced query ${\mathcal{Q}}^*$ may be further decomposed into a series of atomic sub-queries $\{q^*_1, ..., q^*_n\}$. Each sub-query is meticulously formulated by considering both textual and visual contexts, as well as the accumulated knowledge from previous iterations, when relevant. This decomposition allows for a more granular and precise retrieval process, addressing the nuanced dependencies and ambiguities present in real-world queries.
    \end{itemize}
This dual-task approach optimizes information acquisition by minimizing unnecessary retrievals while maximizing the relevance of retrieved content. The structured planning framework significantly enhances the MRAG system's ability to gather comprehensive and accurate information, ensuring computational efficiency.
\end{itemize}

\section{Components \& Technologies of MRAG}
\label{sec:ComponentsTechnologiesMRAG}
In this section, we will sequentially introduce the details of the five key technical components of MRAG: 
Multimodal Document Parsing and Indexing (section\ref{sec:MultimodalDocumentParsingIndexing}), Multimodal Search Planning (section \ref{sec:MultimodalSearchPlanning}), 
Multimodal Retrieval (section \ref{sec:MultimodalRetrieval}), 
Multimodal Generation (section \ref{sec:MultimodalGeneration})

\subsection{Multimodal Document Parsing and Indexing}
\label{sec:MultimodalDocumentParsingIndexing}

MRAG systems significantly enhance the reliability and quality of generated answers, by integrating target multimodal knowledge from external multimodal knowledge bases. 
Target multimodal knowledge can be derived from various granularity in knowledge bases, including localized segments within a single document, cross-segment references within a document, or even cross-document knowledge collections.
Thus, how to effectively parse, index, and organize the multimodal documents in external knowledge bases, can largely affect the model's utilization of target multimodal knowledge, thereby determining end-to-end performance.
In this section, we first classify documents in multimodal knowledge bases according to their structure, then we provide a detailed introduction to the parsing methods and the evolution of these methods for different types of multimodal documents. Specifically, multimodal documents can be categorized into the following three types:
\begin{itemize}[leftmargin=1em, listparindent=\parindent]
\item \textbf{Unstructured Multimodal Data:} refers to various multimodal information that does not have a specific format or schema, such as text, images, videos, and audio.
Among the unstructured data, documents with images are widely studied in MRAG. For example, SlideVQA \cite{tanaka2023slidevqadatasetdocumentvisual} is a typical dataset for visual question-answering task, where all documents are input as images. 
\item \textbf{Semi-structured Multimodal Data:} mainly refers to multimodal information that lacks the rigid schema of traditional relational databases but retains some organizational features, such as PDFs, HTML, XML, and JSON.
In such documents, rule-based methods can directly extract structural characteristics. For instance, in HTML, the title can be identified using the <title> tag.
A common challenge in processing these documents is that their inherent structure, easily interpretable by humans, is often lost during parsing, resulting in information loss.
\item \textbf{Structured Multimodal Data:} refers to multimodal information arranged in a predefined format, typically following a fixed schema, such as relational databases and knowledge graphs.
The primary challenge in handling such data is formulating an accurate structured query language corresponding to natural language.
\end{itemize}

In MRAG scenarios, the primary focus is on processing and leveraging unstructured and semi-structured documents. Document parsing methods in MRAG are broadly categorized into two approaches: extraction-based and representation-based. Each approach has distinct advantages and limitations, with the choice depending on task-specific requirements such as scalability or computational efficiency.
Extraction-based methods involve a two-step process: first, multimodal information is extracted from documents, and second, the extracted data is parsed and structured for storage and downstream use.
In contrast, representation-based methods do not require explicit extraction of multimodal information. Instead, these methods focus on storing document content holistically, often employing representation techniques for document segments. This approach enables a more comprehensive processing of document content.

\subsubsection{Extraction-based}
Early document parsing solutions were entirely extractive. They evolved gradually from plain text extraction to multimodal data extraction, depending on the type of content being extracted. This subsection will present the process in this sequence.
\begin{itemize}[leftmargin=1em, listparindent=\parindent]
\item \textbf{Plain Text Extraction.}
In this phase, only textual information from all modal data in the document was extracted. For example, for tables and images, only their textual content was captured. Semi-structured documents, such as PDFs, XML, and HTML, can be parsed directly according to their structural rules. Numerous open-source tools support such capabilities, including pymupdf \cite{pymupdf} and pdfminer \cite{pdfminer} for PDF parsing, and jsoup \cite{jsoup} for HTML extraction. While this approach enables simple and efficient document parsing, it has limitations: it cannot extract multimodal information (e.g., text within images) and struggles with complex document formats. Additionally, the parsed results often suffer from significant loss of document structure information.

To enhance document parsing accuracy and address the limitations of rule-based methods in handling complex real-world documents, such as in Visual Document Understanding (VDU) tasks, OCR (Optical Character Recognition)-based approaches have become widely adopted. The traditional OCR-based document parsing pipeline consists of three main stages: text detection, text recognition, and text parsing.
Text Detection involves locating and extracting text regions from documents. Early methods primarily relied on Connected Component Analysis (CCA) \cite{chai1999significance, wagner2013ijblob} and Edge Detection algorithms \cite{maini2009study}. For more complex layouts, techniques such as Contour Analysis and Stroke Width Transform (SWT) \cite{epshtein2010detecting, tabassum2015text} were employed to handle multi-oriented text. With advancements in machine learning, hybrid models combining regression-based object detection frameworks (e.g., Faster R-CNN) with semantic segmentation networks were developed to address arbitrary-shaped text instances \cite{liao2016textboxes, zhang2016multiorientedtextdetectionfully, Baek_2019_CVPR}. This stage outputs precise bounding boxes or polygon coordinates around text elements, serving as the basis for subsequent processing.
Text Recognition converts visual text representations into machine-readable text, playing a critical role in digitizing unstructured data. Its evolution can be divided into three phases: The classical phase relied on handcrafted features \cite{pang2011efficient} and statistical models \cite{bose1994connected}, but faced challenges with fragmented processing and limited robustness. The deep learning phase introduced CNNs (Convolutional Neural Networks) for feature extraction and CTC (Connectionist Temporal Classification)/RNNs (Recurrent Neural Networks) for sequence modeling, with breakthroughs like CRNN enabling unified pipelines and improved accuracy on irregular text. The modern phase leverages transformer architectures \cite{li2023trocr}, achieving global context awareness and robustness to arbitrary-shaped text.
Text Parsing reconstructs semantic relationships through three key steps: Layout Analysis segments documents into logical components using rule-based heuristics and graph models based on spatial and typographic cues. Syntactic Parsing extracts structured data from unstructured text using regular expressions and finite-state machines. Post-processing corrects recognition errors through contextual algorithms like language model interpolation (e.g., n-gram models and dictionary lookups). This comprehensive process ensures accurate semantic reconstruction from complex documents.

However, the OCR-dependent approach has critical problems: It is not conducive to parallelization and occupies a large amount of computing resources, besides, errors in the pipeline will propagate downward through the system, affecting the overall performance. In recent years, with the development of Transformer architectures, the aforementioned issues have been effectively addressed. It enhances global context modeling through the self-attention mechanism, significantly improves processing efficiency by leveraging parallel computing, and directly maps images to structured text in an end-to-end training mode. This effectively eliminates the cumulative error issues associated with the multi-stage cascading of traditional OCR systems. LayoutLM \cite{xu2020layoutlm} uses the BERT architecture as the backbone and adds two new input embeddings: a 2-D position embedding and an image embedding to jointly model interactions between text and layout information across scanned document images. LayoutLMv2 \cite{xu2020layoutlmv2} and LayoutLMv3 \cite{huang2022layoutlmv3} further propose a new single multimodal framework to model the interaction among text, layout, and image. DocFormer \cite{appalaraju2021docformer} based on the multimodal transformer architecture proposes a novel multimodal attention layer to fuse text, vision, and spatial features in a document, thereby achieving end-to-end document parsing.
\item \textbf{Multimodal Extraction.} In this phase, the original format of multimodal data is preserved during extraction, allowing downstream tasks to autonomously determine subsequent operations. For semi-structured documents, extraction can be performed similarly using rule-based methods. Relevant multimodal data is identified through specific tags, such as extracting images from HTML files using the "<img>" tag. However, this approach faces similar challenges to plain text extraction.

The pipeline for multimodal document parsing based on OCR consists of three steps: page segmentation, text recognition, and text parsing. Page segmentation, similar to text detection in plain text extraction, locates and extracts target regions while annotating them with semantic labels (e.g., title, table, footnote). This subtask of semantic segmentation commonly employs CNN-based methods, categorized into region-based, FCN-based, and weakly supervised approaches \cite{guo2018review}. Text recognition, similar to plain text extraction, focuses on parsing text data such as titles and page text. Text parsing involves layout analysis and other operations, processing multimodal data according to downstream task requirements. 
In the era of LLMs, multimodal data is often converted into text for utilization, as seen in models like TableNet \cite{paliwal2019tablenet} for tables and UniChart \cite{masry2023unichart} for charts. This necessitates distinct models for extracting captions from different modalities. With the advancement of MLLMs, there is a trend toward unifying these models into a single MLLM framework, leveraging their robust representation capabilities \cite{liu2023mmhqa, lee2023pix2struct}. Further developments in MLLMs enable the direct retention and input of original multimodal data during generation \cite{yu2024visrag, zhang2024mr, BeyondText2024}.
\end{itemize}

\subsubsection{Representation-based}
Although extractive-based methods have been widely adopted, they suffer from several inherent limitations: (1) The parsing process is time-consuming, involves multiple steps, and requires different models for different document types; (2) Critical information, such as document structure, may be lost during extraction; and (3) Parsing errors can propagate to downstream tasks. Recent advancements in MLLMs \cite{Phi-3, PaliGemma, LLaVA} have enabled a novel approach that directly uses document screenshots as primary data for metadata indexing, addressing these issues \cite{DSE, ColPali, lee2024unifiedmultimodalinterleaveddocument, OCRHindersRAG, VisDoM}. To capture both global and local information, DSE \cite{DSE} processes the document screenshot along with its sub-images through a unified encoding framework. Additionally, a late interaction mechanism, inspired by ColBERT \cite{khattab2020colbert}, has been introduced to improve recall efficiency \cite{ColPali}. However, page-level document splitting may hinder the model's ability to capture full context and inter-part relationships. 
To address this problem, a holistic document representation method has been proposed \cite{lee2024unifiedmultimodalinterleaveddocument}, which segments large documents into passages within the token limit of MLLMs. Empirical studies reveal a performance gap between multimodal and text-only retrieval, highlighting differences in effectiveness when using raw multimodal data versus text or combined modalities \cite{OCRHindersRAG, BeyondText2024}. Consequently, a new paradigm has emerged that leverages OCR for text indexing, document screenshots for multimodal indexing, and executes textual and visual RAG in parallel. The results from both streams are then fused through modality integration to produce the final answer \cite{VisDoM}.

\subsection{Multimodal Search Planning}
\label{sec:MultimodalSearchPlanning}
Multimodal search planning refers to the strategies employed by MRAG systems, to effectively retrieve and integrate information from multiple modalities to address complex queries.
The planning can be broadly categorized into two main approaches: fixed planning and adaptive planning.

\subsubsection{Fixed Planning}
Early MRAG systems typically adopt fixed planning strategies for handling multimodal queries, characterized by predetermined processing pipelines that lack flexibility in adapting to diverse query requirements. These approaches can be broadly categorized based on their retrieval modality choices:
\begin{itemize}[leftmargin=1em, listparindent=\parindent]
\item \textbf{Planning for Single-modal Retrieval.} Early fixed planning strategies usually focus on a single modality for retrieval, despite the multimodal nature of input queries. These approaches can be broadly classified into text-centric and image-centric paradigms, reflecting initial efforts to adapt traditional IR query processing techniques \cite{li_ragsurvey, lee2020learning} to the multimodal domain.
    \begin{itemize}[listparindent=\parindent]
    \item \textbf{Text-centric} planning approaches prioritize textual retrieval by transforming multimodal queries into text-only formats. For instance, Plug-and-Play \cite{plugandplay} employs vision-language models to convert the visual component of a query into textual descriptions, followed by text-based retrieval planning. This strategy simplifies the multimodal problem into a traditional text-based RAG pipeline, leveraging established multi-stage query processing techniques from conventional IR systems. However, this approach often introduces a semantic gap between the user's original intent and the generated textual descriptions. The conversion of visual queries to text may fail to precisely capture the user's specific information needs, leading to the retrieval of irrelevant or noisy documents that diverge from the query's focus.
    
    \item \textbf{Image-centric} planning strategies rely solely on image-based retrieval regardless of the query characteristics. Systems such as Wiki-LLaVA \cite{wikillava} demonstrate this paradigm by consistently triggering image retrieval from knowledge bases for multimodal queries. While this approach ensures visual information preservation, it presents practical limitations. Recent empirical studies \cite{hu2024mragbench} highlight that compulsive image retrieval can be counterproductive, particularly when textual information suffices or when retrieved images introduce misleading visual contexts, impairing MLLM performance.
    \end{itemize}
The inflexibility of single-modality planning strategies highlights their inherent limitations: they cannot adapt to the diverse information needs of real-world scenarios. For example, while a text-centric approach may be suitable for queries referencing visual content but focused on factual information, an image-centric strategy is more effective for queries requiring detailed visual comparisons.

\item \textbf{Planning for Multimodal Retrieval.} Recent studies have begun investigating the use of multimodal information retrieval to enhance the performance of MRAG systems. Unlike single-modality approaches, these methods integrate both textual and visual knowledge sources, albeit through fixed processing pipelines. For instance, MMSearch \cite{jiang2024mmsearch} employs a rigid multimodal planning pipeline, mandating Google Lens image searches for all image-containing queries. This is followed by a "Requery" phase, where LLMs reformulate the search query using the original query, image, and Google Lens results. While this structured approach ensures systematic information retrieval, its inflexible design often leads to unnecessary image searches, increasing computational overhead when visual information is irrelevant to the query.
\end{itemize}
Fixed pipeline approaches, whether single-modality or multimodality, exhibit several critical limitations. First, their rigid retrieval strategies struggle to adapt to the diverse nature of real-world queries, where the optimal retrieval modality depends on specific information needs. Second, mandatory retrieval operations often introduce redundant or irrelevant information, particularly when certain knowledge types are unnecessary for addressing the query. Third, these approaches incur significant computational overhead, especially in multimodal pipelines handling large-scale knowledge bases. As highlighted by mR$^2$AG \cite{zhang2024mr}, a more fundamental issue is that not all queries require external knowledge retrieval. Current MRAG systems frequently perform retrieval indiscriminately, resulting in unnecessary computational costs and potential noise in response generation. These limitations emphasize the need to transition from predetermined pipelines to adaptive planning mechanisms that dynamically adjust retrieval strategies based on query characteristics and intermediate results.

\subsubsection{Adaptive Planning}
Recent studies have highlighted two key limitations in fixed pipeline approaches \cite{li2024benchmarkingmultimodalretrievalaugmented}: 
1) Non-adaptive Retrieval Queries: inflexible retrieval strategies that fail to adjust to evolving contexts or intermediate results; and 
2) Overloaded Retrieval Queries: concatenating visual content descriptions with input questions into a single query, leading to ambiguous retrievals and irrelevant knowledge. 
To address these issues, OmniSearch \cite{li2024benchmarkingmultimodalretrievalaugmented} introduces a self-adaptive planning agent for multimodal retrieval, mimicking human problem-solving behavior. Instead of relying on a fixed pipeline, the system dynamically breaks down complex multimodal questions into sub-question chains with retrieval actions. At each step, the agent adapts its next action based on the problem-solving state and retrieved content, enabling deeper understanding of retrieved information and adaptive refinement of retrieval strategies.
CogPlanner \cite{yu2025unveiling} iteratively refines queries and selects retrieval strategies, enabling both parallel and sequential modeling approaches.

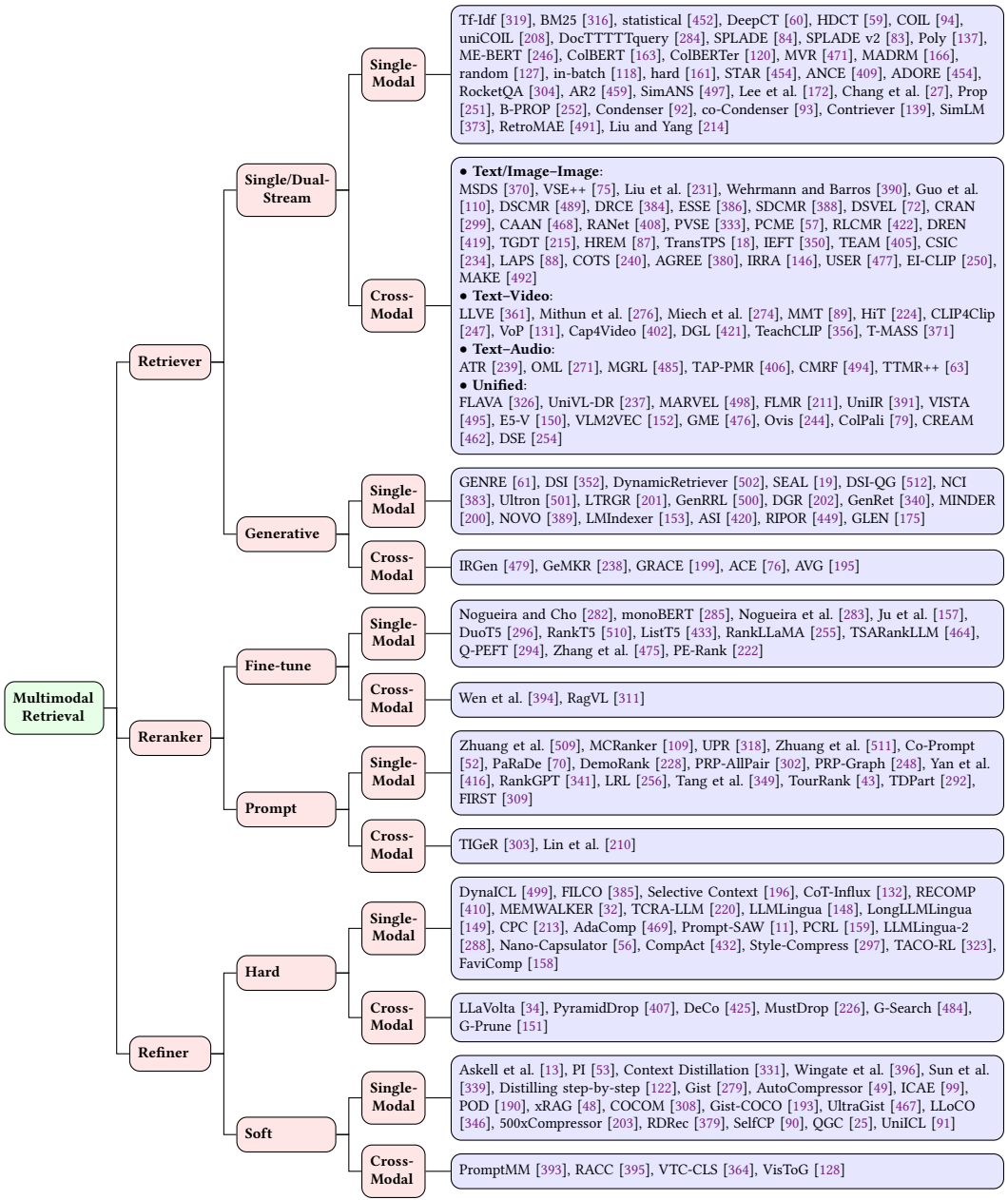
\begin{figure}[htbp]
\centering
\begin{forest}
for tree={
  grow=east, 
  parent anchor=east,
  child anchor=west,
  calign=edge midpoint,
  edge path={
    \noexpand\path[\forestoption{edge}]
      (!u.parent anchor) -- +(5pt,0) |- (.child anchor)\forestoption{edge label};
  },
  l sep=10pt, 
  s sep=5pt,  
  font=\tiny
},
where level=0{
  where n children=0{
      rectangle, rounded corners, draw=black, fill=blue!10, text width=1.15cm, align=center, 
    }{
      rectangle, rounded corners, draw=black, fill=green!10, text width=1.15cm, align=center, 
    }
}{},
where level=1{
  where n children=0{
      rectangle, rounded corners, draw=black, fill=blue!10, text width=0.9cm, align=center, 
    }{
      rectangle, rounded corners, draw=black, fill=red!10, text width=0.9cm, align=center, 
    }
}{},
where level=2{
  where n children=0{
      rectangle, rounded corners, draw=black, fill=blue!10, text width=1.15cm, align=center, 
    }{
      rectangle, rounded corners, draw=black, fill=red!10, text width=1.15cm, align=center, 
    }
}{},
where level=3{
  where n children=0{
      rectangle, rounded corners, draw=black, fill=blue!10, text width=0.65cm, align=center, 
    }{
      rectangle, rounded corners, draw=black, fill=red!10, text width=0.65cm, align=center, 
    }
}{},
where level=4{
  where n children=0{
      rectangle, rounded corners, draw=black, fill=blue!10, text width=0.65cm, align=center, 
    }{
      rectangle, rounded corners, draw=black, fill=red!10, text width=0.65cm, align=center, 
    }
}{},
[\textbf{Multimodal}\\\textbf{Retrieval}
    [\textbf{Refiner}
        [\textbf{Soft}
            [\textbf{Cross-}\\\textbf{Modal}
                [{\parbox{7.5cm}{PromptMM \cite{wei2024promptmm}, RACC \cite{weng2024learning}, VTC-CLS \cite{wang2024cls}, VisToG \cite{huang2024efficient}
                }}, rectangle, rounded corners, draw=black, fill=blue!10, text width=7.5cm, align=left]
            ]
            [\textbf{Single-}\\\textbf{Modal}
                [{\parbox{7.5cm}{\citet{askell2021general}, PI \cite{choi2022prompt}, Context Distillation \cite{snell2022learning}, \citet{wingate2022prompt}, \citet{sun2023instruction}, Distilling step-by-step \cite{hsieh2023distilling}, Gist \cite{mu2023learning}, AutoCompressor \cite{chevalier2023adapting}, ICAE \cite{ge2023context}, POD \cite{li2023prompt}, xRAG \cite{cheng2024xrag}, COCOM \cite{rau2024context}, Gist-COCO \cite{li2024say}, UltraGist \cite{zhang2024compressing}, LLoCO \cite{tan2024lloco}, 500xCompressor \cite{li2024500xcompressor}, RDRec \cite{wang2024rdrec}, SelfCP \cite{gao2024selfcp}, QGC \cite{cao2024retaining}, UniICL \cite{gao2024unifying}
                }}, rectangle, rounded corners, draw=black, fill=blue!10, text width=7.5cm, align=left]
            ]
        ]
        [\textbf{Hard}
            [\textbf{Cross-}\\\textbf{Modal}
                [{\parbox{7.5cm}{LLaVolta \cite{chen2024efficient}, PyramidDrop \cite{xing2024pyramiddrop}, DeCo \cite{yao2024deco}, MustDrop \cite{liu2024multi}, G-Search \cite{zhao2024accelerating}, G-Prune \cite{jiang2025kind}
                }}, rectangle, rounded corners, draw=black, fill=blue!10, text width=7.5cm, align=left]
            ]
            [\textbf{Single-}\\\textbf{Modal}
                [{\parbox{7.5cm}{DynaICL \cite{zhou2023efficient}, FILCO \cite{wang2023learning}, Selective Context \cite{li2023compressing}, CoT-Influx \cite{huang2023fewer}, RECOMP \cite{xu2023recomp}, MEMWALKER \cite{chen2023walking}, TCRA-LLM \cite{liu2023tcra}, LLMLingua \cite{jiang2023llmlingua}, LongLLMLingua \cite{jiang2023longllmlingua}, CPC \cite{liskavets2024prompt}, AdaComp \cite{zhang2024adacomp}, Prompt-SAW \cite{ali2024prompt}, PCRL \cite{jung2024discrete}, LLMLingua-2 \cite{pan2024llmlingua}, Nano-Capsulator \cite{chuang2024learning}, CompAct \cite{yoon2024compact}, Style-Compress \cite{pu2024style}, TACO-RL \cite{shandilya2024taco}, FaviComp \cite{jung2024familiarity}
                }}, rectangle, rounded corners, draw=black, fill=blue!10, text width=7.5cm, align=left]
            ]
        ]
    ]
    [\textbf{Reranker}
        [\textbf{Prompt}
            [\textbf{Cross-}\\\textbf{Modal}
                [{\parbox{7.5cm}{TIGeR \cite{qu2024unified}, \citet{lin2024mm}
                }}, rectangle, rounded corners, draw=black, fill=blue!10, text width=7.5cm, align=left]
            ]
            [\textbf{Single-}\\\textbf{Modal}
                [{\parbox{7.5cm}{
                \citet{zhuang2023beyond}, MCRanker \cite{guo2024generating}, UPR \cite{sachan2022improving}, \citet{zhuang2023open}, Co-Prompt \cite{cho2023discrete}, PaRaDe \cite{drozdov2023parade}, DemoRank \cite{liu2024demorank}, PRP-AllPair \cite{qin2023large}, PRP-Graph \cite{luo2024prp}, \citet{yan2024consolidating}, RankGPT \cite{sun2023chatgpt}, LRL \cite{ma2023zero}, \citet{tang2023found}, TourRank \cite{chen2024tourrank}, TDPart \cite{parry2024top}, FIRST \cite{reddy2024first}
                }}, rectangle, rounded corners, draw=black, fill=blue!10, text width=7.5cm, align=left]
            ]
        ]
        [\textbf{Fine-tune}
            [\textbf{Cross-}\\\textbf{Modal}
                [{\parbox{7.5cm}{\citet{wen2024multimodal}, RagVL \cite{retrievalmllm}
                }}, rectangle, rounded corners, draw=black, fill=blue!10, text width=7.5cm, align=left]
            ]
            [\textbf{Single-}\\\textbf{Modal}
                [{\parbox{7.5cm}{
                \citet{nogueira2019passage}, monoBERT \cite{nogueira2019multi}, \citet{nogueira2020document}, \citet{ju2021text}, DuoT5 \cite{pradeep2021expando}, RankT5 \cite{zhuang2023rankt5}, ListT5 \cite{yoon2024listt5}, RankLLaMA \cite{ma2024fine}, TSARankLLM \cite{zhang2023two}, Q-PEFT \cite{peng2024q}, \citet{zhang2023rank}, PE-Rank \cite{liu2024leveraging}
                }}, rectangle, rounded corners, draw=black, fill=blue!10, text width=7.5cm, align=left]
            ]
        ]
    ]
    [\textbf{Retriever}
        [\textbf{Generative}
            [\textbf{Cross-}\\\textbf{Modal}
                [{\parbox{7.5cm}{IRGen \cite{zhang2024irgen}, GeMKR \cite{long2024generative}, GRACE \cite{li2024generative}, ACE \cite{fang2024ace}, AVG \cite{li2024revolutionizing}
                }}, rectangle, rounded corners, draw=black, fill=blue!10, text width=7.5cm, align=left]
            ]
            [\textbf{Single-}\\\textbf{Modal}
                [{\parbox{7.5cm}{GENRE \cite{de2020autoregressive}, DSI \cite{tay2022transformer}, DynamicRetriever \cite{zhou2023dynamicretriever}, SEAL \cite{bevilacqua2022autoregressive}, DSI-QG \cite{zhuang2022bridging}, NCI \cite{wang2022neural}, Ultron \cite{zhou2022ultron}, LTRGR \cite{li2024learning}, GenRRL \cite{zhou2023enhancing}, DGR \cite{li2024distillation}, GenRet \cite{sun2024learning}, MINDER \cite{li2023multiview}, NOVO \cite{wang2023novo}, LMIndexer \cite{jin2023language}, ASI \cite{yang2023auto}, RIPOR \cite{zeng2024scalable}, GLEN \cite{lee2023glen}
                }}, rectangle, rounded corners, draw=black, fill=blue!10, text width=7.5cm, align=left]
            ]
        ]
        [\textbf{Single/Dual-}\\\textbf{Stream}
            [\textbf{Cross-}\\\textbf{Modal}
                [{\parbox{7.5cm}{$\bullet$ \textbf{Text/Image–Image}:\\
                MSDS \cite{wang2015image}, VSE++ \cite{faghri2017vse++}, \citet{liu2017learning}, \citet{wehrmann2018bidirectional}, \citet{guo2019learning}, DSCMR \cite{zhen2019deep}, DRCE \cite{wang2023dual}, ESSE \cite{wang2024estimating}, SDCMR \cite{wang2024semantics}, DSVEL \cite{engilberge2018finding}, CRAN \cite{qi2018cross}, CAAN \cite{zhang2020context}, RANet \cite{xiong2024reference}, PVSE \cite{song2019polysemous}, PCME \cite{chun2021probabilistic}, RLCMR \cite{yang2021rethinking}, DREN \cite{yang2022dual}, TGDT \cite{liu2023efficient}, HREM \cite{fu2023learning}, TransTPS \cite{bao2023multi}, IEFT \cite{tang2023interacting}, TEAM \cite{xie2022token}, CSIC \cite{liu2022image}, LAPS \cite{fu2024linguistic}, COTS \cite{lu2022cots}, AGREE \cite{wang2023agree}, IRRA \cite{jiang2023cross}, USER \cite{zhang2024user}, EI-CLIP \cite{ma2022ei}, MAKE \cite{zheng2023make}\\
                $\bullet$ \textbf{Text–Video}:\\
                LLVE \cite{torabi2016learning}, \citet{mithun2018learning}, \citet{miech2019howto100m}, MMT \cite{gabeur2020multi}, HiT \cite{liu2021hit}, CLIP4Clip \cite{luo2022clip4clip}, VoP \cite{huang2023vop}, Cap4Video \cite{wu2023cap4video}, DGL \cite{yang2024dgl}, TeachCLIP \cite{tian2024holistic}, T-MASS \cite{wang2024text}\\
                $\bullet$ \textbf{Text–Audio}:\\
                ATR \cite{lou2022audio}, OML \cite{mei2022metric}, MGRL \cite{zhao2023multi}, TAP-PMR \cite{xin2023improving}, CMRF \cite{zhou2024cross}, TTMR++ \cite{doh2024enriching}\\
                $\bullet$ \textbf{Unified}:\\
                FLAVA \cite{singh2022flava}, UniVL-DR \cite{liu2022universal}, MARVEL \cite{zhou2023marvel}, FLMR \cite{lin2023fine}, UniIR \cite{wei2024uniir}, VISTA \cite{zhou2024vista}, E5-V \cite{jiang2024e5}, VLM2VEC \cite{jiang2024vlm2vec}, GME \cite{zhang2024gme}, Ovis \cite{lu2024ovis}, ColPali \cite{faysse2024colpali}, CREAM \cite{zhang2024cream}, DSE \cite{ma2024unifying}
                }}, rectangle, rounded corners, draw=black, fill=blue!10, text width=7.5cm, align=left]
            ]    
            [\textbf{Single-}\\\textbf{Modal}
                [{\parbox{7.5cm}{
                Tf-Idf \cite{salton1988term}, BM25 \cite{robertson1995okapi}, statistical \cite{zhai2008statistical}, DeepCT \cite{dai2020context}, HDCT \cite{dai2020context1}, COIL \cite{gao2021coil}, uniCOIL \cite{lin2021few}, DocTTTTTquery \cite{nogueira2019doc2query}, SPLADE \cite{formal2021splade}, SPLADE v2 \cite{formal2021spladev2}, Poly \cite{humeau2019poly}, ME-BERT \cite{luan2021sparse}, ColBERT \cite{khattab2020colbert}, ColBERTer \cite{hofstatter2022introducing}, MVR \cite{zhang2022multi}, MADRM \cite{kong2022multi}, random \cite{huang2020embedding}, in-batch \cite{henderson2017efficient}, hard \cite{karpukhin2020dense}, STAR \cite{zhan2021optimizing}, ANCE \cite{xiong2020approximate}, ADORE \cite{zhan2021optimizing}, RocketQA \cite{qu2020rocketqa}, AR2 \cite{zhang2021adversarial}, SimANS \cite{zhou2022simans}, \citet{lee2019latent}, \citet{chang2020pre}, Prop \cite{ma2021prop}, B-PROP \cite{ma2021b}, Condenser \cite{gao2021condenser}, co-Condenser \cite{gao2021unsupervised}, Contriever \cite{izacard2021towards}, SimLM \cite{wang2022simlm}, RetroMAE \cite{zheng2022retromae}, \citet{liu2022masked}
                }}, rectangle, rounded corners, draw=black, fill=blue!10, text width=7.5cm, align=left]
            ]
        ]
    ]
]
\end{forest}
\caption{Taxonomy of recent advancements in multimodal retrieval research.}
\label{MultimodalRetrieval}
\end{figure}

\subsection{Multimodal Retrieval}
\label{sec:MultimodalRetrieval}
In this section, we present a comprehensive overview of the three critical components of multimodal retrieval in the MRAG system: retriever (section \ref{sec:RETRIEVER}), reranker, and refiner. Each component plays a distinct yet interconnected role in enhancing the quality and relevance of information retrieval and utilization for LLMs. We summarize the taxonomy of multimodal retrieval research in Figure \ref{MultimodalRetrieval}.

\begin{figure}[htbp]
\centering
\includegraphics[width=0.95\textwidth]{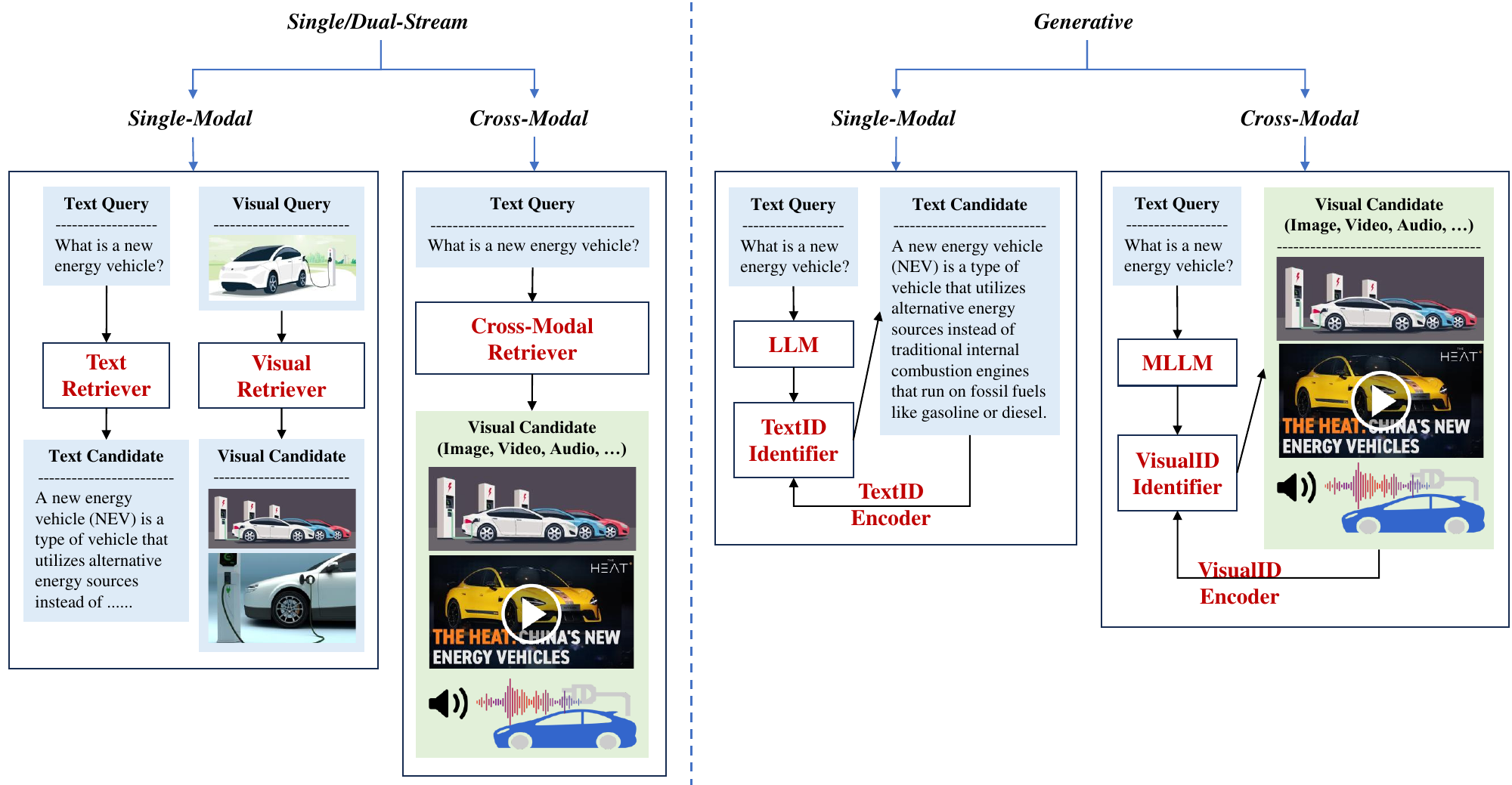}
\caption{The architectures of retriever in multimodal retrieval. } 
\label{retrieval-retriever}
\end{figure}

\subsubsection{RETRIEVER}
\label{sec:RETRIEVER}
The retriever is a core component that sources relevant documents from a large external knowledge base using advanced indexing and search algorithms. It retrieves candidate information aligned with user queries, aiming to provide a broad yet relevant set of documents to support high-quality LLM responses. Its performance is crucial, as it directly influences the quality of the downstream retrieval pipeline. As shown in Figure \ref{retrieval-retriever}, existing retrieval methods fall into two categories based on architecture: Single/Dual-stream Structure and Generative Structure, each involves single-modal (e.g., text, images) and cross-modal information retrieval.
\begin{itemize}[leftmargin=1em, listparindent=\parindent]
\item \textbf{Single/Dual-stream Structure}: The Single-stream Structure integrates multimodal fusion modules to model image-text relationships in a unified semantic space, capturing fine-grained interactions but incurring higher computational costs and slower inference, limiting scalability for large-scale multimodal retrieval tasks in real-world applications. In contrast, the Dual-stream Structure uses separate vision and language streams, leveraging contrastive learning to align global features in a shared semantic space efficiently. However, it lacks explicit multimodal interaction and struggles with feature alignment due to information imbalance, exacerbated by the brevity of dataset captions.
    \begin{itemize}[listparindent=\parindent]
    \item \textbf{Retrieval for Single-Modal.} In MRAG systems, single-modal retrieval focuses on text and image retrieval. Text retrieval uses NLP techniques to extract relevant information from datasets, identifying contextually aligned documents. Image retrieval employs computer vision algorithms and feature extraction methods to encode visual data into high-dimensional vectors for similarity matching. Both modalities are essential for enhancing MRAG system performance.
        \begin{itemize}[listparindent=\parindent]
        \item \textbf{Text-centric.} 
        Text retrieval, a core component of information retrieval (IR), identifies relevant textual information from large corpora or web resources in response to user queries. It is widely used in downstream applications such as question answering \cite{karpukhin2020dense, qu2020rocketqa}, dialogue systems \cite{song2018ensemble, yu2021few, zhang2024usimagent}, web search \cite{gao2021condenser, mei2022learning, mei2023improving}, and retrieval-augmented generation systems \cite{chen2024benchmarking, siriwardhana2023improving, chen2025improving, gong2024cosearchagent}. Recent advancements categorize text retrieval methods into two types: sparse retrieval and dense retrieval.
            \begin{itemize}[listparindent=\parindent]
            \item \textbf{Sparse Text Retrieval.} Early research in text retrieval focused on extracting representative terms from documents, leading to the development of vector space models \cite{salton1975vector} based on the "bag-of-words" assumption, which represents documents and queries as sparse term vectors, ignoring term order. Term weighting methods like tf-idf \cite{salton1988term, aizawa2003information, robertson2004understanding} and BM25 models \cite{robertson1995okapi, robertson2009probabilistic} were introduced to assign weights based on term importance within and across corpora, while inverted indexes \cite{zobel2006inverted} improved retrieval efficiency by organizing corpora into term-document ID pairs. Statistical language modeling \cite{zhai2008statistical} further advanced retrieval by estimating term probability distributions for probabilistic ranking. 
            However, early sparse retrieval methods face limitations, such as assuming term independence and relying on lexical matching, which hinders their ability to capture contextual term importance or semantic relationships between terms. Consequently, these methods struggle to understand deeper textual meanings and contextual relevance between queries and documents.

            Recent advancements in sparse retrieval models have been driven by the integration of pre-trained language models (PLMs). While these approaches leverage PLMs, they remain fundamentally rooted in lexical matching, enabling the reuse of traditional sparse index structures by incorporating auxiliary information such as contextualized embeddings \cite{gao2021coil, lin2021few} and extended tokens \cite{nogueira2019doc2query, formal2021splade, formal2021spladev2}. This research domain focuses on two main approaches: term weighting and term expansion. 
            Term weighting enhances relevance estimation by leveraging context-specific token representations. DeepCT \cite{dai2020context} and HDCT \cite{dai2020context1} use learned token representations to estimate the context-specific importance of terms within passages, while COIL \cite{gao2021coil} and uniCOIL \cite{lin2021few} employ contextualized token representations of exact matching terms to compute relevance via dot products and summed similarity scores.
            Term expansion mitigates vocabulary mismatch by expanding queries or documents using PLMs. For instance, DocTTTTTquery \cite{nogueira2019doc2query} predicts relevant queries for documents to enrich the document’s content, while SPLADE \cite{formal2021splade} and SPLADEv2 \cite{formal2021spladev2} project terms onto vocabulary-sized weight vectors derived from masked language model logits. These vectors, aggregated via methods like summing or max pooling, effectively expand content by incorporating absent terms. Sparsity regularization ensures efficient sparse representations for inverted index usage.

            In summary, sparse retrieval models achieve an optimal balance in cross-domain transfer, retrieval efficiency, and overall effectiveness.
            \item \textbf{Dense Text Retrieval.} Recent advancements in deep learning \cite{cho2014learning, lecun1995convolutional, krizhevsky2012imagenet, abdel2014convolutional, hinton2012deep, graves2012long, cho2014properties}, particularly pre-trained language models (PLMs) \cite{devlin2018bert, liu2019roberta, brown2020language} based on the Transformer architecture \cite{vaswani2017attention, fedus2022switch}, have increasingly adopted dense vector embeddings in low-dimensional Euclidean spaces for modeling semantic relationships between queries and documents. These embeddings enable relevance measurement through Euclidean distances or inner products. Dense retrieval methods have demonstrated strong performance across various information retrieval tasks \cite{karpukhin2020dense, khattab2020colbert, mei2023improving}. Additionally, Approximate Nearest Neighbor Search (ANNS) algorithms \cite{ge2013optimized, jegou2010product, johnson2019billion}, particularly quantization-based methods \cite{jegou2010product, ge2013optimized} and their retrieval-oriented variants \cite{zhan2021jointly, zhan2022learning, zhang2021joint, yamada2021efficient}, enable efficient retrieval of top-ranked documents from large collections using precomputed ANNS indices. Dense retrieval techniques primarily focus on two key aspects: model architecture and training methods.

            For model architecture,  dense retrieval methods employ a two-tower architecture to balance retrieval efficiency and effectiveness by modeling semantic interactions between queries and documents through their representations. These methods vary in representation granularity, primarily falling into two categories: single-vector and multi-vector representations.
            Then, the relevance scores are computed using similarity functions (e.g., cosine similarity, inner product) between these embeddings. A common technique involves placing a special token (e.g., “[CLS]”) at the beginning of a text sequence, with its learned representation capturing the overall semantics. The existing dense retrieval models learn the query and document representations by fine-tuning PLMs like BERT \cite{devlin2018bert}, RoBERTa \cite{liu2019roberta}, or Mamba \citet{gu2023mamba, zhang2024mamba}, or large language models (LLMs) like RepLLaMA \cite{ma2024fine} on annotated datasets (e.g., MSMARCO \cite{nguyen2016ms}, BEIR \cite{thakur2021beir}). However, single-vector bi-encoders struggle to model fine-grained semantic interactions between queries and documents.
            To address this limitation, multi-vector representation enhance text representation and semantic interaction by employing multiple-representation bi-encoders. The Poly-encoder \cite{humeau2019poly} generates multiple context codes to capture text semantics from multiple views. ME-BERT \cite{luan2021sparse} produces $m$ representations for a candidate text using the contextualized embeddings of the first $m$ tokens. ColBERT \cite{khattab2020colbert} maintains per-token contextualized embeddings with a late interaction mechanism. ColBERTer \cite{hofstatter2022introducing} extends ColBERT by combining single- (“[CLS]”) and multi-representation (per-token) mechanisms for better performance. MVR \cite{zhang2022multi} introduces multiple “[VIEW]” tokens to learn diverse representations, with a local loss to identify the best-matched view. MADRM \cite{kong2022multi} learns multiple aspect embeddings for queries and texts, supervised by explicit aspect annotations.

            For training method, to achieve optimal retrieval performance, dense retrieval models are typically trained using two key techniques: negative sampling and pretraining.
            Negative sampling focuses on selecting high-quality negatives to compute the negative log-likelihood loss used for training dense retrieval models. Basic methods include random sampling \cite{huang2020embedding} and in-batch negatives \cite{henderson2017efficient, karpukhin2020dense, qu2020rocketqa}, which increase the number of negatives within memory limits but do not guarantee the inclusion of hard negatives, i.e., irrelevant texts with high semantic similarity to the query. Hard negatives are critical for improving the model's ability to distinguish relevant from irrelevant texts.
            Various approaches have been proposed to incorporate hard negatives. BM25-retrieved documents are used as static hard negatives \cite{gao2021complement, karpukhin2020dense}. STAR \cite{zhan2021optimizing} combines static hard negatives with random negatives, while ANCE \cite{xiong2020approximate} retrieves hard negatives using a warm-up dense retrieval model and refreshes the document index during training. ADORE \cite{zhan2021optimizing} employs an adaptive query encoder to retrieve top-ranked texts as hard negatives, keeping the text encoder and document index fixed. However, hard negatives may include false negatives, introducing noise that can degrade performance. RocketQA \cite{qu2020rocketqa} addresses this by using a cross-encoder to filter out likely false negatives. AR2 \cite{zhang2021adversarial} integrates a dual-encoder retriever with a cross-encoder ranker, jointly optimized through a minimax adversarial objective to produce harder negatives and improve the retriever.
            SimANS \cite{zhou2022simans} introduces the concept of sampling ambiguous negatives, i.e., texts ranked near positives with moderate similarity to the query. These negatives are more informative and less likely to be false negatives, further enhancing model performance.

            Pretraining aims to learn universal semantic representations that generalize to downstream dense retrieval tasks. To enhance the modeling capacity of PLMs, self-supervised pretraining tasks, such as those proposed by \citet{lee2019latent} (selecting random sentences as queries) and \citet{chang2020pre} (leveraging hyperlinks for constructing query-passage pairs), mimic retrieval objectives. Prop \cite{ma2021prop} and B-PROP \cite{ma2021b} use document language models (e.g., unigram, BERT) to sample word sets, training PLMs to predict pairwise preferences. To enhance dense retrieval models, studies focus on improving the “[CLS]” token embedding. Condenser \cite{gao2021condenser} aggregates global text information for masked token recovery, while co-Condenser \cite{gao2021unsupervised} adds a query-agnostic contrastive loss to cluster related text segments while distancing unrelated ones. Contriever \cite{izacard2021towards} generates positive pairs by sampling two spans from the same text and negatives using in-batch and cross-batch texts. 
            Following with an unbalanced architecture (strong encoder, simple decoder), SimLM \cite{wang2022simlm} pretrains the encoder and decoder with replaced language modeling, recovering original tokens after replacement. It further optimizes the retriever through hard negative training and cross-encoder distillation. RetroMAE \cite{zheng2022retromae} utilizes a high masking ratio for the decoder and a standard ratio for the encoder, incorporating an enhanced decoding mechanism with two-stream and position-specific attention masks. \citet{liu2022masked} introduces a two-stage pretraining approach, combining general-corpus pretraining with domain-specific continual pretraining, achieving strong benchmark performance.
            \end{itemize}
        \end{itemize}
    However, single-modal retrieval is inherently limited by its inability to capture cross-modal relationships, which underscores the importance of integrating multimodal retrieval strategies to bridge textual and visual semantics for more comprehensive information retrieval and generation.
    \item \textbf{Retrieval for Cross-modal.} Cross-modal retrieval enables the identification of relevant data in one modality (e.g., images) using a query from another (e.g., text). It enhances MRAG systems by facilitating the retrieval and generation of information across diverse modalities, including text, images, audio, and video.
    \begin{itemize}[listparindent=\parindent]
        \item \textbf{Text–Image Retrieval.} Text–Image Retrieval aims to match images with corresponding textual queries by leveraging multimodal data co-occurrence, such as paired text-image instances or manual annotations, to capture semantic correlations. Existing methods can be categorized into three groups: CNN/RNN-based approaches, Transformer-based techniques, and Vision-Language Pretraining (VLP) model-based methods.

        Early CNN/RNN-based methods \cite{wang2015image, liu2017learning, faghri2017vse++, wehrmann2018bidirectional, lee2018stacked, guo2019learning} extract features from each modality separately using MLP, CNN, and RNN, enforcing cross-modal constraints through positive/negative sample construction. MSDS \cite{wang2015image} uses CNN with a maximum likelihood-based scheme for image-text relevance. VSE++ \cite{faghri2017vse++} combines CNN and RNN with hard sample mining in ranking loss. Advances include residual learning \cite{liu2017learning}, character-level convolution \cite{wehrmann2018bidirectional}, and disentangled representation \cite{guo2019learning} for improved feature mapping and retrieval. DSCMR \cite{zhen2019deep} maps multimodal data into a shared space using modality-specific networks and fully connected layers, leveraging label constraints and pairwise loss for discriminant learning.
        Recent CNN/RNN-based methods improve image-text matching by addressing key challenges. DRCE \cite{wang2023dual} enhances rare content representation and association using a dual-path structure, adaptive fusion, and reranking to mitigate long-tail issues. ESSE \cite{wang2024estimating} tackles one-to-many correspondence by projecting data as sectors with uncertainty apertures. SDCMR \cite{wang2024semantics} employs diverse CNNs for multimodal feature extraction and a dual adversarial mechanism to isolate semantic-shared features, ensuring retrieval consistency. These methods collectively advance cross-modal retrieval robustness and accuracy.
        Spatial attention \cite{huang2017instance, engilberge2018finding, qi2018cross, zhang2020context, xiong2024reference} is widely used in CNN/RNN-based cross-modal retrieval to uncover fine-grained associations by generating weighted masks for local regions, enhancing key features while suppressing irrelevant ones. DSVEL \cite{engilberge2018finding} employs spatial-aware pooling to align image regions with text, while CRAN \cite{qi2018cross} and CAAN \cite{zhang2020context} improve global-local alignment through relation alignment and context-aware selection. RANet \cite{xiong2024reference} refines attention mechanisms with reference attention to reduce incorrect scores and adaptive aggregation to amplify relevant information and minimize redundancy.

        Transformer-based methods \cite{song2019polysemous, chun2021probabilistic, yang2022dual, liu2023efficient, fu2023learning, bao2023multi, yang2021rethinking, zeng2022comprehensive, tang2023interacting} leverage multi-head self-attention to encode multimodal relationships and optimize modality-specific encoders, demonstrating superior performance in multimodal modeling and cross-modal retrieval tasks.
        Recent advancements in multimodal representation learning have focused on enhancing Transformer architectures and feature alignment. PVSE \cite{song2019polysemous} integrates self-attention and residual learning, while PCME \cite{chun2021probabilistic} uses probabilistic embeddings to model one-to-many and many-to-many correlations. RLCMR \cite{yang2021rethinking} tokenizes multimodal data and trains with a unified Transformer encoder for cross-modal semantic correlation. DREN \cite{yang2022dual} refines feature representation through character-level and context-driven augmentation. TGDT \cite{liu2023efficient} unifies coarse- and fine-grained learning with multimodal contrastive loss for feature alignment. HREM \cite{fu2023learning} improves image-text matching by capturing multi-level intra- and inter-modal relationships. TransTPS \cite{bao2023multi} extends Transformers with cross-modal multi-granularity matching and contrastive loss for better feature distinction. IEFT \cite{tang2023interacting} models text-image pairs as unified entities to model their intrinsic correlation.

        With the rapid advancement of pretraining paradigms, Vision-Language Pretraining (VLP) models \cite{li2020oscar, chen2020uniter, li2020unicoder, huang2021seeing, jia2021scaling, radford2021learning, devlin2018bert, dosovitskiy2020image, gu2022wukong}, including both single- and dual-stream architectures, have leveraged large-scale visual-linguistic datasets for joint pretraining. Researchers have utilized the strong representational capabilities \cite{ma2022ei, liu2022image, xie2022token, fu2024linguistic, lu2022cots, wang2023agree, jiang2023cross, zhang2024user, zheng2023make} of VLP models to significantly enhance cross-modal retrieval performance.
        Single-stream models like TEAM \cite{xie2022token} align multimodal token embeddings for token-level matching, while dual-stream approaches such as COTS \cite{lu2022cots} integrate contrastive learning with token- and task-level interactions. Methods like CSIC \cite{liu2022image} and LAPS \cite{fu2024linguistic} improve multimodal alignment by quantifying semantic significance and associating patch features with words, respectively. AGREE \cite{wang2023agree} fine-tunes and reranks cross-modal entities to harmonize their alignment. IRRA \cite{jiang2023cross} employs text-specific mask mechanism to capture fine-grained intra- and inter-modal relationships. USER \cite{zhang2024user}, EI-CLIP \cite{ma2022ei}, and MAKE \cite{zheng2023make} leverage CLIP \cite{radford2021learning} or ALIGN \cite{jia2021scaling} to integrate contrastive learning and keyword enhancement for enriching representations. Overall, VLP models, through strategies such as fine-tuning, reranking, and follow-up training, have become essential for improving cross-modal alignment and interaction. 
        \item \textbf{Text–Video Retrieval.} Text-video retrieval involves matching textual descriptions with corresponding videos, requiring spatiotemporal representations to address temporal dynamics, scene transitions, and precise text-video alignment. This task is more complex than text-image retrieval due to the need to model both visual and sequential information effectively.

        Early CNN/RNN-based methods \cite{torabi2016learning, mithun2018learning, miech2019howto100m, yu2017end, miech2018learning, dong2018predicting} encode videos and texts into a shared latent space for similarity measurement. LLVE \cite{torabi2016learning} employs CNNs and LSTMs to extract latent features from images and texts, with LSTMs further capturing temporal relationships between video frames. Subsequent studies \cite{mithun2018learning, miech2019howto100m} apply mean/max pooling to frame sequences to generate compact video-level representations, prioritizing efficiency over granularity. Later advancements incorporate additional modalities, such as audio and motion, to enhance video semantics \cite{miech2018learning}. For text encoding, simpler methods like Word2Vec, LSTMs, or GRUs are commonly used \cite{mithun2018learning, miech2019howto100m, yu2017end}, with evidence suggesting that combining multiple text encoding strategies improves retrieval performance \cite{dong2018predicting}.

        Transformer-based methods \cite{gabeur2020multi, liu2021hit} utilize self-attention mechanisms to jointly encode videos and texts, enabling cross-modal interaction. MMT \cite{gabeur2020multi} employs mutual attention between video and text modalities, integrating temporal information to enhance feature representation. Inspired by MoCo \cite{he2020momentum}, HiT \cite{liu2021hit} introduces hierarchical cross-modal contrastive matching at both feature and semantic levels. Additionally, these methods \cite{gabeur2020multi, liu2021hit} encode diverse modalities in video data, such as audio and motion, further enriching video representations.

        Recently, VLP-based models \cite{luo2022clip4clip, huang2023vop, wu2023cap4video, yang2024dgl, tian2024holistic, wang2024text} utilize pretrained models like CLIP \cite{radford2021learning} to enhance text-video tasks in text-video retrieval tasks by capturing cross-modal and temporal dependencies. 
        CLIP4Clip \cite{luo2022clip4clip} adapt CLIP for text-video retrieval and captioning, analyzing temporal dependencies. VoP \cite{huang2023vop} introduces prompt tuning and fine-tunes CLIP to model spatiotemporal video aspects, while Cap4Video \cite{wu2023cap4video} leverages zero-shot captioning with CLIP and GPT-2 \cite{radford2019language} for auxiliary captions. DGL \cite{yang2024dgl} proposes dynamic global-local prompt tuning, emphasizing intermodal interaction and global video information through shared latent spaces and attention mechanisms. TeachCLIP \cite{tian2024holistic} improves CLIP4Clip by integrating fine-grained cross-modal knowledge from advanced models, and refining text-video similarity with an frame-feature aggregation block. T-MASS \cite{wang2024text} addresses dataset limitations by enriching text embeddings with stochastic text modeling. 
        \item \textbf{Text–Audio Retrieval.} Text-audio retrieval involves matching textual queries with corresponding audio content, requiring alignment of semantic text information with dynamic acoustic patterns in speech, music, or environmental sounds. The challenge lies in bridging the gap between discrete text and continuous audio signals.

        Early CNN/RNN-based approaches \cite{lou2022audio, mei2022metric, zhao2023multi} focus on encoding text and audio separately and aligning them in a shared space for similarity measurement. ATR \cite{lou2022audio} uses pretrained CNN-based audio networks with NetRVLAD pooling \cite{jegou2010aggregating} to aggregate features into a unified representation. OML \cite{mei2022metric} employs CNNs for robust audio feature extraction and metric learning to enhance audio-text alignment. MGRL \cite{zhao2023multi} leverages CNNs for localized audio features and introduces adaptive aggregation to handle varying text–audio granularities.

        Furthermore, Transformer-based methods \cite{xin2023improving, zhou2024cross, doh2024enriching} utilize multi-head attention mechanisms and fine-tuning to enhance cross-modal interactions. TAP-PMR \cite{xin2023improving} employs scaled dot-product attention to enable text to focus on relevant audio frames, reducing misleading information, while its prior matrix revised loss optimizes dual matching by addressing similarity inconsistencies. CMRF \cite{zhou2024cross} enhances audio-lyrics retrieval through directional cross-modal attention and reinforcement learning to refine multimodal embeddings and interactions. TTMR++ \cite{doh2024enriching} integrates fine-tuned LLMs and rich metadata to generate detailed text descriptions, improving retrieval by addressing musical attributes and user preferences.
        \item \textbf{Unified-Modal Retrieval.} 
        Unified-Modal Retrieval aims to process diverse hybrid-modal data (e.g., text, images, videos) within a unified model architecture, such as transformer-based PLMs, to encode all modalities into a shared feature space. This enables efficient cross-modal retrieval between any pairwise combination of hybrid-modal data. With the growing demand for multimodal applications, there is an increasing need for unified multimodal retrieval models tailored to complex scenarios. Current approaches leverage pre-trained models like CLIP \cite{radford2021learning}, BLIP \cite{li2022blip}, and ALIGN \cite{jia2021scaling} for multimodal embedding.
        For instance, FLAVA \cite{singh2022flava} integrates multiple modalities into a unified framework, leveraging joint pretraining on multimodal data with cross-modal alignment and fusion objectives. 
        Similarly, UniVL-DR \cite{liu2022universal} encodes queries and multimodal resources into a shared embedding space, employing a universal embedding optimization strategy with modality-balanced hard negatives and an image verbalization method to bridge the gap between images and texts.
        MARVEL \cite{zhou2023marvel} addresses the modality gap between images and texts by incorporating visual features into the encoding process. 
        FLMR \cite{lin2023fine} enhances image representations by using a visual model aligned with existing text-based retrievers to supplement the image representation of image-to-text transforms.
        UniIR \cite{wei2024uniir} introduces a unified instruction-guided multimodal retriever, achieving robust generalization through instruction tuning on diverse multimodal-IR tasks.
        VISTA \cite{zhou2024vista} extends image understanding capability by integrating visual token embeddings into a text encoder, supported by high-quality composed image-text data and a multi-stage training algorithm. 
        E5-V \cite{jiang2024e5} fine-tunes MLLMs on single-text or vision-centric relevance data, outperforming traditional image-text pair training. 
        VLM2VEC \cite{jiang2024vlm2vec} proposes a contrastive training framework to convert vision-language models into embedding models using the MMEB dataset \cite{jiang2024vlm2vec}.
        To address modality imbalance, GME \cite{zhang2024gme} trains an MLLM-based dense retriever on the large-scale UMRB dataset\cite{zhang2024gme}. 
        Ovis \cite{lu2024ovis} aligns visual and textual embeddings by integrating a learnable visual embedding table, enabling probabilistic combinations of indexed embeddings for rich visual semantics. 
        ColPali \cite{faysse2024colpali} leverages Vision Language Models and the ViDoRe benchmark \cite{faysse2024colpali} to index documents from their visual features, facilitating efficient query matching with late interaction mechanisms.
        CREAM \cite{zhang2024cream} employs a coarse-to-fine retrieval and ranking approach, combining similarity calculations with large language model-based grouping and attention pooling for MLLM-based multi-page document processing. 
        DSE \cite{ma2024unifying} fine-tunes a large vision-language model on 1.3 million Wikipedia web page screenshots, enabling direct encoding of document screenshots into dense representations.
    \end{itemize}
    \end{itemize}
\item \textbf{Generative Structure}: Traditional information retrieval (IR) methods, which rely on similarity matching to return ranked lists of documents, have long been a cornerstone of information acquisition, dominating the field for decades. However, with the advent of pre-trained language models, generative retrieval (GR) has emerged as a novel paradigm, garnering increasing attention in recent years. GR primarily consists of two fundamental components: model training and document identifier.
Model Training aims to train generative models to effectively index and retrieve documents, while enhancing the model's capacity to memorize information from the document corpus. This is typically achieved through sequence-to-sequence (seq2seq) training, where the model learns to map queries to their corresponding Document Identifiers (DocIDs). The training process emphasizes optimizing the model's understanding of semantic relationships between queries and documents, thereby improving retrieval accuracy.
Document Identifiers (DocIDs) serve as the target output for the generative retrieval model, and unique representations of each document in the corpus. The quality of these identifiers is crucial, as they directly impact the model's ability to memorize and retrieve document information. Effective DocIDs are often generated using dense, low-dimensional embeddings or structured representations that capture the essential content and context of documents, enabling the model to distinguish between documents more accurately and enhancing retrieval performance. By overcoming the limitations of traditional IR in terms of content granularity and relevance matching, GR offers enhanced flexibility, efficiency, and creativity, better aligning with practical demands. 
    \begin{itemize}[listparindent=\parindent]
    \item \textbf{Retrieval for Text-modal.} The recent advancements in generative language models have demonstrated their ability to memorize knowledge from documents and recall knowledge to respond to user queries effectively, which focuses on the use of document identifiers (DocIDs) and their optimization for retrieval tasks. The approaches can be categorized into static DocID-based methods and learnable DocID-based methods.

    Static DocID-based methods rely on pre-defined, fixed document identifiers. They often use unique names, numeric formats, or structured identifiers to represent documents.
    GENRE \cite{de2020autoregressive} generates entity names via constrained beam search using a prefix tree, with document titles serving as DocIDs.
    DSI \cite{tay2022transformer} introduces numeric DocID formats, including unstructured, naively structured, and semantically structured identifiers, trained through indexing and retrieval strategies. 
    DynamicRetriever \cite{zhou2023dynamicretriever} uses unstructured atomic DocIDs and enhances memorization with pseudo queries. 
    SEAL \cite{bevilacqua2022autoregressive} representing documents with N-gram sub-string identifiers, leveraging FM-Index \cite{ferragina2000opportunistic} for retrieval. 
    DSI-QG \cite{zhuang2022bridging} represents documents with generated queries, re-ranked by a cross-encoder. 
    NCI \cite{wang2022neural} generates document identifiers using a seq2seq network with a prefix-aware decoder. It is trained on both labeled and augmented pseudo query-document pairs.
    Ultron \cite{zhou2022ultron} combines URLs and titles as DocIDs to uniquely identify web documents. It encodes documents into a latent semantic space using BERT \cite{devlin2018bert} and compresses vectors via Product Quantization (PQ) \cite{jegou2010product, ge2013optimized}, with PQ codes serving as semantic identifiers. Additional digits ensure DocID uniqueness. 
    LTRGR \cite{li2024learning} focuses on learning to rank passages directly using generative retrieval models, optimizing autoregressive models via rank loss.
    GenRRL \cite{zhou2023enhancing} integrates reinforcement learning for aligning token-level DocID generation with document-level relevance estimation. 
    DGR \cite{li2024distillation} enhances generative retrieval through knowledge distillation, using a cross-encoder as a teacher model to provide fine-grained ranking supervision. 
    Despite these innovations, most approaches rely on static DocIDs, which are not optimized for retrieval tasks, limiting their ability to capture document semantics and relationships, thereby hindering retrieval performance.

    To address this limitation,  Learnable DocID-based methods introduce learnable document representations, where DocIDs are optimized during training to better capture document semantics and improve retrieval performance.
    GenRet \cite{sun2024learning} employs a discrete autoencoder to encode documents into compact DocIDs, minimizing reconstruction error. 
    MINDER \cite{li2023multiview} enhances document representations using multi-view identifiers, including pseudo-queries, titles, and sub-strings. 
    NOVO \cite{wang2023novo} introduces learnable continuous N-gram DocIDs, refining embeddings through query denoising and retrieval tasks. 
    LMIndexer \cite{jin2023language} generates neural sequential discrete IDs via progressive training and contrastive learning, addressing semantic mismatches. 
    ASI \cite{yang2023auto} automates DocID learning, assigning similar IDs to semantically close documents and optimizing end-to-end retrieval using an generative model. 
    RIPOR \cite{zeng2024scalable} improves relevance scoring during sequential DocID generation using dense encoding and Residual Quantization \cite{martinez2014stacked}. 
    GLEN \cite{lee2023glen} employs a dynamic lexical identifier with a two-phase index learning strategy. Firstly, the keyword-based DocID are defined by extracting keywords from documents using self-supervised signals. Secondly, dynamic DocIDs are refined by integrating query-document relevance, enabling efficient inference. 
    The field of generative text retrieval is evolving from static, pre-defined DocIDs to dynamic, learnable DocIDs that better capture document semantics and relationships. Learnable DocIDs, combined with advanced techniques like reinforcement learning, knowledge distillation, and contrastive learning, are driving improvements in retrieval performance.

    \item \textbf{Retrieval for Cross-modal.} Similarly, MLLMs are considered to memorize and retrieve multimodal content, such as images and videos, within their parameters. When presented with a user query for visual content, the MLLM is expected to "recall" the relevant image from its parameters as a response. Achieving this capability presents significant challenges, particularly in developing effective visual memory and recall mechanisms within MLLMs.
    IRGen \cite{zhang2024irgen} employs a seq2seq model to predict discrete visual tokens (image identifiers) from query images. Its key innovation is a semantic image tokenizer that encodes global features into discrete visual tokens, enabling end-to-end differentiable search for improved accuracy and efficiency. 
    GeMKR \cite{long2024generative} integrates LLMs with visual-text features through a generative multimodal knowledge retrieval framework. It first guides multi-granularity visual learning using object-aware prefix tuning techniques to align visual features with LLMs’ text feature space, then adopts a two-step retrieval process: generating knowledge clues relevant to the query and retrieving documents based on these clues.
    GRACE \cite{li2024generative} assigns unique identifier strings to represent images, training MLLMs to memorize and retrieve image identifiers from textual queries. 
    ACE \cite{fang2024ace} combines K-Means and RQ-VAE to construct coarse and fine tokens as multimodal data identifiers, aligning natural language queries with candidate identifiers. 
    AVG \cite{li2024revolutionizing} introduces autoregressive voken (i.e., visual token) generation, tokenizing images into vokens that serve as image identifiers while preserving visual and semantic alignment. By framing text-to-image retrieval as a token-to-voken generation task, AVG bridges the gap between generative training and retrieval objectives through discriminative training, refining the learning direction during token-to-voken generation.
    \end{itemize}
\end{itemize}

\subsubsection{RERANKER}
\label{sec:RERANKER}
Reranker, as a critical second-stage component in multimodal retrieval, is designed to re-rank a multimodal document list initially retrieved by a first-stage retriever. It achieves this by employing advanced relevance scoring mechanisms, such as cross-attention models, which enable more contextual interactions between queries and documents. Based on the utilization of large models, including LLMs and MLLMs, existing reranking methods can be categorized into two primary paradigms: fine-tuning-as-reranker and prompting-as-reranker.
\begin{itemize}[leftmargin=1em, listparindent=\parindent]
\item \textbf{Fine-tuning-as-Reranker}: The fine-tuning-as-reranker paradigm adapts PLMs to domain-specific reranking tasks through supervised fine-tuning on domain-specific datasets, addressing their inherent lack of ranking awareness and inability to effectively measure query-document relevance. 
    \begin{itemize}[listparindent=\parindent]
    \item \textbf{Reranking for Text-Modal}: According the development of large models' architecture, reranker can be divided to three categories: encoder-only, encoder-decoder, and decoder-only.

    Encoder-only rerankers have advanced document ranking by fine-tuning PLMs (e.g., BERT \cite{devlin2018bert}) to achieve precise relevance estimation. Key examples include \citet{nogueira2019passage} and monoBERT \cite{nogueira2019multi}, which format query-document pairs as query-document sequences. The relevance score is derived from the “[CLS]” token’s representation via a linear layer, with optimization achieved through negative sampling and cross-entropy loss.

    Existing research on encoder-decoder rerankers primarily formulates document ranking as a generation task \cite{nogueira2020document, ju2021text, pradeep2021expando, zhuang2023rankt5}, fine-tuning models like T5 to generate classification tokens (e.g., “true” or “false”) for query-document pairs, with relevance scores derived from token logits \cite{nogueira2020document}. Extensions include multi-view learning approaches \cite{ju2021text} that simultaneously generate classification tokens for query-document pairs and queries conditioned on documents, and DuoT5 \cite{pradeep2021expando}, which compares the classification tokens of document pairs to determine relative relevance. 
    Beyond these approaches, studies have explored alternative training losses and architectures. Contrast with previous methods that rely on text generation losses, RankT5 \cite{zhuang2023rankt5} directly produces numerical relevance scores for each query-document pair, optimizing with ranking losses instead of generation losses. 
    ListT5 \cite{yoon2024listt5} further advances this by processing multiple documents simultaneously, directly generating reranked lists using the Fusion-in-Decoder architecture.

    Recent studies \cite{ma2024fine, zhang2023two, peng2024q, zhang2023rank, liu2024leveraging} have explored fine-tuning decoder-only models like LLaMA for document reranking. RankLLaMA \cite{ma2024fine} formats query-document pairs into prompts and uses the last token representation for relevance scoring. TSARankLLM \cite{zhang2023two} employs a two-stage training approach: continuous pretraining on web-sourced relevant text pairs to align LLMs with ranking tasks, followed by fine-tuning with supervised data and tailored loss functions. Q-PEFT \cite{peng2024q} introduces query-dependent parameter-efficient fine-tuning to generate accurate queries from documents. In contrast, listwise approaches like those in \cite{zhang2023rank} and PE-Rank \cite{liu2024leveraging} focus on directly outputting reranked document lists. 
    \citet{zhang2023rank} highlight the limitations of point-wise datasets with binary labels, and instead use ranking outputs from existing systems as gold standards to train a listwise reranker. 
    PE-Rank \cite{liu2024leveraging} compresses documents into single embeddings, reducing input length and improving reranking efficiency.
    \item \textbf{Reranking for Cross-Model}: The multi-modal reranking uses the multi-modal question and multi-modal knowledge items to obtain the relevance score, as reranking have already shown its importance in various knowledge-intensive tasks. 
    \citet{wen2024multimodal} fine-tunes a pretrained MLLM to facilitate cross-item interaction between questions and knowledge items. The reranker is trained on the same dataset as the answer generator, using distant supervision by checking whether answer candidates appear in the knowledge text.
    RagVL \citet{retrievalmllm} introduces a novel framework featuring knowledge-enhanced reranking and noise-injected training. The approach involves instruction-tuning the MLLM with a simple yet effective template to enhance its ranking capability, enabling it to serve as a reranker for accurately filtering the top-$k$ retrieved images.
    \end{itemize}
    In summary, these approaches leverages the representational capacity of large models while optimizing them for task-specific relevance signals, often achieving high reranking accuracy. However, it requires substantial computational resources and labeled training data, resulting in increased costs.

\item \textbf{Prompting-as-Reranker}: In contrast, the prompting-as-reranker paradigm leverages large models in a zero-shot or few-shot manner by designing prompts that direct the model to generate relevance scores or rankings directly. This approach exploits the inherent knowledge and reasoning capabilities of large models, eliminating the need for extensive fine-tuning and offering greater flexibility and resource efficiency. Researchers have explored prompting LLMs and MLLMs to perform ranking tasks on multimodal documents, with prompting strategies generally categorized into three types: point-wise, pair-wise, and list-wise methods.
    \begin{itemize}[listparindent=\parindent]
    \item \textbf{Reranking for Text-Model}: LLMs are increasingly employed in text-modal reranking tasks, leveraging their advanced capabilities to optimize the ranking of textual documents.

    Point-wise methods evaluate the relevance between a query and individual documents, reranking them based on relevance scores.. 
    \citet{zhuang2023beyond} integrates fine-grained relevance labels into prompts for better document distinction.
    MCRanker \cite{guo2024generating} addresses biases in existing point-wise rerankers by generating relevance scores based on multi-perspective criteria. 
    UPR \cite{sachan2022improving} re-scores retrieved passages using a zero-shot question generation model.
    \citet{zhuang2023open} show that LLMs pre-trained without supervised instruction fine-tuning (e.g., LLaMA) also exhibit strong zero-shot ranking capabilities. Despite their effectiveness, these methods often rely on suboptimal handcrafted prompts. 
    To improve prompts for ranking tasks, Co-Prompt \cite{cho2023discrete} introduces a discrete prompt optimization method for improving prompt generation in reranking tasks. 
    PaRaDe \cite{drozdov2023parade} proposes a difficulty-based approach to select the most challenging in-context demonstrations for prompts, though experiments reveal that this method does not significantly outperform random selection. 
    To improve demonstration selection, DemoRank \cite{liu2024demorank} advances demonstration selection with a dependency-aware demonstration reranker, optimizing top-ranked examples through efficient training sample construction and a novel list-pairwise loss.

    Pair-wise methods involve presenting LLMs with a query and a document pair, instructing them to identify the more relevant document. PRP-AllPair \cite{qin2023large} generates all possible pairs, assigns discrete relevance judgments, and aggregates these into a final relevance score per document. PRP-Graph \cite{luo2024prp} improves this by using judgment generation probabilities and a graph-based aggregation for scoring relevance. Additionally, a post-processing technique \cite{yan2024consolidating} refines LLM-generated labels by aligning them with pairwise preferences while minimizing deviations from original values.

    Listwise methods directly rank document lists by incorporating queries and documents into prompts, instructing LLMs to output reranked document identifiers. 
    RankGPT \cite{sun2023chatgpt} introduces instructional permutation generation and a sliding window strategy to address context length limits, while LRL \cite{ma2023zero} reorders document identifiers for candidate documents. However, these methods face challenges: (1) performance is highly sensitive to document order, revealing positional bias, and (2) the sliding window strategy limits the number of documents ranked per iteration. 
    Recent advancements have attempted to address these issues: \citet{tang2023found} propose permutation self-consistency to mitigate bias. TourRank \cite{chen2024tourrank} introduces a tournament mechanism, parallelizing reranking to minimize the impact of initial document order. TDPart \cite{parry2024top} employs a top-down partitioning algorithm, which processes documents to depth using a pivot element. FIRST \cite{reddy2024first} leverages the output logits of the first generated identifier to directly obtain a ranked ordering of candidates.
    \item \textbf{Reranking for Cross-Model}: Prompt-Based Multimodal Reranker uses prompts to guide a MLLM in reranking items. TIGeR \cite{qu2024unified} proposes a framework leveraging multimodal LLMs for zero-shot reranking via a generative retrieval approach. However, their method is limited to text-only query retrieval tasks. In contrast, \citet{lin2024mm} extends this scope by utilizing multimodal LLMs to address diverse multimodal reranking tasks, supporting queries and documents in text, image, or interleaved text-image formats.

    \end{itemize}
    In summary, these approaches leverages the pre-existing knowledge and reasoning capabilities of LLMs, reducing the need for extensive task-specific fine-tuning. Consequently, it provides greater flexibility and resource efficiency, particularly in scenarios with limited labeled data or computational resources. However, its effectiveness depends heavily on the quality and design of the prompts, as well as the model's ability to generalize its pre-trained knowledge to the specific demands of the target task.
\end{itemize}

\subsubsection{REFINER}
\label{sec:REFINER}
Theoretically, LLMs improves with more comprehensive task-relevant knowledge in the retrieved and reranked context. However, unlimited input length poses practical deployment challenges: 
(1) Limited Context Window: LLMs have a fixed input length determined during pre-training, and any text exceeding this limit is truncated, leading to loss of contextual semantics. 
(2) Catastrophic Forgetting: Insufficient cache space can cause LLMs to forget previously learned knowledge when processing long sequences. 
(3) Slow Inference Speed. 
Consequently, refined prompts are crucial for optimizing LLM performance.

The refiner is an optional yet highly impactful component that optimizes retrieved and reranked information before its utilization by the LLM. It performs advanced processing tasks, such as summarization, distillation, or contextualization, to condense and refine content into a more digestible and actionable format. By extracting key insights, eliminating redundancies, and aligning information with the query's context, the refiner enhances the utility of the retrieved data, enabling the LLM to generate more coherent, accurate, and contextually relevant responses.

Prompt refinement can be achieved through two primary approaches: hard prompt methods and soft prompt methods. Hard prompt methods involve filtering out unnecessary or low-information content, still using natural language tokens and resulting in less fluent but generalizable prompts that can be used across LLMs with different embedding configurations. Soft prompt methods, in contrast, encode prompt information into continuous representations, producing latent vectors (special tokens) that are not human-readable but optimized for model performance.
\begin{itemize}[leftmargin=1em, listparindent=\parindent]
\item \textbf{Hard Prompt Refiner}: Hard prompts consist of natural language tokens from the LLM/MLLM's vocabulary, representing specific words or sub-words, and can be generated by humans or models. 
    \begin{itemize}[listparindent=\parindent]
    \item \textbf{Refining for Text-Model}: Recent advancements in prompt compression and context distillation aim to optimize the efficiency of LLMs. 
    DynaICL \cite{zhou2023efficient} employs a meta controller to dynamically allocate in-context demonstrations based on input complexity and computational constraints. 
    FILCO \cite{wang2023learning} distills retrieved documents using lexical and information-theoretic methods—String Inclusion, Lexical Overlap, and CXMI—training both context filtering and generation models for RAG tasks. 
    CPC \cite{liskavets2024prompt} preserves semantic integrity by using a context-aware encoder to remove irrelevant sentences, while AdaComp \cite{zhang2024adacomp} dynamically selects optimal documents via a compression-rate predictor.
    LLMLingua \cite{jiang2023llmlingua} introduces a coarse-to-fine approach, compressing prompt components (instructions, questions, demonstrations) using a small language model (SLM) to measure token informativeness via perplexity (PPL). LongLLMLingua \cite{jiang2023longllmlingua} extends this to long documents, employing a linear scheduler, reordering mechanism, and contrastive perplexity to retain question-relevant tokens while ensuring key information integrity. 
    CoT-Influx \cite{huang2023fewer} compresses GPT-4-generated Chain-of-Thought (CoT) prompts using a shot-pruner and token-pruner, both implemented as MLPs trained via reinforcement learning. These methods collectively improve performance while reducing useless CoT examples and redundant tokens.
    Selective Context \cite{li2023compressing} evaluates lexical unit informativeness using a causal language model and a percentile-based filtering method to remove redundancy. It calculates token self-information by predicting next-token probabilities, aggregating these at phrase and sentence levels. 
    Prompt-SAW \cite{ali2024prompt} preserves syntactic and semantic structures by extracting key tokens via relation-aware graphs, integrating them into compressed prompts.
    PCRL \cite{jung2024discrete} treats prompt compression as a binary classification task, using a frozen pre-trained policy language model with trainable MLP layers. The compression policy labels tokens as include or exclude, optimizing a reward function that balances faithfulness and prompt length reduction.
    LLMLingua-2 \cite{pan2024llmlingua} employs a bidirectional encoder-only model with a linear classification layer for compression, determining token retention or removal. 
    RECOMP \cite{xu2023recomp}  employs extractive and abstractive compressors to generate query-focused summaries, leveraging contrastive learning and knowledge distillation. 
    Nano-Capsulator \cite{chuang2024learning} optimizes compression using reward feedback from response differences and enforces strict length constraints. 
    MEMWALKER \cite{chen2023walking} uses interactive prompting to build and navigate a memory tree for context summarization. CompAct \cite{yoon2024compact} sequentially compresses document segments for long-context question-answering, achieving high compression rates. 
    Style-Compress \cite{pu2024style} iteratively refines prompts using diverse styles and task-specific examples, evaluated by larger LLMs. 
    TCRA-LLM \cite{liu2023tcra} combines summarization and semantic compression to reduce token size. 
    TACO-RL \cite{shandilya2024taco} employs reinforcement learning for task-aware prompt compression, ensuring low latency. 
    FaviComp \cite{jung2024familiarity} enhances evidence familiarity by combining token probabilities from compression and target models, reducing perplexity.
    \item \textbf{Refining for Cross-Model}: Recent advancements in visual token compression for MLLMs focus on enhancing efficiency without significant performance loss. 
    LLaVolta \cite{chen2024efficient} introduces a method to reduce the number of visual tokens, enhancing training and inference efficiency without compromising performance. To minimize information loss during compression while maintaining training efficiency, it employs a lightweight, staged training scheme. This scheme progressively compresses visual tokens from heavy to light compression during training, ensuring no information loss during testing.
    PyramidDrop \cite{xing2024pyramiddrop} is a visual redundancy reduction strategy for MLLMs, designed to improve efficiency in both inference and training with negligible performance loss. It partitions the MLLM into several stages and drops a predefined ratio of image tokens at the end of each stage.
    DeCo \cite{yao2024deco} proposes the principle of "Decouple Compression from Abstraction," which involves compressing visual tokens at the patch level using projectors while allowing the LLM to handle visual semantic abstraction entirely.
    MustDrop \cite{liu2024multi} measures the importance of each token throughout its lifecycle, including the vision encoding, prefilling, and decoding stages. During vision encoding, it merges spatially adjacent tokens with high similarity and establishes a key token set to retain vision-critical tokens. In the prefilling stage, it further compresses vision tokens guided by text semantics using a dual-attention filtering strategy. In the decoding stage, an output-aware cache policy reduces the size of the KV cache. By employing tailored strategies across these stages, MustDrop achieves an optimal balance between performance and efficiency.
    G-Search \cite{zhao2024accelerating} proposes a greedy search algorithm to determine the minimum number of vision tokens to retain at each layer, from shallow to deep. Based on this strategy, a parametric sigmoid function (P-Sigmoid) is designed to guide token reduction at each layer of the MLLM, with parameters optimized using Bayesian Optimization.
    G-Prune \cite{jiang2025kind} introduces a graph-based method for training-free visual token pruning. It treats visual tokens as nodes and constructs connections based on semantic similarities. Information flow is propagated through weighted links, and the most important tokens are retained for MLLMs after iterations.
    \end{itemize}
Although interpretable and transparent, the inherent ambiguity of hard prompts often hinders the precise expression of intent, limiting their effectiveness in diverse or complex scenarios. Crafting accurate and impactful hard prompts demands significant human effort and may require model-based refinement or optimization. Moreover, even minor variations in hard prompts can lead to inconsistent LLM performance for identical tasks.
\item \textbf{Soft Prompt Refiner}: Soft prompts are trainable, continuous vectors that match the dimensionality of token embeddings in LLM's vocabulary. Unlike hard prompts, which rely on discrete tokens from a predefined vocabulary, soft prompts are optimized through training to capture nuanced meanings that discrete tokens cannot express. When fine-tuned on diverse datasets, soft prompts enhance the LLM's performance across various tasks.
    \begin{itemize}[listparindent=\parindent]
    \item \textbf{Refining for Text-Model}: Language models convert text prompts into vectors for denser representation, enabling compression of discrete text into continuous vectors within the model. These vectors can serve as internal parameters (internalization) or additional soft prompts (encoding). Such compression extends the context window and enhances inference speed, particularly with repeated prompt usage.

    Early work focused on system prompt internalization.
    \citet{askell2021general} used Knowledge Distillation to align models with human values, while \citet{choi2022prompt} introduced Pseudo-Input Generation, generating pseudo-inputs from prompts and distilling knowledge between teacher and student models to avoid redundant inference computations.
    Later research compressed user prompt contexts. 
    \citet{snell2022learning} distilled abstract instructions, reasoning, and examples into prompts with distinct distribution differences, enabling task execution without explicit prompts. 
    \citet{sun2023instruction} internalized ranking techniques for zero-shot relevance tasks, while Distilling Step-by-Step \cite{hsieh2023distilling} improved reasoning tasks by distilling rationales as additional supervision. 
    In retrieval-augmented generation, xRAG \cite{cheng2024xrag} integrated compressed document embeddings via a plug-and-play projector, using self-distillation for robustness.
    For context compression, COCOM \cite{rau2024context} reduced long contexts to few embeddings, balancing trade-offs between decoding time and answer quality. 
    LLoCO \cite{tan2024lloco} learned offline compressed representations for efficient QA retrieval. 
    QGC \cite{cao2024retaining} retained key information under high compression using query-guided dynamic strategies. 
    UniICL \cite{gao2024unifying} unified demonstration selection, compression, and generation within a single frozen LLM, projecting demonstrations and inputs into virtual tokens for semantic-based processing.

    Recent advancements in prompt compression for LLMs focus on encoding hard prompts into reusable soft prompts to enhance efficiency and generalization across tasks. 
    Early work by \citet{wingate2022prompt} distilled complex hard prompts into concise soft prompts by minimizing output distribution differences, reducing inference costs. 
    A series of works aim to enhance generalization across diverse prompts. 
    Gist \cite{mu2023learning} used meta-learning to encode multi-task instructions into gist tokens, while Gist-COCO \cite{li2024say} employed an encoder-decoder architecture to compresses original prompts into shorter gist prompts, via the Minimum Description Length principle. 
    UltraGist \cite{zhang2024compressing} optimized cross-attention for compressing ultra-long contexts into near-lossless UltraGist tokens. 
    AutoCompressor \cite{chevalier2023adapting} iteratively compressed contexts segments into summary vectors using a Recurrent Memory Transformer, reducing computational load. 
    Other approaches, like ICAE \cite{ge2023context} and 500xCompressor \cite{li2024500xcompressor}, fine-tuned LoRA-adapted LLMs for context encoding and prompt compression. 
    For LLM-based recommendations, POD \cite{li2023prompt} distilled discrete prompt templates into continuous prompt vectors  with an whole-word embedding to integrate the item ID, while RDRec \cite{wang2024rdrec} synthesizes training data and internalizes rationales into a smaller model.
    SelfCP \cite{gao2024selfcp} balances training cost, inference efficiency, and generation quality by compressing over-limit prompts asynchronously using frozen LLMs as the compressor and generator and trainable linear layers to project hidden states into LLM-acceptable memory tokens.
    \item \textbf{Refining for Cross-Model}: PromptMM \cite{wei2024promptmm} tackles overfitting and side information inaccuracies in multi-modal recommenders by using Multi-modal Knowledge Distillation with prompt-tuning. It compresses models by distilling user-item relationships and multi-modal content from complex teacher models to lightweight student models, eliminating extra parameters. Soft prompt-tuning bridges the semantic gap between multi-modal context and collaborative signals, enhancing robustness. Additionally, a disentangled multi-modal list-wise distillation with modality-aware re-weighting addresses multimedia data inaccuracies.
    RACC \cite{weng2024learning} compresses and aggregates retrieved knowledge for image-question pairs, generating a compact Key-Value (KV) cache modulation to adapt downstream frozen MLLMs for efficient inference.
    VTC-CLS \cite{wang2024cls} uses the prior knowledge of the association between the [CLS] token and visual tokens in the visual encoder to evaluate visual token importance, enabling Visual Token Compression and shortening visual context.
    VisToG \cite{huang2024efficient} introduces a grouping mechanism using pretrained vision encoders to group similar image segments without segmentation masks. Semantic tokens represent image segments after linear projection and before input into the vision encoder. Isolated attention identifies and eliminates redundant visual tokens, reducing computational demands.
    \end{itemize}
However, as dataset size increases, so do the computational resource requirements. Additionally, soft prompts are less interpretable than hard prompts, as their continuous vectors are not directly readable or explainable by humans.
\end{itemize}

\begin{figure}[htbp]
\centering
\begin{forest}
for tree={
  grow=east, 
  parent anchor=east,
  child anchor=west,
  calign=edge midpoint,
  edge path={
    \noexpand\path[\forestoption{edge}]
      (!u.parent anchor) -- +(5pt,0) |- (.child anchor)\forestoption{edge label};
  },
  l sep=10pt, 
  s sep=5pt,  
  font=\tiny
},
where level=0{
  text width=1.1cm,
  where n children=0{
      rectangle, rounded corners, draw=black, fill=blue!10, align=center, 
    }{
      rectangle, rounded corners, draw=black, fill=green!10, align=center, 
    }
}{},
where level=1{
  text width=1.35cm, 
  where n children=0{
      rectangle, rounded corners, draw=black, fill=blue!10, align=center, 
    }{
      rectangle, rounded corners, draw=black, fill=red!10, align=center, 
    }
}{},
where level=2{
  text width=10.0cm, 
  where n children=0{
      rectangle, rounded corners, draw=black, fill=blue!10, align=left, 
    }{
      rectangle, rounded corners, draw=black, fill=red!10, align=left, 
    }
}{}
[\textbf{Multimodal}\\\textbf{Generation}
    [\textbf{Modality}\\\textbf{Augmentation}
        [{\parbox{10cm}{InternLM-XComposer \cite{InternLM-XComposer}, InternLM-XComposer2 \cite{InternLM-XComposer2}, InternLM-XComposer-2.5 \cite{InternLM-XComposer-2.5}, MuRAR \cite{MuRAR2024}, $M^2RAG$ \cite{mmRAG2024}
        }}]
    ]
    [\textbf{MLLM}
        [{\parbox{10cm}{$\bullet$ \textbf{Image $\oplus$ Text $\rightarrow$ Text}\\
        BLIP-2 \cite{li2023blip}, ChatSpot \cite{zhao2023chatspot}, OpenFlamingo \cite{awadalla2023openflamingo}, ASM \cite{wang2023all}, Qwen-VL \cite{bai2023qwen}, Kosmos-2.5 \cite{lv2023kosmos}, InternLM-XComposer \cite{InternLM-XComposer}, JAM \cite{aiello2023jointly}, Kosmos-1 \cite{huang2023language}, PaLM-E \cite{driess2023palm}, ViperGPT \cite{suris2023vipergpt}, PandaGPT \cite{su2023pandagpt}, PaLI-X \cite{chen2023pali}, LLaVA-Med \cite{li2023llava}, LLaVAR \cite{zhang2023llavar}, mPLUG-DocOwl \cite{ye2023mplugwl}, P-Former \cite{jian2023bootstrapping}, MiniGPT-v2 \cite{chen2023minigpt}, LLaVA \cite{liu2023visual}, MiniGPT-4 \cite{zhu2023minigpt}, mPLUG-Owl \cite{ye2023mplug}, Otter \cite{li2023mimic}, MultiModal-GPT \cite{gong2023multimodal}, CogVLM \cite{wang2024cogvlm}, mPLUG-Owl2 \cite{ye2024mplug}, Monkey \cite{li2024monkey}, DocPedia \cite{feng2024docpedia}, ShareGPT4V \cite{chen2024sharegpt4v}, mPLUG-PaperOwl \cite{hu2024mplug}, RLHF-V \cite{yu2024rlhf}, Silkie \cite{li2023silkie}, Lyrics \cite{lu2023lyrics}, VILA \cite{lin2024vila}, CogAgent \cite{hong2024cogagent}, Volcano \cite{lee2023volcano}, DRESS \cite{chen2024dress}, LION \cite{chen2024lion}, Osprey \cite{yuan2024osprey}, LLaVA-MoLE \cite{chen2024llava}, VLGuard \cite{zong2024safety}, MobileVLM V2 \cite{chu2024mobilevlm}, ViGoR \cite{yan2024vigor}, V* \cite{wu2024v}, MobileVLM \cite{chu2023mobilevlm}, TinyGPT-V \cite{yuan2023tinygpt}, DocLLM \cite{wang2023docllm}, LLaVA-Phi \cite{zhu2024llava}, KAM-CoT \cite{mondal2024kam}, InternLM-XComposer2 \cite{InternLM-XComposer2}, InternLM-XComposer-2.5 \cite{InternLM-XComposer-2.5}, MoE-LLaVA \cite{lin2024moe}, VisLingInstruct \cite{zhu2024vislinginstruct} \\
        $\bullet$ \textbf{Image $\oplus$ Text $\rightarrow$ Image $\oplus$ Text}\\
        Visual ChatGPT \cite{wu2023visual}, DetGPT \cite{pi2023detgpt}, FROMAGe \cite{koh2023grounding}, Shikra \cite{chen2023shikra}, GPT4RoI \cite{zhanggpt4roi}, SEED \cite{ge2023planting}, LISA \cite{lai2024lisa}, GILL \cite{koh2023generating}, Kosmos-2 \cite{peng2023kosmos}, DreamLLM \cite{dong2023dreamllm}, MiniGPT-5 \cite{zheng2023minigpt}, Kosmos-G \cite{pan2023kosmos}, VisCPM \cite{hu2023large}, CM3Leon \cite{yu2023scaling}, LaVIT \cite{jin2023unified}, GLaMM \cite{rasheed2024glamm}, RPG \cite{yang2024mastering}, Vary-toy \cite{wei2024small}, CogCoM \cite{qi2024cogcom}, SPHINX-X \cite{liu2024sphinx}, LLaVA-Plus \cite{liu2024llava}, PixelLM \cite{ren2024pixellm}, VL-GPT \cite{zhu2023vl}, CLOVA \cite{gao2024clova}, Emu-2 \cite{sun2024generative}, MM-Interleaved \cite{tian2024mm}, DiffusionGPT \cite{qin2024diffusiongpt}\\
        $\bullet$ \textbf{Video $\oplus$ Text $\rightarrow$ Text}\\
        Video-ChatGPT \cite{maaz2023video}, VideoChat \cite{li2023videochat}, Dolphins \cite{ma2024dolphins}\\
        $\bullet$ \textbf{Video $\oplus$ Text $\rightarrow$ Video $\oplus$  Text}\\
        Video-LaVIT \cite{jin2024video}\\
        $\bullet$ \textbf{Unified $\rightarrow$ Text}\\
        Flamingo \cite{alayrac2022flamingo}, X-LLM \cite{chen2023x}, LanguageBind \cite{zhu2023languagebind}, InstructBLIP \cite{liu2023visual}, MM-REACT \cite{yang2023mm}, X-InstructBLIP \cite{panagopoulou2023x}, EmbodiedGPT \cite{mu2023embodiedgpt}, Video-LLaMA \cite{zhang2023video}, Lynx \cite{zeng2024matters}, LLaMA-VID \cite{li2024llama}, InternVL \cite{chen2024internvl}, AnyMAL \cite{moon2024anymal}\\
        $\bullet$ \textbf{Unified $\rightarrow$ Image $\oplus$ Text}\\
        BuboGPT \cite{zhao2023bubogpt}, Emu \cite{sun2023emu}, GroundingGPT \cite{li2024lego}\\
        $\bullet$ \textbf{Unified $\rightarrow$ Unified}\\
        TEAL \cite{yang2023teal}, GPT-4 \cite{achiam2023gpt}, Gemini \cite{team2023gemini}, HuggingGPT \cite{shen2023hugginggpt}, CoDi-2 \cite{tang2024codi}, AudioGPT \cite{huang2024audiogpt}, ModaVerse \cite{wang2024modaverse}, MLLM-Tool \cite{wang2024tool}, ControlLLM \cite{liu2024controlllm}, NExT-GPT \cite{wu2024next}
        }}]
    ]
]
\end{forest}
\caption{Taxonomy of recent advancements in multimodal generation research.}
\label{MultimodalGeneration}
\end{figure}
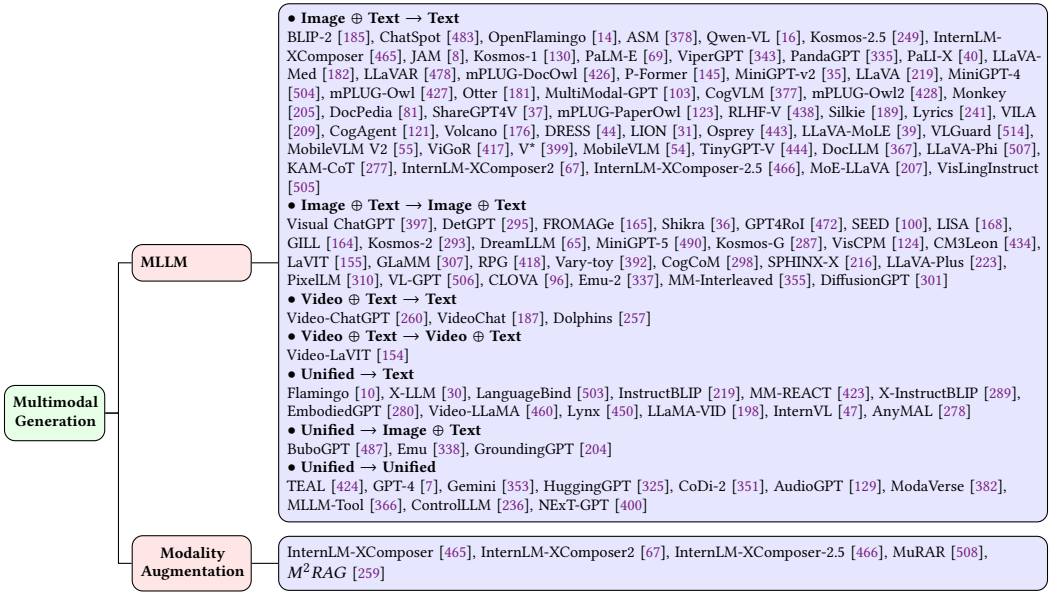

\subsection{Multimodal Generation}
\label{sec:MultimodalGeneration}
Multimodal generation based on Multimodal Large Language Models (MLLMs) represents a significant advancement, enabling the generation of content across multiple modalities such as text, images, audio, and video. These models leverage the strengths of large language models (LLMs) and extend them to handle and integrate diverse data types, creating rich, coherent, and contextually relevant outputs. We classify MLLMs from generative perspectives of inputs and outputs, and summarize the related researches in Figure \ref{MultimodalGeneration}.

\subsubsection{MODALITY INPUT}
With the rapid advancement of large language models in the domain of textual knowledge comprehension and question-answering, researchers try to explore how to enable these models to understand and process inputs from a broader range of modalities, thereby facilitating more extensive multimodal question-answering tasks. Initial efforts focused on incorporating image modality into the input of large models. 
For instance, Blip-2 \cite{li2023blip} proposes a generic and efficient pre-training strategy that bootstraps vision-language pre-training from off-the-shelf frozen pre-trained image encoders and frozen large language models. BLIP-2 bridges the modality gap with a lightweight Querying Transformer, which is pre-trained in two stages. The first stage bootstraps vision-language representation learning from a frozen image encoder. The second stage bootstraps vision-to-language generative learning from a frozen language model.
Internlm-xcomposer2 \cite{dong2024internlm} proposes a vision-language model excelling in free-form text-image composition and comprehension. This model goes beyond conventional vision-language understanding, adeptly crafting interleaved text-image content from diverse inputs like outlines, detailed textual specifications, and reference images, enabling highly customizable content creation.
DiffusionGPT \cite{qin2024diffusiongpt} leverages Large Language Models (LLM) to offer a unified generation system capable of seamlessly accommodating various types of prompts and integrating domain-expert models. DiffusionGPT constructs domain-specific Trees for various generative models based on prior knowledge. When provided with an input, the LLM parses the prompt and employs the Trees-of-Thought to guide the selection of an appropriate model.

As the variety of modal data continues to expand, more complex modalities, such as video, have been integrated into the inputs of large models. 
For instance, Video-ChatGPT \cite{maaz2023video} proposes a multimodal model that merges a video-adapted visual encoder with an LLM. The resulting model is capable of understanding and generating detailed conversations about videos.
Video-LaVIT \cite{jin2024video} address spatiotemporal dynamics limitations in video-language pretraining with an efficient video decomposition that represents each video as keyframes and temporal motions. These are then adapted to an LLM using well-designed tokenizers that discretize visual and temporal information as a few tokens, thus enabling unified generative pre-training of videos, images, and text. At inference, the generated tokens from the LLM are carefully recovered to the original continuous pixel space to create various video content. The proposed framework is both capable of comprehending and generating image and video content.

Recently, the input for multimodal large models has evolved from specialized modal data to a unified input that can handle arbitrary modal data. 
For instance, InstructBLIP \cite{panagopoulou2023x} conduct a vision-language instruction tuning based on the pretrained BLIP-2 models. Additionally, we introduce an instruction-aware Query Transformer, which extracts informative features tailored to the given instruction.
InternVL \cite{chen2024internvl} design a large-scale vision-language foundation model (InternVL) which scales up the vision foundation model to 6 billion parameters and progressively aligns it with the LLM using web-scale image-text data from various sources.
GPT-4 \cite{achiam2023gpt} proposes a large-scale, multimodal model which can accept image and text inputs and produce text outputs. While less capable than humans in many real-world scenarios, GPT-4 exhibits human-level performance on various professional and academic benchmarks.
HuggingGPT \cite{shen2023hugginggpt} proposes an LLM-powered agent that leverages LLMs (eg, ChatGPT) to connect various AI models in machine learning communities (eg, Hugging Face) to solve AI tasks. Specifically, we use ChatGPT to conduct task planning when receiving a user request, select models according to their function descriptions available in Hugging Face, execute each subtask with the selected AI model, and summarize the response according to the execution results. By leveraging the strong language capability of ChatGPT and abundant AI models in Hugging Face, HuggingGPT can tackle a wide range of sophisticated AI tasks spanning different modalities and domains.
NExT-GPT \cite{wu2024next} present an end-to-end general-purpose any-to-any MM-LLM system. NExT-GPT connect an LLM with multimodal adaptors and different diffusion decoders, enabling NExT-GPT to perceive inputs and generate outputs in arbitrary combinations of text, image, video, and audio. By leveraging the existing well-trained high-performing encoders and decoders, NExT-GPT is tuned with only a small amount of parameter (1\%) of certain projection layers, which not only benefits low-cost training but also facilitates convenient expansion to more potential modalities. Moreover, we introduce a modality-switching instruction tuning (MosIT) and manually curate a high-quality dataset for MosIT, based on which NExT-GPT is empowered with complex cross-modal semantic understanding and content generation.

\subsubsection{MODALITY OUTPUT}
With the explosive growth in the capabilities of MLLMs, the ability to answer questions based on multimodal inputs and generate multimodal outputs has also seen a qualitative improvement. There is also increasing attention from researchers on VQA scenarios that shift from generating text results to generating multimodal results that include text.In this section, we are discussing multimodal outputs that are not scenarios like text-to-image or text-to-video, which only generate a single modality, but rather scenarios where the answers includes text and at least one other modality of data, such as text-image output, or image-video output.In the basic VQA task, MIMOQA\cite{singh-etal-2021-mimoqa}  was the first to propose the concept of multimodal output, which achieved the capability of multimodal output by transforming questions into an image-text matching task. It constructed a dual-tower model called MExBERT. The text stream, based on BERT, takes in the query and related documents to output the final text answer. The visual stream, based on VGG-19, receives images related to the query and documents, outputting a relevance score between the image and text. The final insertion of the image is determined by this relevance score. Its groundbreaking introduction to multimodal output research has, however, certain limitations: 1) It is necessary to screen out images related to the question. The model only needs to select and output images from the small number of screened ones. The task is relatively simple. 2) Multimodality is still limited to the image modality. 3) To simplify the issue, it is still limited to scenarios where the input images must include at least one relevant image. Based on the aforementioned limitations, the latest research has made corresponding improvements\cite{InternLM-XComposer,InternLM-XComposer2,NExT-GPT,InternLM-XComposer-2.5,MuRAR2024,mmRAG2024}. 

A common workflow paradigm for implementing multimodal output is to first conduct position identification after generating a text answer to determine where to insert multimodal data. Subsequently, based on the surrounding context of the corresponding positions, candidate multimodal data is retrieved. Finally, a relevance matching model is utilized to determine the final data to be inserted. InternLM-XComposer \cite{InternLM-XComposer} achieves multimodal output of text and images. After generating each paragraph of text, it calls a model to determine whether to insert an image. If it is determined that an image needs to be inserted, it will generate a caption of the image to be inserted and search the web for candidate images, eventually allowing the model to select the most relevant image from candidate set for insertion. InternLM-XComposer2 and 2.5 \cite{InternLM-XComposer2,InternLM-XComposer-2.5} allow users to directly input a set of candidate images on the basis of the above. 
MuRAR \cite{MuRAR2024} has also implemented multi - modal output in RAG scenarios based on this paradigm, but it has innovated the methods of position identification and candidate set recall in RAG scenarios. It uses source attribution to confirm the correspondence between the generated snippet and the retrieved snippet from the large model input, thereby determining the insertion point, and the candidate set directly uses the multimodal data associated with the retrieved snippet, simplifying the recall operation. In addition, it has expanded the multimodal data from images to include tables and videos. $M^2RAG$ \cite{mmRAG2024} employs an alternative paradigm to achieve multimodal output in the RAG scenario. It uses the user's query to simultaneously recall associated text elements and images. Then, based on the associations of the images and text elements in the original document, they are refined. Subsequently, MLLMs are employed to vectorize the images or convert them into descriptions, which are input into the generative model in the form of placeholders. The output generates answer text and a simple description placeholder for the associated image. Finally, through a chain-of-thought(COT) process, the placeholders are converted into actual images. NExT-GPT \cite{NExT-GPT} employs an entirely different and novel paradigm. It directly trains a unified multimodal large model, unifying the reasoning and generation process, and directly generates multimodal data including text, images, videos, etc., through the model \cite{NExT-GPT}.

\section{Dataset for MRAG}
\label{sec:datasetMRAG}
To evaluate the general capabilities of MRAG systems in real-world multimodal understanding and knowledge-based question-answering tasks, we curated a collection of existing datasets designed to comprehensively evaluate the MRAG pipeline. These datasets are categorized into two classes: 
(1) Retrieval \& Generation-Joint Components, which evaluate the synergy of retrieval and generation by requiring systems to retrieve external knowledge and generate accurate responses; and 
(2) Generation, focusing solely on the model's ability to produce contextually accurate outputs without external retrieval. 
This categorization enables a detailed evaluation of MRAG systems' strengths and limitations in diverse scenarios.

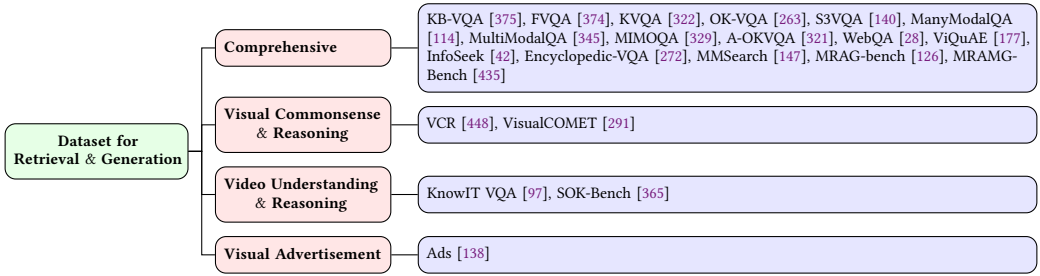
\begin{figure}[htbp]
\centering
\begin{forest}
for tree={
  grow=east, 
  parent anchor=east,
  child anchor=west,
  calign=edge midpoint,
  edge path={
    \noexpand\path[\forestoption{edge}]
      (!u.parent anchor) -- +(5pt,0) |- (.child anchor)\forestoption{edge label};
  },
  l sep=10pt, 
  s sep=5pt,  
  font=\tiny
},
where level=0{
  text width=2.2cm,
  where n children=0{
      rectangle, rounded corners, draw=black, fill=blue!10, align=center, 
    }{
      rectangle, rounded corners, draw=black, fill=green!10, align=center, 
    }
}{},
where level=1{
  text width=2.1cm, 
  where n children=0{
      rectangle, rounded corners, draw=black, fill=blue!10, align=center, 
    }{
      rectangle, rounded corners, draw=black, fill=red!10, align=center, 
    }
}{},
where level=2{
  text width=8.0cm, 
  where n children=0{
      rectangle, rounded corners, draw=black, fill=blue!10, align=left, 
    }{
      rectangle, rounded corners, draw=black, fill=red!10, align=left, 
    }
}{}
[\textbf{Dataset for}\\\textbf{Retrieval} \& \textbf{Generation}
    [\textbf{Visual Advertisement}
        [{\parbox{8cm}{Ads \cite{hussain2017automatic}
        }}]
    ]
    [\textbf{Video Understanding}\\\& \textbf{Reasoning}
        [{\parbox{8cm}{KnowIT VQA \cite{garcia2020knowit}, SOK-Bench \cite{wang2024sok}
        }}]
    ]
    [\textbf{Visual Commonsense}\\\& \textbf{Reasoning}
        [{\parbox{8cm}{VCR \cite{zellers2019recognition}, VisualCOMET \cite{park2020visualcomet}
        }}]
    ]
    [\textbf{Comprehensive}
        [{\parbox{8cm}{KB-VQA \cite{wang2015explicit}, FVQA \cite{wang2017fvqa}, KVQA \cite{shah2019kvqa}, OK-VQA \cite{marino2019ok}, S3VQA \cite{jain2021select}, ManyModalQA \cite{hannan2020manymodalqa}, MultiModalQA \cite{talmor2021multimodalqa}, MIMOQA \cite{singh2021mimoqa}, A-OKVQA \cite{schwenk2022okvqa}, WebQA \cite{chang2022webqa}, ViQuAE \cite{lerner2022viquae}, InfoSeek \cite{chen2023can}, Encyclopedic-VQA \cite{mensink2023encyclopedic}, MMSearch \cite{jiang2024mmsearch}, MRAG-bench \cite{hu2024mrag}, MRAMG-Bench \cite{yu2025mramg}
        }}]
    ]
]
\end{forest}
\caption{Categories of MRAG dataset for retrieval \& generation.}
\label{DatasetRetrievalGeneration}
\end{figure}
\begin{table}[htbp]
\scriptsize
\centering
\caption{Summary of dataset for retrieval \& generation.}
\begin{tabular}{|p{60pt}|p{15pt}|p{290pt}|}
\hline
\textbf{Dataset} & \textbf{Time} & \textbf{Statistics} \\ \hline
\multicolumn{3}{|l|}{\textbf{Comprehensive}} \\ \hline 
KB-VQA \cite{wang2015explicit} & 2015 & 2,402 questions, 700 images, 1 knowledge bases. \\ \hline 
FVQA \cite{wang2017fvqa} & 2017 & 5,826 questions, 2,190 images, 3 knowledge bases. \\ \hline 
KVQA \cite{shah2019kvqa} & 2019 & 183,007 question-answer pairs about 18,880 unique entities contained within 24,602 images. \\ \hline
OK-VQA \cite{marino2019ok} & 2019 & 14,055 questions, 10 scenarios.  \\ \hline
S3VQA \cite{jain2021select} & 2021 & 6,765 question-image pairs. \\ \hline
ManyModalQA \cite{hannan2020manymodalqa} & 2020 & 10,190 questions with 2,873 image, 3,789 text, and 3,528 table.  \\ \hline
MultiModalQA \cite{talmor2021multimodalqa} & 2021 & 29,918 questions that requires knowledge from text, tables, and images (35.7\% require cross-modality reasoning). \\ \hline
MIMOQA \cite{singh2021mimoqa} & 2021 & 56,693 QA pairs, with 401,182 images. \\ \hline
A-OKVQA \cite{schwenk2022okvqa} & 2022 & 24,903 multiple-choice questions.  \\ \hline
WebQA \cite{chang2022webqa} & 2022 & 24,929 image-based and 24,343 text-based questions. \\ \hline
ViQuAE \cite{lerner2022viquae} & 2022 & 3.7K questions paired with images. A Knowledge base composed of 1.5M Wikipedia articles paired with images. \\ \hline
InfoSeek \cite{chen2023can} & 2023 & 8.9K human-written and 1.3M semi-automated questions, 9 image classification and retrieval datasets. \\ \hline
Encyclopedic-VQA \cite{mensink2023encyclopedic} & 2023 &  1M Image-Question-Answer triplets derived from 221k textual QA pairs from 16.7k different categories. Each QA pair is combined with (up to) 5 images. 514k unique images. 15k textual single-hop questions, 25k multi-answer questions, and 22k two-hop questions. \\  \hline
MMSearch \cite{jiang2024mmsearch} & 2024 & 2,901 unique images, 300 manually collected queries spanning 14 subfields. \\ \hline
MRAG-bench \cite{hu2024mrag} & 2024 & 1,353 multiple-choice questions, 16,130 images, 9 scenarios. \\ \hline
MRAMG-Bench \cite{yu2025mramg} & 2025 & 4,800 QA pairs across three distinct domains, containing 4,346 documents and 14,190 images, with tasks categorized into three difficulty levels. \\ \hline
\multicolumn{3}{|l|}{\textbf{Visual Commonsense Reasoning}} \\ \hline
VCR \cite{zellers2019recognition} & 2019 & 290k multiple choice QA problems derived from 110k movie scenes.  \\ \hline
VisualCOMET \cite{park2020visualcomet} & 2020 & 1,465,704 commonsense inferences over 59,356 images, and 139,377
distinct events. \\ \hline
\multicolumn{3}{|l|}{\textbf{Video Understanding \& Commonsense Reasoning}} \\ \hline
KnowIT VQA \cite{garcia2020knowit} & 2020 & 24,282 human-generated QA pairs about a popular sitcom. \\ \hline
SOK-Bench \cite{wang2024sok} & 2024 & 44K QA pairs covers over 12 types of questions, sourcing from about 10K situations. Each question is accompanied by two types of answers: a direct answer and a set of four multiple-choice options.  \\ \hline
\multicolumn{3}{|l|}{\textbf{Visual Advertisement}} \\ \hline
Ads \cite{hussain2017automatic} & 2017 &  202,090 questions from 64,832 image ads and 3,477 video ads. \\ \hline
\end{tabular}
\label{table:datasetRA}
\end{table}

\subsection{Dataset for Retrieval \& Generation}
Datasets for Retrieval \& Generation in MRAG are designed to evaluate end-to-end systems capable of retrieving relevant knowledge from multimodal sources (e.g., text, images, videos) and generating accurate responses. These datasets evaluate the synergistic integration of retrieval and generation, focusing on the system's ability to dynamically utilize external knowledge to improve response quality and relevance.
In this section, we introduce key benchmarks designed for diverse evaluation of Retrieval \& Generation tasks. Figure \ref{DatasetRetrievalGeneration} provides an overview of existing benchmarks, while Table \ref{table:datasetRA} summarizes the statistics of selected representative datasets.

Early knowledge-based datasets include KB-VQA \cite{wang2015explicit} and FVQA \cite{wang2017fvqa}, which rely on closed knowledge. FVQA, for instance, uses a fixed knowledge graph, making questions straightforward once the knowledge is known, with minimal reasoning required.
KVQA \cite{shah2019kvqa} focuses on images in Wikipedia articles, primarily testing named entity recognition and Wikipedia knowledge retrieval rather than commonsense reasoning. 
OK-VQA \cite{marino2019ok} and A-OKVQA \cite{schwenk2022okvqa} evaluate multimodal reasoning using external knowledge, with A-OKVQA introducing "rationale" annotations to better evaluate knowledge acquisition and reasoning. 
S3VQA \cite{jain2021select} extends OK-VQA by requiring object detection and web queries, but like OK-VQA, it often reduces to single retrieval tasks rather than complex reasoning.
MultiModalQA \cite{talmor2021multimodalqa} pioneers complex questions requiring reasoning across snippets, tables, and images, focusing on cross-modal knowledge extraction. However, its template-based questions simplify the task to filling in blanks with modality-specific answering mechanisms. 
ManyModalQA \cite{hannan2020manymodalqa} also uses snippets, images, and tables but emphasizes answer modality choice over knowledge aggregation. 
MIMOQA \cite{singh2021mimoqa} introduces “Multimodal Input Multimodal Output”, requiring both text and image selections to enhance understanding.
WebQA \cite{chang2022webqa} is a manually crafted, multi-hop multimodal QA dataset that retrieves visual content but provides only textual answers, relying solely on MLLMs for reasoning, making it unsuitable for models dependent on linguistic context. 
ViQuAE \cite{lerner2022viquae} focuses on answering questions about named entities grounded in a visual context using a Knowledge Base. 
InfoSeek \cite{chen2023can} and Encyclopedic-VQA \cite{mensink2023encyclopedic} target knowledge-based questions beyond common sense knowledge, with Encyclopedic-VQA using model-generated annotations.
MMSearch \cite{jiang2024mmsearch} evaluates MLLMs as multimodal search engines, focusing on image-to-image retrieval. 
Compared with previous works, MRAG-bench \cite{hu2024mrag} evaluates MLLMs in utilizing vision-centric retrieval-augmented knowledge, identifying scenarios where visual knowledge outperforms textual knowledge. 
MRAMG-Bench \cite{yu2025mramg} evaluates answers combining text and images, leveraging multimodal data within a corpus.
Additionally, VCR \cite{zellers2019recognition} and VisualCOMET \cite{park2020visualcomet}, derived from movie scenes, evaluate Visual Commonsense Reasoning. 
KnowIT VQA \cite{garcia2020knowit} and SOK-Bench \cite{wang2024sok} focus on video understanding and reasoning task, combining visual, textual, and temporal reasoning with knowledge-based questions. 
Ads \cite{hussain2017automatic} proposes an automatic advertisement understanding task, featuring rich annotations on topics, sentiments, and persuasive reasoning.

\subsection{Dataset for Generation}
\begin{figure}[htbp]
\centering
\begin{forest}
for tree={
  grow=east, 
  parent anchor=east,
  child anchor=west,
  calign=edge midpoint,
  edge path={
    \noexpand\path[\forestoption{edge}]
      (!u.parent anchor) -- +(5pt,0) |- (.child anchor)\forestoption{edge label};
  },
  l sep=10pt, 
  s sep=5pt,  
  font=\tiny
},
where level=0{
  text width=1.1cm,
  where n children=0{
      rectangle, rounded corners, draw=black, fill=blue!10, align=center, 
    }{
      rectangle, rounded corners, draw=black, fill=green!10, align=center, 
    }
}{},
where level=1{
  text width=2.1cm, 
  where n children=0{
      rectangle, rounded corners, draw=black, fill=blue!10, align=center, 
    }{
      rectangle, rounded corners, draw=black, fill=red!10, align=center, 
    }
}{},
where level=2{
  text width=9.0cm, 
  where n children=0{
      rectangle, rounded corners, draw=black, fill=blue!10, align=left, 
    }{
      rectangle, rounded corners, draw=black, fill=red!10, align=left, 
    }
}{}
[\textbf{Dataset for}\\\textbf{Generation}
    [\textbf{Multidisciplinary}
        [{\parbox{9cm}{ScienceQA \cite{lu2022learn}, MMMU \cite{yue2024mmmu}, CMMU \cite{he2024cmmu}, CMMMU \cite{zhang2024cmmmu}, MMMU-Pro \cite{yue2024mmmupro}
        }}]
    ]
    [\textbf{Conversational QA}
        [{\parbox{9cm}{SparklesDialogue \cite{huang2023sparkles}, SciGraphQA \cite{li2023scigraphqa}, ConvBench \cite{liu2024convbench}, MMDU \cite{liu2024mmdu}
        }}]
    ]
    [\textbf{Industry}
        [{\parbox{9cm}{MME-Industry \cite{yi2025mme}
        }}]
    ]
    [\textbf{Video Understanding}
        [{\parbox{9cm}{TGIF-QA \cite{jang2017tgif}, ActivityNet-QA \cite{yu2019activitynet}, EgoSchema \cite{mangalam2023egoschema}, Video-MME \cite{fu2024video}, MVBench \cite{li2024mvbench}, MMBench-Video \cite{fang2024mmbench}, MLVU \cite{zhou2024mlvu}, LVBench \cite{wang2024lvbench}, Event-Bench \cite{du2024towards}, VNBench \cite{zhao2024needle}, TempCompass \cite{liu2403tempcompass}, MovieChat \cite{song2024moviechat}
        }}]
    ]
    [\textbf{Mathematics}
        [{\parbox{9cm}{MathVista \cite{lu2023mathvista}, We-Math \cite{qiao2024we}, Math-Vision \cite{wang2024measuring}, Olympiadbench \cite{he2024olympiadbench}, MathVerse \cite{zhang2025mathverse}
        }}]
    ]
    [\textbf{Structural Document}
        [{\parbox{9cm}{FigureQA \cite{kahou2017figureqa}, DocVQA \cite{mathew2021docvqa}, VisualMRC \cite{tanaka2021visualmrc}, ChartQA \cite{masry2022chartqa}, InfographicVQA \cite{mathew2022infographicvqa}, ChartBench \cite{xu2023chartbench}, SciGraphQA \cite{li2023scigraphqa}, MMC-Benchmark \cite{liu2023mmc}, MP-DocVQA \cite{tito2023hierarchical}, ChartX \cite{xia2024chartx}, DocGenome \cite{xia2024docgenome}, CharXiv \cite{wang2024charxiv}, MMLongBench-Doc \cite{ma2024mmlongbench}, ComTQA \cite{zhao2024tabpedia}, Web2Code \cite{yun2024web2code}, VisualWebBench \cite{liu2024visualwebbench}, SciFIBench \cite{roberts2024scifibench}
        }}]
    ]
    [\textbf{Optical Character}\\\textbf{Recognition}
        [{\parbox{9cm}{TextVQA \cite{singh2019towards}, OCR-VQA \cite{mishra2019ocr}, WebSRC \cite{chen2021websrc}, OCRBench \cite{liu2024ocrbench}, VCR \cite{zhang2024vcr}, SEED-Bench-2-Plus \cite{li2024seed2plus}
        }}]
    ]
    [\textbf{Comprehensive}
        [{\parbox{9cm}{VQA v2 \cite{goyal2017making}, NLVR2 \cite{suhr2018corpus}, VizWiz \cite{gurari2018vizwiz}, MME \cite{Chaoyou2023Mme}, Visit-Bench \cite{bitton2023visit}, Touchstone \cite{bai2023touchstone}, MM-Vet \cite{yu2023mm}, InfiMM-Eval \cite{han2023infimm}, Q-Bench \cite{wu2023q}, Seed-Bench \cite{li2023seed}, SEED-Bench-2 \cite{li2024seed}, SEED-Bench-2-Plus \cite{li2024seed2plus}, LVLM-eHub \cite{xu2024lvlm}, LAMM \cite{yin2024lamm}, MMT-Bench \cite{ying2024mmt}, RealWorldQA \cite{realworldQA}, WV-Bench \cite{lu2024wildvision}, MME-RealWorld \cite{zhang2024mme}, MMStar \cite{chen2024we}, CV-Bench \cite{tong2024cambrian}, MDVP \cite{lin2024draw}, FOCI \cite{geigle2024african}, MMVP \cite{tong2024eyes}, V*-Bench \cite{wu2024v}, MME-RealWorld \cite{zhang2024mme}, Visual COT \cite{shao2024visual}, Mementos \cite{wang2024mementos}, MIRB \cite{zhao2024benchmarking}, ReMI \cite{kazemi2024remi}, MuirBench \cite{wang2024muirbench}, VEGA \cite{zhou2024vega}, MMBench \cite{liu2025mmbench}, BLINK \cite{fu2025blink}
        }}]
    ]
]
\end{forest}
\caption{Categories of MRAG dataset for generation.}
\label{figure:DatasetGeneration}
\end{figure}
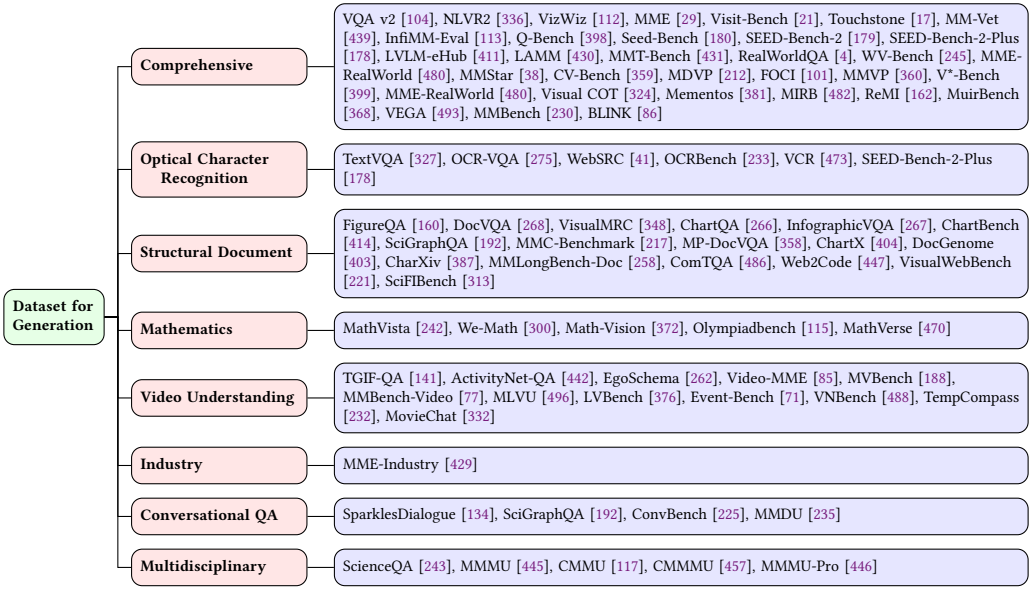

The Generation category evaluates a model's intrinsic capacity to generate contextually accurate outputs based solely on its pre-trained knowledge and internal reasoning, without external retrieval. This evaluation isolates the generation component, providing insights into the model's foundational language understanding capabilities. It enables a detailed analysis of MRAG systems' strengths and limitations across diverse scenarios.
In this section, we provide an overview of representative benchmarks developed for various evaluation of Generation tasks. The existing benchmarks are systematically organized in Figure \ref{figure:DatasetGeneration}, and the statistics of selected representative benchmarks are summarized in Table \ref{table:generation}.

\subsubsection{Comprehensive}
To rigorously evaluate the capabilities of MLLMs, a diverse range of evaluation benchmarks has been developed. These benchmarks are designed to test various dimensions of model performance. By utilizing these benchmarks, researchers can systematically quantify the strengths and limitations of MLLMs, ensuring their alignment with real-world applications and user expectations. This evaluation framework not only supports the iterative improvement of MLLMs but also provides a standardized basis for comparing models in terms of perceptual and reasoning abilities.

VQA v2 \cite{goyal2017making}, an early benchmark with 453K annotated QA pairs, focuses on open-ended questions with concise answers. VizWiz \cite{gurari2018vizwiz}, introduced around the same time, includes 8K QA pairs from visually impaired individuals' daily lives, addressing real-world needs of disabled users. NLVR2 \cite{suhr2018corpus} explores multi-image vision capabilities by evaluating captions against image pairs. However, these benchmarks often fail to assess modern MLLMs' emergent capabilities, such as advanced reasoning. Recent efforts like LVLM-eHub \cite{xu2024lvlm}, MDVP \cite{lin2024draw}, and LAMM \cite{yin2024lamm} compile extensive datasets for comprehensive evaluation, revealing that while MLLMs excel in commonsense tasks, they lag in image classification, OCR, VQA, large-scale counting, fine-grained attribute differentiation, and precise object localization. Fine-tuning can mitigate some of these limitations.

Researchers are developing specialized benchmarks to address the limitations of traditional evaluations for MLLMs. 
Notable examples include MME \cite{Chaoyou2023Mme}, which covers 14 perception and cognition tasks; MMBench \cite{liu2025mmbench}, featuring 20 ability dimensions, including object localization and social reasoning; and SEED-Bench \cite{li2023seed}, which focuses on multiple-choice questions. SEED-Bench-2 \cite{li2024seed} expanded the scope to 24K QA pairs, including the evaluation of both text and image generation.
MMT-Bench \cite{ying2024mmt} further scaled up to 31K QA pairs across diverse scenarios. 
Common findings reveal that model performance improves with scale, but challenges persist in fine-grained perception tasks (e.g., spatial localization), chart and visual mathematics comprehension, and interleaved image-text understanding. Open-source MLLMs have shown rapid progress, often matching or surpassing closed-source models.

Real-world usage scenarios are critical for evaluating model performance in practical applications. 
Benchmarks like RealWorldQA \cite{realworldQA} evaluates spatial understanding capabilities sourced from real-life scenarios, while BLINK \cite{fu2025blink} highlights tasks such as visual correspondence and multi-view reasoning that challenge current MLLMs despite being intuitive for humans. 
WV-Bench \cite{lu2024wildvision} and Visit-Bench \cite{bitton2023visit} emphasize human preferences and instruction-following capabilities, whereas V*-Bench \cite{wu2024v} evaluates high-resolution image processing and correct visual details through attribute recognition and spatial reasoning tasks. 
MME-RealWorld \cite{zhang2024mme} enhances quality and difficulty with extensive annotated QA pairs and high-resolution images. These benchmarks reveal that fine-grained perception tasks remain challenging for models, while artistic style recognition and relative depth perception are relatively stronger. Although closed-source models like GPT-4o outperform others, human performance still surpasses general models significantly.

Many studies simplify evaluation into binary or multi-choice problems for easier quantification, but this approach overlooks the importance of the reasoning process, which is critical for understanding model capabilities. 
To address this, some works use open-ended generation and LLM-based evaluators, though these face challenges with inaccurate LLM scoring. 
For instance, MM-Vet \cite{yu2023mm} employs diverse question formats to assess integrated vision-language capabilities, while Touchstone \cite{bai2023touchstone} emphasizes real-world dialogue evaluation, arguing that multiple-choice questions are insufficient for evaluating multimodal dialogue capabilities. 
InfiMM-Eval \cite{han2023infimm} evaluates models on deductive, abductive, and analogical reasoning across tasks, including intermediate reasoning steps, aligning with practical scenarios like mathematical problem-solving. 
These benchmarks highlight the strengths and limitations of MLLMs in complex tasks. Closed-source models excel in reasoning but struggle with complex localization, structural relationships, charts, and visual mathematics. High-resolution data improves recognition of small objects, dense text, and fine-grained details. While Chain-of-Thought (CoT) strategies significantly boost reasoning in closed-source models, their impact on open-source models remains limited.

The development of multimodal benchmarks emphasizes continuous refinement to accurately assess model capabilities. 
MMStar \cite{chen2024we} addresses data leakage by curating 1.5K visually-dependent QA pairs, while CV-Bench \cite{tong2024cambrian} tackles the scarcity of vision-centric benchmarks with 2.6K manually-inspected samples for 2D/3D understanding. 
FOCI \cite{geigle2024african} evaluates MLLMs using domain-specific subsets and supplementary classification datasets, revealing challenges in fine-grained perception. 
MMVP \cite{tong2024eyes} identifies 9 distinct patterns in CLIP-based models, showing MLLMs' struggles with visual details, with only Gemini and GPT-4V performing above random guessing. 
Q-Bench \cite{wu2023q} evaluates low-level attribute perception, highlighting GPT-4V's near-human performance. 
VisualCOT \cite{shao2024visual} introduces visual chain-of-thought prompts to enhance MLLMs' focus on specific image regions.
To further upgrading vision capabilities on multiple image understanding, Mementos \cite{wang2024mementos} evaluates sequential image understanding, while MIRB \cite{zhao2024benchmarking} focuses on multi-image reasoning across perception, visual knowledge, and multi-hop reasoning tasks. 
ReMI \cite{kazemi2024remi} designs 13 tasks with diverse image relationships and input formats, and MuirBench \cite{wang2024muirbench} includes 12 multi-image understanding tasks with unanswerable variants for robust assessment. 
VEGA \cite{zhou2024vega} is specifically designed to evaluate interleaved image-text comprehension. The task requires models to identify relevant images and text while filtering out irrelevant information to arrive at the correct answer. Evaluation results reveal that even advanced proprietary MLLMs, such as GPT-4V and Gemini 1.5 Pro, achieve only modest performance, highlighting significant room for improvement in interleaved information processing capabilities.

\subsubsection{Optical Character Recognition (OCR)}
Multimodal benchmarks are increasingly focusing on the evaluation of Optical Character Recognition (OCR) tasks, driving progress in document understanding. Early benchmarks like TextVQA \cite{singh2019towards} and OCR-VQA \cite{mishra2019ocr} evaluated standard text recognition, while WebSRC \cite{chen2021websrc} introduces advanced structural reasoning tasks like web page layout interpretation. 
SEED-Bench-2-Plus \cite{li2024seed2plus} and OCRBench \cite{liu2024ocrbench} expanded evaluation to diverse data types, including charts, maps, and web pages, showing models achieving near state-of-the-art performance in recognizing various OCR text. 
VCR \cite{zhang2024vcr} addresses OCR task with partially obscured text embedded in images, requiring content reconstruction. 
Despite advancements, many MLLMs struggle with fine-grained OCR tasks. While models like GPT-4V perform well, they lag behind specialized OCR models. Performance varies significantly by data type, with knowledge graphs and maps posing greater challenges than simpler formats like charts, suggesting potential improvements through data-specific optimization or dedicated OCR integration.

\subsubsection{Structural Document}
Structural documents, including charts, HTML web content, and various document formats, play a critical role in practical applications due to their ability to efficiently convey complex information. These data types are characterized by their highly structured nature and information density, distinguishing them from natural images. Unlike images, which rely on visual patterns and textures, structural documents require models to comprehend intricate layouts, spatial relationships, and semantic connections between embedded elements such as text, tables, and graphical components.

To advance models capable of understanding and reasoning with such data, several benchmarks have been proposed for different types of structural documents. 
Early dataset FigureQA \cite{kahou2017figureqa} introduces a visual reasoning corpus with synthetic images and scientific-style figures, focusing on relationships between plot elements. 
ChartQA \cite{masry2022chartqa} emphasizes VQA with charts, ranging from tasks that require both data extraction and math reasoning. 
ChartX \cite{xia2024chartx} collects a comprehensive dataset with 22 topics, 18 chart types, and 7 tasks, incorporating multiple modalities. 
VisualMRC \cite{tanaka2021visualmrc} targets visual machine reading comprehension, emphasizing natural language understanding and generation. 
ChartBench \cite{xu2023chartbench} evaluates chart comprehension and data reliability through complex reasoning. 
MMC-Benchmark \cite{liu2023mmc} provides a human-annotated benchmark to assess MLLMs on visual chart understanding tasks like chart information extraction, reasoning, and classification. 
Web2Code \cite{yun2024web2code} introduces a webpage-to-code dataset for instruction tuning and an evaluation framework to assess MLLMs' webpage understanding and HTML code translation capabilities. 
VisualWebBench \cite{liu2024visualwebbench} evaluates MLLMs on various web tasks at website, element, and action levels. 
Many charts lack data point annotations, necessitating MLLMs to infer values using chart elements. 
ComTQA \cite{zhao2024tabpedia} introduces a table VQA benchmark for perception and comprehension tasks, while DocVQA \cite{mathew2021docvqa} focuses on document image QA with an emphasis on information extraction tasks. 
InfographicVQA \cite{mathew2022infographicvqa} targets understanding infographics images, which are designed to present information concisely.
Infographics exhibit diverse layouts and structures, requiring basic reasoning and arithmetic skills. As MLLMs advance, benchmarks now focus on complex chart and document understanding. 
For instance, DocGenome \cite{xia2024docgenome} analyzes scientific papers, covering tasks like information extraction, layout detection, VQA, and code generation. 
CharXiv \cite{wang2024charxiv} targets challenging charts from scientific papers, while MP-DocVQA \cite{tito2023hierarchical} extends DocVQA to multi-page scenario, where questions are constructed based on multi-page documents instead of single page.
MMLongBench-Doc \cite{ma2024mmlongbench} focuses on long document understanding, averaging 47.5 pages. 
SciGraphQA \cite{li2023scigraphqa} is a synthetic dataset with 295K QA dialogues about academic graphs, generated using Palm-2 from CS/ML ArXiv papers. 
SciFIBench \cite{roberts2024scifibench} benchmarks scientific figure interpretation, using adversarial filtering for negative examples and human verification for quality assurance.

Despite advancements, a performance gap persists between proprietary and open-source models on conventional benchmarks. Current MLLMs continue to face challenges in reasoning tasks and long-context document comprehension, particularly in interpreting extended multimodal contexts, which remains a critical limitation.

\subsubsection{Mathematics}
Visual math problem-solving is key to evaluating MLLMs, leading to the development of specialized benchmarks. 
MathVista \cite{lu2023mathvista} pioneered this effort by aggregating 28 existing datasets and introducing 3 new ones, featuring diverse tasks like logical, algebraic, and scientific reasoning with various visual inputs. 
Subsequent benchmarks, such as Math-Vision \cite{wang2024measuring} and OlympiadBench \cite{he2024olympiadbench}, introduced more complex tasks and fine-grained evaluation methods. 
We-Math \cite{qiao2024we} decomposes problems into sub-problems to assess fundamental understanding, while MathVerse \cite{zhang2025mathverse} further evaluates MLLMs' comprehension of math diagrams by transforming problems into six versions with varying proportions of visual and textual content.

Despite promising results from MLLMs, significant challenges remain. existing MLLMs often struggle with interpreting complex diagrams, rely heavily on textual cues, and address composite problems through memorization rather than underlying reasoning. These limitations highlight the need for further development in MLLM capabilities.

\subsubsection{Video Understanding}
Traditional video-QA benchmarks like TGIF-QA \cite{jang2017tgif} and ActivityNet-QA \cite{yu2019activitynet} are domain-specific, focusing on tasks related to human activities. 
With advancements in MLLMs, new benchmarks have emerged to address more complex video understanding challenges. 
Video-MME \cite{fu2024video} explores diverse video domains with multimodal inputs and manual annotations, while MVBench \cite{li2024mvbench} reannotates existing datasets using ChatGPT. 
MMBench-Video \cite{fang2024mmbench} features free-form questions for short to medium-length videos. 
Benchmarks like MLVU \cite{zhou2024mlvu}, LVBench \cite{wang2024lvbench}, Event-Bench \cite{du2024towards}, and VNBench \cite{zhao2024needle} emphasize long-video understanding, testing models on extended multimodal contexts. 
VNBench \cite{zhao2024needle} introduces a synthetic framework for evaluating tasks like retrieval and ordering, by inserting irrelevant images or text into videos.
Specialized benchmarks like EgoSchema \cite{mangalam2023egoschema} focus on egocentric videos. 
TempCompass \cite{liu2403tempcompass} evaluates fine-grained temporal perception, and MovieChat \cite{song2024moviechat} targets long videos but often reduces tasks to short-video problems. 
Current MLLMs, especially open-source ones, face challenges with long-context processing and temporal perception, underscoring the need for improved capabilities in these areas.

\subsubsection{Industry}
The absence of a comprehensive benchmark for evaluating MLLMs across diverse industry verticals has limited understanding of their applicability in specialized real-world scenarios. 
To address this gap, MME-Industry \cite{yi2025mme} was developed specifically for industrial applications, covering over 21 industrial sectors such as power generation, electronics manufacturing, textile production, steel, and chemical processing. 
Domain experts from each sector meticulously annotated and validated test cases, ensuring the benchmark's reliability, accuracy, and practical relevance. MME-Industry thus serves as a robust tool for assessing MLLMs in industrial contexts.

\subsubsection{Conversational QA}
Current MLLMs are primarily designed for multi-round chatbot interactions, yet most benchmarks focus on single-round QA tasks. To better align with real-world conversational scenarios, multi-round QA benchmarks have been developed to simulate human-AI interactions with extended contextual histories.
SparklesDialogue \cite{huang2023sparkles} evaluates conversational proficiency across multiple images and dialogue turns, featuring flexible text-image interleaving with two rounds and four images per instance. 
SciGraphQA \cite{li2023scigraphqa} constructs multi-turn QA conversations based on scientific graphs from Arxiv papers, emphasizing complex scientific discourse. 
ConvBench \cite{liu2024convbench} assesses perception, reasoning, and creation capabilities across individual rounds and overall conversations, revealing that MLLMs' reasoning and creation failures often stem from inadequate fine-grained perception. 
MMDU \cite{liu2024mmdu} engages models in multi-turn, multi-image conversations, with up to 20 images and 27 turns, highlighting that the performance gap between open-source and closed-source models is largely due to limited conversational instruction tuning data.
These benchmarks collectively enhance the evaluation of MLLMs in complex, real-world interaction scenarios.

\subsubsection{Multidisciplinary}
The mastery of multidisciplinary knowledge is a key indicator of a model's expertise, and several benchmarks have been developed to evaluate this capability. 
ScienceQA \cite{lu2022learn} comprises scientific questions annotated with lectures and explanations, designed to facilitate chain-of-thought evaluation. It spans grade-level knowledge across diverse domains. 
MMMU \cite{yue2024mmmu} presents a more challenging college-level benchmark across diverse subjects, including engineering, art and design, business, science, humanities, social science, and medicine. Its question format extends beyond single image-text pairs to include interleaved text and images. 
Similarly, CMMU \cite{he2024cmmu} and CMMMU \cite{zhang2024cmmmu} provide Chinese domain-specific benchmarks for grade-level and college-level knowledge, respectively. 
MMMU-Pro \cite{yue2024mmmupro} enhances MMMU with a more robust version for advanced evaluation.

\begingroup
\scriptsize
\begin{longtable}{|p{60pt}|p{15pt}|p{290pt}|}
\caption{Summary of dataset for generation.} \label{table:generation} \\ 
\hline
\textbf{Dataset} & \textbf{Time} & \textbf{Statistics} \\ \hline
\multicolumn{3}{|l|}{\textbf{Comprehensive}} \\ \hline
VQA v2 \cite{goyal2017making} & 2017 & contains more than 443K training, 214K validation and 453K test image-question pairs. \\ \hline
NLVR2 \cite{suhr2018corpus} & 2018 & contains 107,292 examples of English sentences paired with web photographs, including 29,680 unique sentences and 127,502 images. The task is to determine whether a natural language caption is true about a pair of photographs. \\ \hline
VizWiz \cite{gurari2018vizwiz} & 2018 & contains 20,000 training, 3,173 validation, and 8,000 test sets of visual questions originating from blind people. \\ \hline
MME \cite{Chaoyou2023Mme} & 2023 & measures both perception and cognition abilities on a total of 14 subtasks \\ \hline
Visit-Bench \cite{bitton2023visit} & 2023 & comprising 592 instances and 1,159 public images. The instances are either from 45 newly assembled instruction families or reformatted from 25 existing datasets. 10 instruction families cater to multi-image query scenarios. \\ \hline
Touchstone \cite{bai2023touchstone} & 2023 & 908 questions covering 27 subtasks.  The highest proportion of questions pertains to recognition, accounting for about 44.1\%, followed by comprehension questions at 29.6\%. The proportions of the other categories are 15.3\% for basic descriptive ability, 7.4\% for visual storytelling ability, and 3.6\% for multi-image analysis ability.  \\ \hline
MM-Vet \cite{yu2023mm} & 2023 & defines 6 core vision-language capabilities and examines the 16 integrations of interest derived from their combinations. It contains 200 images, and 218 questions (samples), all paired with their respective ground truths. \\ \hline
InfiMM-Eval \cite{han2023infimm} & 2023 & It consists of 279 manually curated reasoning questions, associated with a total of 342 images. The dataset is categorized into three reasoning paradigms: deductive, abductive, and analogical reasoning. 49 questions pertain to abductive reasoning, 181 require deductive reasoning, and 49 involve analogical reasoning. Furthermore, the dataset is divided into two folds based on reasoning complexity, with 108 classified as “High” reasoning complexity and 171 as “Moderate” reasoning complexity. \\ \hline
Q-Bench \cite{wu2023q} & 2023 & consists of 2,990 diverse-sourced images, each equipped with a human-asked question focusing on its low-level attributes. \\ \hline
Seed-Bench \cite{li2023seed} & 2023 & consists of 19K multiple-choice questions with accurate human annotations, which spans 12 evaluation dimensions including the comprehension of both the image and video modality.  \\ \hline
SEED-Bench-2 \cite{li2024seed} & 2024 & comprises 24K multiple-choice questions with accurate human annotations, which span 27 dimensions, including the evaluation of both text and image generation. \\ \hline
LVLM-eHub \cite{xu2024lvlm} & 2024 & contains 42 datasets in our LVLM-eHub. The sizes of specific datasets are 109.8K, 29.5K, 177.2K, 67.3K, and 8.9K for visual perception, knowledge acquisition, reasoning, commonsense, and object hallucination, respectively. \\ \hline
LAMM \cite{yin2024lamm} & 2024 & evaluate 9 common image tasks, using a total of 11 datasets with over 62,439 samples, and 3 common point cloud tasks, by utilizing 3 datasets with over 12,788 data samples. \\ \hline
MMT-Bench \cite{ying2024mmt} & 2024 & comprises 31,325 meticulously curated multi-choice visual questions from various multimodal scenarios, covering 32 core meta-tasks and 162 subtasks in multimodal understanding. \\ \hline
RealWorldQA \cite{realworldQA} & 2024 & consists of 765 images, with a question and easily verifiable answer for each image. \\ \hline
WV-Bench \cite{lu2024wildvision} & 2024 & constructed by selecting 500 high-quality samples from 8,000 user submissions in WV-ARENA. \\ \hline
MME-RealWorld \cite{zhang2024mme} & 2024 & constructed by collecting more than 300K images from public datasets and the Internet, filtering 13,366 high-quality images for annotation and contributing to 29,429 question-answer pairs that cover 43 subtasks across 5 real-world scenarios. \\ \hline
MMStar \cite{chen2024we} & 2024 & contains 1,500 challenging samples, each rigorously validated by humans. It identify 6 core capabilities (i.e., coarse perception, fine-grained perception, instance reasoning, logical reasoning, science \& technology, mathematics) along with 18 specific dimensions. \\ \hline
CV-Bench \cite{tong2024cambrian} & 2024 & provides 2,638 manually-inspected examples, and formulate natural language questions that evaluates 2D understanding via spatial relationships \& object counting, and 3D understanding via depth order \& relative distance. \\ \hline
MDVP \cite{lin2024draw} & 2024 & contains 1.6M unique image-visual prompt-text instruction-following samples, including natural images, document images, OCR images, mobile screenshots, web screenshots, and multi-panel images. \\ \hline
FOCI \cite{geigle2024african} & 2024 & constructed from 5 popular classification datasets for different domains: 1) aircraft contains images of 100 different aircraft types; 2) flowers contains images of 102 different flower species; 3) food covers 101 dishes; 4) pets contains images of 37 cat and dog breeds. 5) cars covers 196 car models. Additionally, FOCI create 4 domain subsets for animals (1322 classes), plants (957 classes), food (563 classes), and man-made objects (2631 classes). \\ \hline
MMVP \cite{tong2024eyes} & 2024 & summarizes 9 prevalent patterns of the CLIP-blind pairs, such as “orientation”, “counting”, and “viewpoint”. Utilizing the collected CLIP-blind pairs, MMVP design 150 pairs with 300 questions. \\ \hline
V*-Bench \cite{wu2024v} & 2024 & It is built based on 191 high-resolution images with an average image resolution of 2246×1582. V*-Bench contains two sub-tasks: attribute recognition and spatial relationship reasoning. The attribute recognition task has 115 samples. The spatial relationship reasoning task has 76 samples. \\ \hline
MME-RealWorld \cite{zhang2024mme} & 2024 & contains 29,429 question-answer pairs that cover 43 subtasks across 5 real-world scenarios, where each one has at least 100 questions. \\ \hline
Visual COT \cite{shao2024visual} & 2024 & 438k visual chain-of-thought question-answer pairs spans across five distinct domains, each consisting of a question, an answer, and an intermediate bounding box as CoT contexts. About 98k question-answer pairs include extra detailed reasoning steps. \\ \hline
Mementos \cite{wang2024mementos} & 2024 & It consists of 4,761 image sequences with varying episode lengths, encompassing diverse scenarios from dailylife, robotics tasks, and comic-style storyboards. Each sequence is paired with a human-annotated description of the primary objects and their behaviors within the sequence. \\ \hline
MIRB \cite{zhao2024benchmarking} & 2024 & comprises 925 samples with average image number of 3.78, constructed from four distinct categories of multi-image understanding: perception, visual world knowledge, reasoning, and multi-hop reasoning. \\ \hline
ReMI \cite{kazemi2024remi} & 2024 & It consists of 13 tasks that span a range of domains and properties. The tasks require reasoning over up to six images, with all tasks requiring reasoning over at least two images. The images comprise a variety of heterogeneous image types. \\ \hline
MuirBench \cite{wang2024muirbench} & 2024 & It consists of 12 distinctive multi-image understanding tasks that involve 10 categories of multi-image relations, comprising 11,264 images and 2,600 multiple-choice questions, with an average of 4.3 images per instance.  \\ \hline
VEGA \cite{zhou2024vega} & 2024 & contains 50k scientific literature entries, over 200k question-and-answer pairs, and a rich trove of 400k images. It includes the Interleaved Image-Text Comprehension subset, which is segmented into two categories based on token length: one supports up to 4,000 tokens, while the other extends to 8,000 tokens. Here, images are equated to 256 tokens each. Both categories offer roughly 200k training instances and approximately 700 test samples.  \\ \hline
MMBench \cite{liu2025mmbench} & 2025 & contains 3,217 multiple-choice questions covering a diverse spectrum of 20 fine-grained skills. \\ \hline
BLINK \cite{fu2025blink} & 2025 & reformats 14 classic computer vision tasks, and contains 3,807 multiple-choice questions across 7.3K images, paired with single or multiple images and visual prompting. \\ \hline
\multicolumn{3}{|l|}{\textbf{Optical Character Recognition}} \\ \hline
TextVQA \cite{singh2019towards} & 2019 & contains 45,336 questions asked by humans on 28,408 images that require reasoning about text to answer. Each question-image pair has 10 ground truth answers provided by humans. \\ \hline
OCR-VQA \cite{mishra2019ocr} & 2019 & comprises of 207,572 images of book covers and contains more than 1 million question-answer pairs about visual question answering by reading text in images. \\ \hline
WebSRC \cite{chen2021websrc} & 2021 & It consists of 400K question-answer pairs, which are collected from 6.4K web pages. Along with the QA pairs, corresponding HTML source code, screenshots, and metadata are also provided in the dataset. Each question in WebSRC requires a certain structural understanding of a web page to answer. \\ \hline
OCRBench \cite{liu2024ocrbench} & 2024 & includes 1000 question-answer pairs, which is consist of five components: text recognition, scene text-centric VQA, document-oriented VQA, key information extraction, and handwritten mathematical expression recognition.  \\ \hline
VCR \cite{zhang2024vcr} & 2024 & It comprise 2.11M English and 346K Chinese entities, featuring captions in both languages across ‘easy’ and ‘hard’ difficulty levels. \\ \hline
SEED-Bench-2-Plus \cite{li2024seed2plus} & 2024 & comprises 2.3K multiple-choice questions with precise human annotations, spanning three broad categories: Charts, Maps, and Webs. \\ \hline
\multicolumn{3}{|l|}{\textbf{Structural Document}} \\ \hline
FigureQA \cite{kahou2017figureqa} & 2017 & its training set contains 100,000 images with 1.3 million questions; the validation and test sets each contain 20,000 images with over 250, 000 questions. The images are synthetic, scientific-style figures from five classes: line plots, dot-line plots, vertical and horizontal bar graphs, and pie charts. \\ \hline
DocVQA \cite{mathew2021docvqa} & 2021 & contains 50,000 question-answer pairs with 12,767 document images sourced from documents in UCSF Industry Documents Library. \\ \hline
VisualMRC \cite{tanaka2021visualmrc} & 2021 &  It contains 30562 pairs of a question and an abstractive answer for 10,197 document images sourced from multiple domains of webpages. \\ \hline
ChartQA \cite{masry2022chartqa} & 2022 & consists of 20,882 charts curated from four different online sources. It covers 9,608 human-written questions focusing on logical and visual reasoning questions, and generates another 23,111 questions automatically from human-written chart summaries.  \\ \hline
InfographicVQA \cite{mathew2022infographicvqa} & 2022 & comprises 30,035 questions over 5,485 images. Questions in the dataset include questions grounded on tables, figures and visualizations and questions that require combining multiple cues. \\ \hline
ChartBench \cite{xu2023chartbench} & 2023 & includes over 68k charts and more than 600k high-quality instruction data, covering 9 major categories and 42 subcategories of charts. 5 chart question-answering tasks to assess the models’ cognitive and perceptual abilities. \\ \hline
SciGraphQA \cite{li2023scigraphqa} & 2023 & generate 295K samples of open-vocabulary multi-turn question-answering dialogues about the graphs. As context, it provided the text-only Palm-2 with paper title, abstract, paragraph mentioning the graph, and rich text contextual data from the graph itself, obtaining dialogues with an average 2.23 question-answer turns for each graph. \\ \hline
MMC-Benchmark \cite{liu2023mmc} & 2023 & consists of 2k QA pairs with 1,063 unique images, accompanied by 1,275 multiple-choice questions and 851 free-form questions. The average length of the questions is 15.6. \\ \hline
MP-DocVQA \cite{tito2023hierarchical} & 2023 & contains 46K questions and 6K documents, with a total of 48K pages (images). On average, each question is associated with 8.27 pages. \\ \hline
ChartX \cite{xia2024chartx} & 2024 & collected 48K multi-modal chart data covering 22 topics, 18 chart types, and 7 tasks. Each chart data within this dataset includes 4 modalities, including image, Comma-Separated Values (CSV), python code, and text description. 7 chart tasks is classified into perception tasks and cognition tasks. \\ \hline
DocGenome \cite{xia2024docgenome} & 2024 & constructed by annotating 500K scientific documents from 153 disciplines in the arXiv open-access community. It contains structure data from all modalities including 13 layout attributes along with their LATEX source codes. It provides 6 logical relationships between different entities within each scientific document. It covers various document-oriented tasks. \\ \hline
CharXiv \cite{wang2024charxiv} & 2024 & involves 2,323 real-world charts handpicked from scientific papers spanning 8 major subjects published on arXiv, and produces more than 10K questions. \\ \hline
MMLongBench-Doc \cite{ma2024mmlongbench} & 2024 & comprising 1,082 expert-annotated questions. It is constructed upon 135 lengthy PDF-formatted documents with an average of 47.5 pages and 21,214 textual tokens. 494 questions are single-page questions (with one evidence page). 365 questions are cross-page questions requiring evidence across multiple pages. 223 questions are designed to be unanswerable for detecting potential hallucinations. \\ \hline
ComTQA \cite{zhao2024tabpedia} & 2024 & comprises a total of 9,070 QA pairs across 1,591 images. It contains challenging questions, such as multiple answers, mathematical calculations, and logical reasoning. \\ \hline
Web2Code \cite{yun2024web2code} & 2024 & contains a total of 1179.7k webpage based instruction-response pairs. \\ \hline
VisualWebBench \cite{liu2024visualwebbench} & 2024 & consists of 7 tasks, and comprises 1.5K human-curated instances from 139 real websites, covering 87 sub-domains. \\ \hline
SciFIBench \cite{roberts2024scifibench} & 2024 & consists of 2000 multiple-choice scientific figure interpretation questions split between two tasks across 8 categories. The questions are curated from arXiv paper figures and captions. \\ \hline
\multicolumn{3}{|l|}{\textbf{Mathematics}} \\ \hline
MathVista \cite{lu2023mathvista} & 2023 & incorporates 28 existing multimodal datasets, including 9 math-targeted question answering (MathQA) datasets and 19 VQA datasets. In addition, it creates three new datasets (i.e., IQTest, FunctionQA, PaperQA) which are tailored to evaluating logical reasoning on puzzle test figures, algebraic reasoning over functional plots, and scientific reasoning with academic paper figures, respectively. It consists of 6,141 examples, with 736 of them being newly curated. \\ \hline
We-Math \cite{qiao2024we} & 2024 & It collect and categorize 6.5K visual math problems, spanning 67 hierarchical knowledge concepts and 5 layers of knowledge granularity. \\ \hline
Math-Vision \cite{wang2024measuring} & 2024 & comprises 3,040 mathematical problems within visual contexts across 12 grades, selected from 19 math competitions. It contains 1,532 problems in an open-ended format and 1,508 in a multiple-choice format. All problems encompass 16 subjects over 5 levels of difficulty. \\ \hline
Olympiadbench \cite{he2024olympiadbench} & 2024 & an Olympiad-level bilingual multimodal scientific benchmark, featuring 8,476 problems from Olympiad-level mathematics and physics competitions, including the Chinese college entrance exam. Each problem is detailed with expert-level annotations for step-by-step reasoning. \\ \hline
MathVerse \cite{zhang2025mathverse} & 2025 & It contains 2,612 math problems from three fundamental math subjects, i.e., plane geometry (1,746), solid geometry (332), and functions (534). Each problem is then transformed by human annotators into six distinct versions, each offering varying degrees of information content in multimodality, contributing to 15K test samples in total. \\ \hline
\multicolumn{3}{|l|}{\textbf{Video Understanding}} \\ \hline
TGIF-QA \cite{jang2017tgif} & 2017 & consists of 103,919 QA pairs collected from 56,720 animated GIFs. TGIF-QA includes four task types: repetition count, repeating action, state transition, frame QA. \\ \hline
ActivityNet-QA \cite{yu2019activitynet} & 2019 & It exploits 5,800 videos from the ActivityNet dataset, which contains about 20,000 untrimmed web videos representing 200 action classes. Each video is annotated with ten question-answer pairs using crowdsourcing to finally obtain 58,000 question-answer pairs. The maximum question length is 20 and the maximum answer length is 5. The average question length is 8.67 and average answer length is 1.85. \\ \hline
EgoSchema \cite{mangalam2023egoschema} & 2023 & consists of over 5000 human curated multiple choice question answer pairs, spanning over 250 hours of real video data. For each question, it requires the correct answer to be selected between five given options based on a three-minute-long video clip. \\ \hline
Video-MME \cite{fu2024video} & 2024 & It contains a annotated set of 2,700 high-quality multiple-choice questions (3 per video) from 900 videos, 744 subtitles and 900 audio files across various scenarios. For diversity in video types, it spans 6 visual domains, with 30 subfields. For duration in temporal dimension, it encompasses both short-, medium-, and long-term videos, ranging from 11 seconds to 1 hour. \\ \hline
MVBench \cite{li2024mvbench} & 2024 & covers 20 video temporal understanding tasks that cannot be effectively solved with a single frame. Each task produces 200 multiple-choice QA pairs by leveraging ChatGPT to automatically reannotate existing video datasets with their original annotations. \\ \hline
MMBench-Video \cite{fang2024mmbench} & 2024 & incorporates approximately 600 web videos from YouTube, spanning 16 major categories. Each video ranges in duration from 30 seconds to 6 minutes. The benchmark includes roughly 2,000 original question-answer pairs, contributed by volunteers, covering a total of 26 fine-grained capabilities. \\ \hline
MLVU \cite{zhou2024mlvu} & 2024 & consists of 3,102 questions across 9 categories with 2,593 questions for dev set and 509 questions for test set. It is made up of videos of diversified lengths, spanning from 3 min to more than 2 hours. Besides, each video is further partitioned as incremental segments, e.g., the first 3 min, the first 6 min, and the entire video. \\ \hline
LVBench \cite{wang2024lvbench} & 2024 & gathers an initial collection of 500 videos, each with a minimum duration of 30 minutes. Finally, these videos is annotated to select a subset of 103 videos. \\ \hline
Event-Bench \cite{du2024towards} & 2024 & includes 6 event-related tasks and 2,190 test instances. \\ \hline
VNBench \cite{zhao2024needle} & 2024 & 1,350 samples with 9 sub-tasks. \\ \hline
TempCompass \cite{liu2403tempcompass} & 2024 & collects a total of 410 videos and 500 pieces of meta-information, with 9 content categories. \\ \hline
MovieChat \cite{song2024moviechat} & 2024 & 1K long videos and 13K manual question-answering pairs. \\ \hline
\multicolumn{3}{|l|}{\textbf{Industry}} \\ \hline
MME-Industry \cite{yi2025mme} & 2025 & encompasses 21 distinct domain, comprising 1050 question-answer pairs with 50 questions per domain. \\ \hline
\multicolumn{3}{|l|}{\textbf{Conversational QA}} \\ \hline
SparklesDialogue \cite{huang2023sparkles} & 2023 & SparklesDialogueCC comprises 4.5K dialogues, each consisting of at least two images spanning two conversational turns. SparklesDialogueVG includes 2K dialogues, each with at least three distinct images across two turns. \\ \hline
SciGraphQA \cite{li2023scigraphqa} & 2023 & selected 290,000 Computer Science or Machine Learning ArXiv papers, and then used Palm-2 to generate 295K samples of open-vocabulary multi-turn question-answering dialogues about the graphs. As context, it provided the text-only Palm-2 with paper title, abstract, paragraph mentioning the graph, and rich text contextual data from the graph itself, obtaining dialogues with an average 2.23 question-answer turns for each graph. \\ \hline
ConvBench \cite{liu2024convbench} & 2024 & comprises 577 image-instruction pairs tailored for multi-round conversations. Each pair is structured around three sequential instructions, each targeting a distinct cognitive skill—beginning with perception, followed by reasoning, and culminating in creation. Encompassing 215 tasks, the benchmark is divided into 71 tasks focused on perception, 65 on reasoning, and 79 on creation. \\ \hline
MMDU \cite{liu2024mmdu} & 2024 & comprises 110 multi-image multi-turn dialogues with more than 1600 questions, each accompanied by detailed long-form answers. The questions in MMDU involve 2 to 20 images, with an average image\&text token length of 8.2k tokens, a maximum turn length of 27, and a maximum image\&text length reaching 18K tokens. \\ \hline
\multicolumn{3}{|l|}{\textbf{Multidisciplinary}} \\ \hline
ScienceQA \cite{lu2022learn} & 2022 & multiple-choice science question dataset containing 21,208 examples. It covers diverse topics across three subjects: natural science, social science, and language science. \\ \hline
MMMU \cite{yue2024mmmu} & 2024 & includes 11.5K multimodal questions from college exams, quizzes, and textbooks, covering 6 core disciplines. These questions span 30 subjects and 183 subfields, comprising 30 highly heterogeneous image types. \\ \hline
CMMU \cite{he2024cmmu} & 2024 & It consists of 3,603 questions in 7 subjects, covering knowledge from primary to high school. The questions can be categorized into 3 types: multiple-choice, multiple-response, and fill-in-the-blank. \\ \hline
CMMMU \cite{zhang2024cmmmu} & 2024 & A Chinese Multi-discipline multimodal Understanding, including 12k manually collected multimodal questions from college exams, quizzes, and textbooks, covering 6 core disciplines. These questions span 30 subjects and comprise 39 highly heterogeneous image typesbenchmark. \\ \hline
MMMU-Pro \cite{yue2024mmmupro} & 2024 & 3460 questions in total (1730 samples are in the standard format and the other 1730 are in the screenshot or photo form) \\ \hline
\end{longtable}
\endgroup

\section{Evaluation Metrics of MRAG}
\label{sec:EvaluationMetricsMRAG}



Multimodal RAG systems generally consist of four core components: document parsing, search planning, retrieval, and generation, which collectively influence their end-to-end performance. Accurate and comprehensive evaluation of these components is essential, leveraging available multimodal benchmarks. In practice, three common evaluation strategies are typically employed: human evaluation, rule-based evaluation, and LLM/MLLM-based evaluation. Each strategy offers distinct advantages and disadvantages in calculating evaluation metrics.
\begin{itemize}[leftmargin=1em]
\item \emph{\textbf{Human evaluation}}: Human evaluation is widely regarded as the gold standard for assessing MRAG systems, as their effectiveness is ultimately determined by human users. This method is extensively used in research to ensure the reliability and relevance of model outputs. For instance, Bingo \cite{cui2023holistic} employs human annotators to assess the accuracy of GPT-4V’s responses, with a focus on identifying and analyzing model biases. In hallucination detection, M-HalDetect \cite{gunjal2024detecting} demonstrates that human evaluation outperforms model-based methods in detecting subtle inaccuracies, highlighting its precision. Additionally, WV-Arena \cite{lu2024wildvision} uses a human voting system combined with Elo ratings to rank and compare multiple models, providing a robust benchmarking framework.
However, human evaluation presents challenges, including increased time and labor costs, which limit its scalability for large-scale assessments. The reliability of results can also be affected by the limited number of evaluators, as individual biases may influence outcomes. To address these issues, some studies employ diverse evaluator pools and cross-validation techniques to enhance the balance and representativeness of assessments. Nonetheless, the trade-off between evaluation accuracy and resource expenditure remains a critical consideration in designing RAG model evaluation methodologies.
\item \emph{\textbf{Rule-based evaluation}}: Rule-based evaluation metrics \cite{chen2021websrc, zhang2024vcr, yin2024lamm} are essential for assessing the performance of MRAG systems. These metrics rely on standardized evaluation tools, enabling objective, reproducible assessments with minimal human intervention. Compared to subjective human evaluations, deterministic metrics offer significant advantages, including reduced time consumption, lower susceptibility to bias, and greater consistency across multiple assessments. Such consistency is particularly crucial for large-scale evaluations or when comparing different systems or model iterations.
\item \emph{\textbf{LLM/MLLM-based evaluation}}: For evaluation of MRAG systems, LLMs/MLLMs are employed to compare reference answers with generated outputs or to directly score responses. For example, MM-Vet \cite{yu2023mm} uses GPT-4 to automate evaluation, generating scores for each sample based on the input question, ground truth, and model output. Similarly, TouchStone \cite{bai2023touchstone} and LLaVA-bench \cite{liu2023visual} leverage GPT-4 to directly compare generated answers with reference answers, simplifying the evaluation process.
While integrating LLMs/MLLMs in evaluation reduces human effort, it has limitations. This approach is prone to systematic biases, such as sensitivity to the order of response presentation. Additionally, evaluation outcomes are heavily influenced by the inherent capabilities and limitations of the LLMs/MLLMs themselves, leading to potential inconsistencies, as different models may produce divergent results for the same task. These challenges underscore the need for careful model selection and evaluation design to mitigate biases and ensure reliable assessments.
\end{itemize}

\subsection{Metrics of Retrieval \& Generation}
The evaluation of MRAG systems is essential for ensuring their effectiveness and reliability in processing complex, multimodal data. Evaluation metrics can be broadly classified into rule-based and LLM/MLLM-based approaches.
\begin{itemize}[leftmargin=1em]
\item \textbf{Rule-based Metrics}: Rule-based metrics evaluate the performance of MRAG systems using predefined criteria and heuristics. These metrics are generally interpretable, transparent, and computationally efficient, making them well-suited for tasks with well-defined benchmarks. Examples of common rule-based metrics include:
    \begin{itemize}[leftmargin=1em]
    \item \textbf{Exact Match (EM)}: This metric evaluates whether the model’s output exactly matches the ground truth, offering a clear and unambiguous performance measure. It is especially valuable in tasks requiring high accuracy and fidelity to reference data, such as question answering, fact verification, and information retrieval. While exact match (EM) provides a straightforward and interpretable evaluation, it may fall short in scenarios where semantically equivalent but lexically divergent responses are acceptable.
    \item \textbf{ROUGE-N (N-gram Recall)}: The ROUGE metric is a widely used framework for evaluating text summarization and generation tasks. ROUGE-N measures the overlap of N-grams (contiguous sequences of N words) between generated text and one or more reference texts, with a strong emphasis on recall. This metric assesses how well the generated text captures the essential content of the reference. For example, ROUGE-1 evaluates unigram overlap, ROUGE-2 focuses on bigrams, and higher-order N-grams (e.g., ROUGE-3) capture more complex linguistic structures. While ROUGE-N provides a quantitative measure of lexical similarity, it is often supplemented by other metrics to account for semantic coherence, fluency, and relevance, particularly in multimodal contexts where textual and non-textual data interact. 
    \item \textbf{BLEU}: BLEU is a widely used metric in NLP for evaluating the quality of machine-generated text by assessing its similarity to one or more reference texts. Initially designed for machine translation, BLEU has been adapted to various NLP tasks, including multimodal generation. In multimodal settings, BLEU can evaluate the alignment between generated text and associated modalities (e.g., images, videos) by comparing the output to reference descriptions or captions. However, while BLEU offers a quantitative measure of n-gram overlap, it has limitations in capturing semantic depth, contextual coherence, and multimodal consistency, which are essential for comprehensive evaluation in MRAG systems.
    \item \textbf{Mean Reciprocal Rank (MRR)}: MRR is a widely used metric for evaluating the performance of systems that produce ranked lists of results, such as search engines, recommendation systems, or retrieval-augmented models. MRR measures the rank position of the first relevant item in the returned list, reflecting the system's ability to surface correct or useful information quickly. It is calculated as the average of the reciprocal ranks of the first relevant result across multiple queries or tasks. A higher MRR indicates better performance, as it demonstrates the system's effectiveness in prioritizing relevant results at the top of the list.
    \item \textbf{CIDEr (Consensus-based Image Description Evaluation)}: CIDEr is specifically designed to measure the agreement between machine-generated captions and human-authored reference captions. It utilizes a TF-IDF weighting mechanism to quantify the similarity between generated and reference texts.
    \item \textbf{SPICE (Semantic Propositional Image Caption Evaluation)}: The evaluation of MRAG systems frequently utilizes the SPICE metric to assess the quality of generated captions. SPICE prioritizes semantic fidelity by parsing captions into structured scene graphs, which depict objects, attributes, and relationships within the text. These generated scene graphs are subsequently compared to reference graphs derived from ground-truth captions. By emphasizing semantic similarity over lexical overlap, SPICE offers a robust measure of how well the generated content aligns with the intended meaning. This makes it particularly well-suited for evaluating multimodal systems that integrate visual and textual information, ensuring a nuanced and contextually accurate assessment of MRAG outputs.
    \item \textbf{BERTScore}: Evaluation of MRAG focuses on assessing the quality and relevance of outputs in contexts integrating both textual and non-textual data (e.g., images, audio). A key metric for evaluating textual components is BERTScore, which utilizes contextual embeddings from BERT to measure semantic similarity between generated and reference texts. Unlike traditional metrics such as BLEU or ROUGE, which depend on exact word matches or n-gram overlap, BERTScore captures deeper semantic relationships by aligning tokens based on their contextual embeddings.
    \item \textbf{Perplexity}: It measures the model's ability to predict the next word in a sequence, with lower perplexity values indicating greater confidence and accuracy in predictions. This reflects a stronger understanding of the underlying data distribution.
    \end{itemize}
Rule-based metrics offer objective and reproducible outcomes but frequently lack the adaptability needed to capture nuanced semantic or contextual understanding, especially in multimodal environments where text, images, and other data types interact.
\item \textbf{LLM/MLLM-based Metrics}: The emergence of LLMs and MLLMs has transformed evaluation paradigms, enabling the use of their advanced reasoning and comprehension capabilities. LLM/MLLM-based metrics now provide more holistic and context-aware assessments of MRAG systems, with key approaches including:
    \begin{itemize}[leftmargin=1em]
    \item \textbf{Answer Precision}: This metric measures the degree to which the knowledge in a model-generated answer is supported or entailed by the ground truth. It assesses the accuracy and relevance of retrieved information by evaluating the overlap between the model's output and the factual or contextual basis provided by the ground truth. High answer precision indicates that the model effectively utilizes retrieved multimodal data to produce responses aligned with the expected factual content. This metric is crucial for evaluating the reliability and factual consistency of multimodal RAG systems, ensuring that generated outputs are both contextually appropriate and informationally accurate.
    \item \textbf{Ground Truth Recall}: This metric measures the degree to which the knowledge in the ground truth is accurately captured and reflected in the model-generated response. It assesses the model's ability to retrieve and integrate relevant information from the provided knowledge base or multimodal sources, ensuring the output aligns with the factual or contextual details in the reference data. It is particularly crucial for evaluating retrieval-augmented systems, as it directly quantifies the fidelity of the model's output to the intended knowledge. Higher scores indicate stronger alignment with the ground truth, reflecting enhanced retrieval and generation capabilities.
    \item \textbf{Retrieved Context Precision}: This metric measures the alignment between the knowledge in the retrieved context and the information in the ground truth response. It evaluates the proportion of relevant and accurate information in the retrieved context that is directly supported or entailed by the ground truth, assessing the retrieval system's precision and contextual appropriateness in generating accurate responses. This metric is especially vital in multimodal RAG systems, where integrating diverse data types (e.g., text, images, audio) requires robust evaluation of relevance and precision across modalities.
    \item \textbf{Retrieved Context Recall}: This metric measures the degree to which the retrieved context aligns with and encompasses the knowledge necessary to generate ground truth responses. It evaluates the proportion of relevant information from the ground truth captured within the retrieved context, serving as a key indicator of the retrieval system's effectiveness in supporting accurate and comprehensive response generation. High values indicate that the retrieval mechanism effectively identifies and incorporates essential knowledge, thereby enhancing the overall performance of the MRAG system.
    \item \textbf{Faithfulness}: This metric evaluates the extent to which generated text maintains factual consistency with the information in the retrieved documents, ensuring the output accurately reflects the source material and minimizes hallucinations or deviations from the evidence. In MRAG systems, it also ensures alignment with multimodal retrieved content, including textual, visual, and auditory elements, maintaining consistency across modalities.
    \item \textbf{Hallucination}: This metric measures the proportion of generated outputs containing hallucinated content, such as unsupported claims, fabricated information, or inaccuracies not substantiated by the retrieved data. It is essential for evaluating the reliability and factual consistency of the model's responses.
    \end{itemize}
LLM/MLLM-based metrics are highly effective at capturing complex semantic relationships and contextual nuances, making them particularly suitable for multimodal RAG systems. However, they may inherit biases from the underlying models and demand substantial computational resources.
\item \textbf{Metric Calculation}: When evaluating multimodal retrieval-augmented generation systems, implementation methods for the same metric can vary significantly, primarily categorized into coarse-grained and fine-grained approaches. These methodologies differ in their granularity and the depth of analysis applied to assess the quality of model-generated responses against reference answers.
    \begin{itemize}[leftmargin=1em]
    \item \textbf{Coarse-Grained Evaluation:} Coarse-grained evaluation utilizes LLMs or MLLMs to compare model-generated responses with reference answers. This method involves inputting both the generated output and the reference into the LLM/MLLM, which evaluates the overall semantic alignment, coherence, and relevance between the two. The model assesses whether the generated content captures the core meaning and intent of the reference, providing a holistic score or qualitative feedback.
    This approach is computationally efficient and scalable, making it suitable for rapid benchmarking and high-level quality checks in large-scale applications. However, its broad focus may overlook fine-grained inaccuracies, such as subtle factual errors, logical inconsistencies, or nuanced contextual mismatches. Consequently, coarse-grained evaluation is best used as an initial screening tool or in scenarios where high-level semantic fidelity is prioritized over detailed precision. For more rigorous evaluation, it is often supplemented by fine-grained metrics that address specific aspects of content quality.
    In summary, coarse-grained evaluation offers a pragmatic balance between efficiency and effectiveness, particularly in applications requiring quick assessments or large-scale model comparisons.
    \item \textbf{Fine-Grained Evaluation:} Fine-grained evaluation, such as RAGChecker \cite{ru2024ragchecker} and RAGAS \cite{es2024ragas}, offers a nuanced and detailed approach to assessing MRAG systems, surpassing the limitations of coarse-grained methods. This approach involves decomposing both model-generated responses and reference answers into granular knowledge points or semantic units, which are individually evaluated based on criteria such as accuracy, relevance, and alignment with the reference. By analyzing responses at this level of detail, the method enables precise identification of a model's strengths and weaknesses, particularly in capturing and reproducing intricate information.
    The fine-grained approach is especially valuable for diagnosing performance issues in handling complex or nuanced content. However, it is computationally intensive, requiring robust mechanisms for extracting, matching, and evaluating multiple semantic units. Careful design of these mechanisms is essential to ensure evaluation consistency and reliability. Despite its challenges, this method provides a rigorous and comprehensive framework for advancing the development and refinement of MRAG systems, making it a critical tool in the field.
    \end{itemize}
The choice between coarse-grained and fine-grained evaluation depends on the assessment objectives. Coarse-grained methods are ideal for obtaining quick, high-level insights, whereas fine-grained approaches are better suited for detailed analysis and iterative model refinement. Integrating both strategies can provide a balanced perspective, combining the efficiency of coarse-grained evaluation with the precision of fine-grained analysis to comprehensively assess MRAG systems.
\end{itemize}

\section{Challenges of MRAG}
\label{sec:ChallengesMRAG}
In this section, we delineate the challenges associated with various modules in a MRAG system. These challenges span multiple critical components, including document parsing and indexing, search planning, retrieval, generation, dataset, and evaluation. Each module presents unique complexities that must be addressed to ensure the system's effectiveness and robustness.

\subsection{Document Parsing and Indexing}
Document parsing and indexing has established the data foundation based on MRAG, which plays a crucial role in the entire system. The relevant technologies extensively studied even before the advent of LLMs, have seen significant advancements in the LLM era. However, they continue to face several challenges that necessitate further exploration and refinement.
\begin{itemize}[leftmargin=1em, listparindent=\parindent]
\item \textbf{Challenges in Data Accuracy and Completeness:} As the primary input source, the accuracy and completeness of upstream data are critical. Errors or omissions in the upstream data can propagate and amplify downstream, significantly degrading system performance. For example, while MRAG systems have enhanced document information preservation—such as capturing per-page screenshots—they still face challenges in maintaining inter-page relationships. This limitation is particularly problematic in long documents with associated segments. Preserving these relationships is essential for ensuring contextually accurate outputs.
\item \textbf{Balancing Multimodal and Textual Data:} The document parsing module has grown increasingly complex as modern MRAG systems must handle multimodal data, including images, tables, and text. To address this, contemporary approaches preserve the original multimodal inputs to minimize information loss, while also converting them into textual captions or descriptions. Although retaining the original data reduces information loss, relying solely on it has proven suboptimal. Recent studies highlight the benefits of leveraging textual representations derived from multimodal data. For example, \citet{BeyondText2024} showed that models generate higher-quality responses using textual captions from images rather than processing raw images directly. Similarly, \citet{mmRAG2024} found that LLMs outperform MLLMs in text generation tasks, revealing a performance gap between multimodal and text-focused systems. This gap highlights the limitations of current multimodal systems in effectively integrating diverse data types, necessitating additional components in document parsing pipelines. These enhancements, while improving functionality, increase system complexity and expand the volume of data requiring processing, storage, and management.
\end{itemize}

\subsection{Multimodal Search Planning}
The challenges in multimodal search planning can be more effectively understood through a hierarchical framework similar to leveled RAG systems, where queries span a spectrum from simple factual retrievals to complex, creative tasks. This framework highlights three critical challenges that must be addressed to advance the field.
\begin{itemize}[leftmargin=1em, listparindent=\parindent]
\item \textbf{Intelligent Adaptive Planning Mechanisms:} The primary challenge is developing intelligent adaptive planning mechanisms that can dynamically adjust to the diversity and complexity of queries. Current systems often rely on predetermined pipelines, which fail to accommodate variations in query characteristics or computational constraints, leading to inefficient resource allocation and suboptimal performance \cite{hu2024mragbench, zhang2024mr}. While fixed strategies may suffice for homogeneous query types, real-world applications handle heterogeneous query patterns that demand dynamic adjustment of retrieval strategies. For example, complex queries involving multi-hop reasoning or creative problem-solving could greatly benefit from a multi-agent collaborative approach \cite{wang2025pikeragspecializedknowledgerationale}. In such a framework, specialized agents could explore parallel reasoning paths, propose complementary retrieval strategies, and collaboratively synthesize findings to construct comprehensive search plans. This collaborative paradigm not only simulates diverse perspectives but also facilitates intricate interactions between knowledge sources and reasoning steps. By evaluating search plans from multiple angles, such systems can balance effectiveness and efficiency, ensuring robust performance across diverse query types.
\item \textbf{Query Reformulation and Semantic Alignment:} A second major challenge is query reformulation, particularly in maintaining semantic alignment between the original multimodal query intent and the reformulated queries \cite{li2024benchmarkingmultimodalretrievalaugmented}. As queries become more sophisticated, accurately capturing and maintaining their intent grows increasingly complex. This challenge is amplified in multimodal contexts, where queries may integrate text, images, audio, or other data types, each requiring precise interpretation. To address this, multi-perspective reformulation strategies could be employed, leveraging diverse interpretations of the query to generate reformulations that better align with the original intent. Such strategies might integrate contextual understanding, domain-specific knowledge, and cross-modal alignment techniques to ensure semantic consistency with the user's intent.
\item \textbf{Comprehensive Evaluation Benchmarks:} The third critical challenge is the absence of comprehensive evaluation benchmarks capable of assessing planning mechanisms across diverse query complexities and scenarios. Existing benchmarks often focus on narrow performance aspects, failing to capture the full spectrum of real-world applications. To address this gap, future benchmarks should evaluate systems across multiple dimensions, including adaptability to query diversity, robustness in handling complex queries, and efficiency in resource utilization. These benchmarks should incorporate a wide range of query types, from simple factual retrievals to multi-hop reasoning and creative tasks, ensuring rigorous testing under realistic conditions. Additionally, benchmarks should incorporate metrics for semantic alignment in query reformulation, computational efficiency, and scalability.
\end{itemize} 
These interconnected challenges highlight the need for future research to develop adaptive planning mechanisms capable of addressing both query diversity and complexity. This could involve multi-agent coordination for advanced cases, alongside robust query reformulation and comprehensive evaluation frameworks.

\subsection{Retrieval}
Multimodal retrieval has made significant progress but continues to face challenges that can be categorized into methodological and practical issues. These challenges arise from the inherent complexity of integrating and retrieving information across diverse data modalities such as text, images, audio, and video. Below, we outline the key challenges in this field:
\begin{itemize}[leftmargin=1em]
\item \textbf{Heterogeneity of Cross-Modal Data}: The heterogeneity of data across modalities poses a significant challenge in multimodal retrieval and representation learning. Text, being sequential and discrete, relies on syntactic and semantic structures best captured by language models, while images, being spatial and continuous, require convolutional or transformer-based architectures to extract hierarchical visual features. This structural divergence complicates cross-modal alignment and comparison, as each modality demands specialized feature extraction techniques tailored to its unique characteristics.
Extracting meaningful and comparable features from each modality is non-trivial, requiring domain-specific expertise and sophisticated models capable of capturing nuanced data properties. For instance, while transformers excel in processing sequential data like text, their adaptation to spatial data like images often necessitates architectural modifications, such as vision transformers (ViTs), to handle pixel arrays. Aligning these features into a unified representation space that preserves cross-modal semantic relationships remains a major challenge. Current approaches, including cross-modal transformers and MLLMs, often fail to create a common embedding space that adequately captures the semantic richness of each modality while ensuring inter-modal consistency.
\item \textbf{Cross-modal components (reranker, refiner)}: While the dual-tower architecture has made significant strides in first-stage retrieval by efficiently encoding and aligning multimodal data (e.g., text and images), developing advanced reranking models that enable fine-grained multimodal interaction remains a challenge. Additionally, refining external multimodal knowledge post-retrieval and reranking remains underexplored, despite its potential to enhance result accuracy and relevance. Addressing these gaps requires innovative methodologies that leverage MLLMs and LLMs to enable sophisticated cross-modal understanding and reasoning. 
\end{itemize}

\subsection{Generation}
The multimodal module in MRAG achieves human-aligned sensory representation through diversified modality integration, which significantly enhances user experience and system usability. However, achieving these enhancement objectives entails addressing shared challenges across both QA systems and multimodal generation:
\begin{itemize}[leftmargin=1em]
\item \textbf{Multimodal Input}: Multimodal systems face the challenge of integrating diverse data structures and representations across modalities such as text, images, audio, and video.
As multimodal models evolve, they are increasingly required to process arbitrary combinations of modalities (e.g., text+image, text+video, image+audio). This necessitates a highly flexible and adaptive framework capable of dynamically accommodating diverse input configurations. Such frameworks must be modality-agnostic, enabling seamless integration of any input combination without predefined structures or extensive retraining. Achieving this flexibility involves designing architectures that generalize across modalities, extract relevant features, and fuse them meaningfully, regardless of input composition. 
\item \textbf{Multimodal Output}: 
    \begin{itemize}[leftmargin=1em]
    \item \textbf{Coherent and Contextually Relevant Generation}: Ensuring consistency across different modalities in the output presents a significant challenge. For instance, in a text-image pair, the image must accurately reflect the scene or object described in the text, while the text should precisely convey the visual content. 
    \item \textbf{Positioning and Integration of Multimodal Elements}: In multimodal outputs, such as text with embedded images or videos, the model must intelligently determine where to integrate non-textual elements. This requires an understanding of the narrative flow and the identification of optimal insertion points to enhance coherence and readability.
    Additionally, the model should dynamically generate or retrieve relevant multimodal content based on context. For instance, when creating a text-image pair, the model may need to generate an image caption, search for relevant images, and select the most appropriate one. This process must be efficient and seamless to ensure the final output is both relevant and high-quality.
    \item \textbf{Diversity of Outputs}: In some applications, generating diverse outputs—such as multiple images or videos corresponding to a given text description—is essential. However, balancing diversity with relevance and quality poses a significant challenge. The model must explore a broad range of possibilities while ensuring each output remains contextually appropriate and adheres to high-quality standards.
    \end{itemize}
\end{itemize}

\subsection{Dataset \& Evaluation}
The advancement of MLLMs has heightened the need for comprehensive evaluation. Despite the introduction of over a hundred benchmarks by both academic and industrial communities, several challenges remain in the current evaluation landscape.
First, there is a lack of a universally accepted, standardized capability taxonomy, with existing benchmarks often defining their own disparate ability dimensions. Second, current benchmarks exhibit significant gaps in critical areas such as instruction following, complex multimodal reasoning, multi-turn dialogue, and creativity assessment. Third, task-specific evaluations for MLLMs are insufficient, particularly in commercially relevant domains like invoice recognition, multimodal knowledge base comprehension, and UI understanding and industry. Finally, while existing multimodal benchmarks primarily focus on image and video modalities, there is a notable deficit in assessing capabilities related to audio and 3D representations. Addressing these challenges is essential for developing more robust and comprehensive evaluation methodologies for MLLMs in the future.

Despite rapid advancements, current evaluations of MLLMs remain insufficiently comprehensive, primarily focusing on perception and reasoning abilities through objective questions. This creates a significant gap between evaluation methodologies and real-world applications. Moreover, optimizing models based on objective assessments often leads developers to prioritize objective question corpora during instruction tuning, potentially degrading the quality of dialogue experiences. Although subjective multimodal evaluation platforms like WildVision and OpenCompass MultiModal Arena have emerged, further research is needed to develop assessment methods that better align with practical usage scenarios. Current evaluation strategies predominantly rely on curated or crafted questions to assess specific capabilities, yet complex multimodal tasks typically require the integration of multiple skills. For instance, a chart-related question may involve OCR, spatial relationship recognition, reasoning, and calculations. The absence of decoupled assessments for these distinct capabilities represents a major limitation in existing frameworks. Additionally, crucial abilities such as instruction following remain under-evaluated. Multiturn dialogue, the primary mode of human interaction with multimodal models, remains a weakness for most models, and corresponding evaluations, are still in their infancy. In the realm of complex multimodal reasoning, current evaluations predominantly focus on mathematical and examination problems, necessitating improvements in both difficulty and relevance to everyday use cases. Notably, the evaluation of multimodal creative tasks, a key application area for these models—such as text generation based on image and textual prompts—remains largely unexplored, highlighting a critical gap in the current evaluation landscape.

MLLMs are still in the early stages of development, with limited business applications to date. As a result, current evaluations primarily focus on assessing foundational capabilities rather than real-world performance. Moving forward, it is critical to develop evaluation frameworks that measure MLLM performance on specific tasks with commercial value, such as large-scale document processing, multimodal knowledge base comprehension, anomaly detection, and industrial visual inspection. When designing task-specific evaluations, it is essential to consider not only performance metrics but also computational costs and inference speeds, benchmarking them against traditional computer vision methods like OCR, object detection, and action recognition to determine practical applicability. Additionally, a key potential of MLLMs lies in their ability to plan and interact with environments as agents to solve complex problems. Developing diverse virtual environments for MLLMs to demonstrate agent-based problem-solving capabilities will likely become a critical component of future evaluations. Current efforts in this domain remain nascent, highlighting a promising area for future research in multimodal AI assessment.

\section{Future Directions}
\label{sec:FutureWork}
In this chapter, we propose several suggestions to the future development of multimodal Retrieval-Augmented Generation (MRAG) systems, informed by related research and identified challenges. These recommendations collectively aim to overcome existing limitations and unlock the full potential of MRAG in complex, real-world scenarios.

\subsection{Documents Parsing}
Multimodal document parsing has become a crucial element in MRAG systems, particularly with the emergence of large language models (LLMs) and multimodal large models (MLLMs). The fusion of text, images, and other data types into a cohesive framework presents both transformative opportunities and notable challenges. This paper provides a detailed analysis of future directions in this evolving field.
\begin{itemize}[leftmargin=1em]
\item \textbf{Enhancing Data Accuracy and Completeness:}
    \begin{itemize}[leftmargin=1em]
    \item \textbf{Contextual Relationship Preservation:} To improve the accuracy and coherence of multimodal document parsing, especially for long and complex documents, advanced algorithms are needed to capture and preserve both inter-page and intra-document relationships. Techniques such as graph-based representations and hierarchical document modeling can help maintain contextual coherence across the document. These methods enable the system to understand structural and semantic dependencies between sections, tables, figures, and other elements, ensuring the preservation of the document's logical flow.
    Additionally, cross-referencing mechanisms are essential for linking related content across pages. These mechanisms dynamically connect sections, tables, and figures, facilitating seamless retrieval and utilization of contextual relationships in downstream tasks like information extraction, summarization, or question answering. By integrating these approaches, the system can better handle the complexities of long documents, ensuring accurate maintenance and leveraging of contextual relationships for enhanced performance in multimodal document understanding tasks. This is particularly relevant when combining Optical Character Recognition (OCR), LLMs, and MLLMs to process and interpret documents with diverse content types.
    \item \textbf{Error Detection and Correction:} To improve the accuracy and reliability of multimodal document parsing, integrating advanced error detection and correction mechanisms is crucial. Leveraging LLMs and MLLMs, systems can validate extracted text against the original document, identifying and correcting inaccuracies or omissions. These models can be enhanced with consistency-checking algorithms to ensure coherence and accuracy across multimodal data, including text, images, and tables.
    For critical documents, a human-in-the-loop (HITL) approach is advisable. This involves human reviewers verifying and refining parsed data, especially in cases where systems may struggle with complex layouts, ambiguous content, or domain-specific nuances. By combining the strengths of LLMs, MLLMs, and human expertise, this hybrid approach ensures high accuracy and reliability, making it suitable for precision-demanding applications such as legal, medical, or financial document processing. 
    \end{itemize}
\item \textbf{Improving Multimodal Data Integration:} 
    \begin{itemize}[leftmargin=1em]
    \item \textbf{Unified Multimodal Representation:} Advancing multimodal document parsing requires the development of unified representation frameworks that seamlessly integrate diverse data types, such as text, images, and tables, into a cohesive structure. Such frameworks enable robust, context-aware analysis by leveraging multimodal transformers—like CLIP, Flamingo, or other state-of-the-art models—to encode disparate modalities into a shared embedding space. This interoperability enhances downstream tasks, including information extraction, question answering, and summarization.
    A promising approach involves hybrid strategies that combine raw multimodal data with textual representations. For instance, raw images can support visual tasks (e.g., object detection or layout analysis), while textual captions or OCR-derived text can improve text generation tasks (e.g., summarization or translation). This dual methodology leverages the strengths of each modality, ensuring both accuracy and efficiency in processing complex documents, as demonstrated by recent research.
    Additionally, integrating LLMs and MLLMs with Optical Character Recognition (OCR) systems can enhance the parsing of scanned or image-based documents. By aligning OCR outputs with multimodal embeddings, these systems improve the handling of noisy or unstructured data, enabling more accurate interpretation and contextual understanding.
    \item \textbf{Advanced Captioning and Description Generation:} To improve multimodal data integration, particularly in document parsing, enhancing automated captioning and description generation for non-textual elements like images, tables, and charts is critical. Leveraging state-of-the-art vision-language models (VLMs) and MLLMs can boost the accuracy and contextual relevance of textual descriptions. These models bridge the gap between visual and textual data, enabling more comprehensive document understanding.
    Integrating domain-specific knowledge into captioning models is essential for generating accurate and contextually tailored descriptions. This can be achieved by fine-tuning pre-trained models on domain-specific datasets or incorporating external knowledge bases. Such an approach ensures that descriptions align with the document's content, enhancing the utility of multimodal data integration.
    \end{itemize}
\item \textbf{Leveraging LLMs and MLLMs for Enhanced Parsing:} 
    \begin{itemize}[leftmargin=1em]
    \item \textbf{LLM/MLLM-Driven Parsing and Indexing:} LLMs and MLLMs can be fine-tuned on domain-specific corpora to improve their ability to parse and interpret complex document structures. Leveraging their advanced multimodal understanding, these models can accurately identify and extract key information—such as legal clauses, scientific hypotheses, or technical specifications—even from dense or unstructured text. Fine-tuning enhances their proficiency in recognizing domain-specific terminology, relationships, and contextual nuances.
    Furthermore, LLMs and MLLMs can generate metadata, tags, and summaries. By automatically annotating documents with relevant keywords, classifications, or concise summaries, these models streamline the organization and accessibility of large document repositories. This capability is particularly valuable in applications like legal case management, academic research, and enterprise knowledge bases.
    In multimodal contexts, MLLMs extend these capabilities by integrating and interpreting data from diverse sources, such as text, images, tables, and diagrams. This enables a more comprehensive parsing process, where visual and textual elements are jointly analyzed to extract richer, more accurate information. For example, in scientific documents, MLLMs can parse and correlate data from textual descriptions and accompanying charts, facilitating a deeper understanding of the content.
    \item \textbf{Bridging the Gap Between LLMs and MLLMs:} A promising approach involves hybrid architectures that combine the strengths of MLLMs and LLMs. MLLMs process raw multimodal inputs (e.g., images, audio, video) to extract meaningful representations, while LLMs generate coherent and contextually accurate textual outputs. This division of labor optimizes performance, as MLLMs excel in multimodal feature extraction and LLMs in linguistic precision. For example, in document parsing, MLLMs analyze visual layouts, tables, or embedded graphics, while LLMs synthesize this information into structured textual formats.
    \end{itemize}
\end{itemize}

\subsection{Multimodal Search Planning}
The future of multimodal search planning should focus on addressing three key challenges within the hierarchical framework: intelligent adaptive planning mechanisms, query reformulation and semantic alignment, and comprehensive evaluation benchmarks. Below are targeted suggestions for advancing each area.
\begin{itemize}[leftmargin=1em]
\item \textbf{Intelligent Adaptive Planning Mechanisms:}
    \begin{itemize}[leftmargin=1em]
    \item \textbf{Multi-Agent Collaborative Systems:} To address the challenges of multimodal search and complex query resolution, multi-agent collaborative systems can be designed to leverage specialized agents working in tandem. These systems enhance efficiency, adaptability, and robustness in handling multi-hop, creative, or cross-modal queries. Key mechanisms include: 
    1) Parallel Reasoning Paths: Specialized agents can simultaneously explore multiple reasoning trajectories, enabling faster and more comprehensive solutions. This approach is particularly effective for multi-hop queries, where intermediate reasoning steps are critical, or for creative tasks requiring diverse perspectives. By evaluating multiple pathways in parallel, the system can identify optimal solutions while mitigating the risk of local optima.
    2) Complementary Retrieval Strategies: Agents can employ diverse retrieval methodologies, such as keyword-based, semantic, or cross-modal retrieval, to address different aspects of a query. For instance, one agent might focus on extracting structured data, while another leverages semantic embeddings or visual-textual alignment for multimodal contexts. The synthesis of these strategies ensures robust and contextually relevant search outcomes, enhancing the system's ability to handle heterogeneous data sources.
    3) Dynamic Resource Allocation: Agents can monitor computational resources and system constraints in real-time, dynamically adjusting retrieval strategies to optimize performance. For example, under limited computational bandwidth, agents might prioritize lightweight retrieval methods or redistribute tasks to balance load. This adaptive mechanism ensures efficient resource utilization while maintaining high-quality query resolution.
    4) Integration with MLLMs and LLMs: The collaborative multi-agent framework can be seamlessly integrated with MLLMs and LLMs to enhance their capabilities. MLLMs can serve as central orchestrators, interpreting multimodal inputs and coordinating agent tasks, while LLMs provide deep contextual understanding and reasoning support. This integration enables the system to handle complex, multimodal queries with greater precision and adaptability.
    \item \textbf{Hierarchical Planning Frameworks:} To address the challenges of ambiguous or highly creative queries in multimodal search and planning, integrating human feedback into the decision-making process is essential. Human-in-the-loop (HITL) systems facilitate iterative refinement by leveraging user expertise to guide and validate intermediate results. These interactive systems enable users to dynamically adjust search parameters, prioritize modalities, or correct misinterpretations, ensuring more accurate and contextually relevant outcomes. By combining the strengths of multimodal large language models (MLLMs) with human intuition, HITL systems enhance adaptability, build trust, and improve the robustness of intelligent planning frameworks. This collaborative approach is particularly valuable in domains requiring nuanced understanding, creativity, or domain-specific knowledge.
    \item \textbf{Reinforcement Learning for Adaptation:} Intelligent adaptive planning mechanisms can be developed using reinforcement learning (RL) to enable dynamic, context-aware decision-making. By modeling the search process as a sequential decision problem, RL agents can be trained to optimize resource allocation and retrieval accuracy. These agents adapt their strategies by receiving rewards for minimizing computational overhead, reducing latency, and delivering precise results tailored to query characteristics, such as modality, complexity, and user intent.
    \end{itemize}
\item \textbf{Query Reformulation and Semantic Alignment:} 
    \begin{itemize}[leftmargin=1em]
    \item \textbf{Multi-Perspective Reformulation:} To address the complexity of multimodal search, query reformulation strategies must generate diverse interpretations while preserving the original intent across modalities. This involves:
    1) Contextual Understanding: Leveraging contextual embeddings (e.g., from transformer-based models) to capture semantic nuances and contextual dependencies, ensuring reformulated queries retain the richness of the original input.
    2) Cross-Modal Alignment: Employing advanced techniques like contrastive learning to align representations across text, images, and audio modalities. By embedding queries and multimodal data into a shared latent space, this ensures consistent interpretation and retrieval across diverse data types.
    3) Domain-Specific Knowledge Integration: Incorporating domain-specific ontologies or knowledge graphs to enhance reformulation accuracy, particularly in specialized fields. This leverages structured domain knowledge to improve the relevance and precision of reformulated queries.
    \item \textbf{Interactive Query Refinement:} A pivotal strategy is interactive query refinement, which allows users to iteratively adjust queries based on intermediate results. Intelligent systems can facilitate this process by suggesting alternative query formulations, identifying ambiguities, and providing contextual feedback to better align queries with user intent. By integrating user feedback loops and real-time semantic analysis, these systems dynamically bridge the gap between user input and multimodal data, ensuring more precise and contextually relevant search outcomes. 
    \item \textbf{Explainable Reformulation:} A critical aspect of query reformulation is its explainability. By offering clear and concise explanations for how queries are transformed, users gain insight into the system's reasoning and decision-making processes. For example, when a user submits a vague or ambiguous query, the system can generate a reformulated version while detailing the rationale behind the changes, such as term disambiguation, incorporation of contextual cues, or alignment with multimodal data (e.g., text, images, or audio). This transparency fosters user trust, enables validation of the system's interpretation, and enhances user control and satisfaction. Furthermore, explainable reformulation underscores the importance of semantic alignment, where the system bridges the gap between user intent and the underlying data representation, ensuring the reformulated query accurately reflects the user's needs.
    \end{itemize}
\item \textbf{Comprehensive Evaluation Benchmarks:} 
    \begin{itemize}[leftmargin=1em]
    \item \textbf{Diverse Query Datasets:} It is crucial to establish robust benchmarks. These benchmarks should incorporate diverse query datasets spanning a wide range of query types, from simple factual retrievals to complex multi-hop reasoning and creative tasks. The datasets must reflect real-world heterogeneity in query patterns and modalities, capturing the intricacies of user interactions across text, image, audio, and video inputs. By integrating such diversity, benchmarks can more accurately assess model performance, generalization capabilities, and adaptability to varied real-world applications. 
    \item \textbf{Multi-Dimensional Metrics:} To ensure the effectiveness and reliability of Multimodal Search Planning systems, robust evaluation frameworks must be established. These frameworks should employ multi-dimensional metrics to comprehensively assess system performance across diverse operational scenarios. Key dimensions include:
    1) Adaptability: The system's ability to handle a broad spectrum of query types, from simple to highly complex, while integrating multiple modalities (e.g., text, images, audio). This metric evaluates the model's flexibility in addressing varied user needs and its capacity to generalize across domains.
    2) Robustness: The system's resilience under challenging conditions, such as computational constraints, noisy or incomplete inputs, and adversarial scenarios. Robustness ensures consistent performance in real-world applications, where ideal conditions are seldom present.
    3) Efficiency: The optimization of resource utilization (e.g., memory, processing power) and response time. This metric is critical for scalability and user satisfaction, especially in time-sensitive or resource-constrained environments.
    4) Semantic Alignment: The system's accuracy in preserving and interpreting the intent of user queries during reformulation or multimodal integration. This ensures that outputs remain contextually and semantically aligned with the user's original request.
    \end{itemize}
\end{itemize}

\subsection{Retrieval}
The challenges in multimodal retrieval underscore the complexity of integrating and retrieving information across diverse data types. To address these issues and advance the field, future research should prioritize the following directions:
\begin{itemize}[leftmargin=1em]
\item \textbf{Unified Cross-Modal Representation Learning}: The primary objective is to develop robust and unified representation learning frameworks that effectively align and compare data across diverse modalities, including text, images, audio, and video. A key element of this framework is the enhancement of cross-modal attention mechanisms to better model complex interactions between modalities. Cross-modal attention layers, inspired by transformer architectures, are central to capturing fine-grained relationships. These mechanisms enable one modality to focus on relevant features in another, allowing the model to dynamically prioritize the most informative aspects of the data. For example, text can guide attention over visual regions, or audio cues can emphasize relevant temporal segments in video data. Techniques such as multi-head cross-modal attention and hierarchical attention further refine this process, ensuring robust and context-aware representations.
\item \textbf{Cross-Modal Context}: The primary objective is to improve the ability of models to perform fine-grained interactions between modalities, particularly in the reranking and refinement stages.
    \begin{itemize}[leftmargin=1em]
    \item \textbf{Reranker:} Multi-Modal Reranking Models rerank retrieved document list by incorporating detailed cross-modal interactions, such as text-image, text-audio, or video-text relationships. By integrating LLMs and MLLMs, reranking models enhance their ability to capture nuanced semantic alignments between modalities.
    \item \textbf{Refiner:} Leveraging their cross-modal reasoning abilities, MLLMs refine retrieval results through knowledge-enhanced refinement, yielding more accurate, contextually relevant, and semantically rich outputs.
    This refinement process utilizes MLLMs' contextual understanding and multimodal alignment to re-rank, filter, or augment retrieved content. 
    \end{itemize}
\end{itemize}

\subsection{Generation}
The future of multimodal generation should focus on overcoming existing challenges on multimodal input and output. Below are key suggestions for advancing multimodal generation systems.
\begin{itemize}[leftmargin=1em]
\item \textbf{Flexible and Adaptive Multimodal Input Frameworks:} To address the growing complexity of multimodal data, it is crucial to develop modality-agnostic architectures that dynamically adapt to diverse and arbitrary input modality combinations, such as text+image, text+video, or image+audio. These frameworks should process inputs without relying on predefined structures or extensive retraining. 
\item \textbf{Coherent and Contextually Relevant Multimodal Output:} Achieving coherent and contextually relevant outputs in multimodal generation necessitates the development of advanced models capable of maintaining consistency across modalities. For example, in text-image generation tasks, the generated image must precisely align with the textual description, and the text should accurately reflect the visual content of the image. This cross-modal consistency is essential for ensuring the reliability and usability of multimodal systems.
\item \textbf{Intelligent Positioning and Integration of Multimodal Elements:} To seamlessly integrate non-textual elements (e.g., images, videos, audio) into a narrative, models must be trained to identify optimal insertion points. This requires a deep understanding of the content's structure, flow, and contextual nuances to ensure coherence, readability, and enhanced user engagement. Advanced techniques, such as attention mechanisms, can analyze the narrative's semantic and syntactic structure, enabling the model to determine where multimodal elements can complement or enrich the text.
Modern multimodal systems must dynamically retrieve or generate contextually relevant non-textual content. For example, when generating a text-image pair, the model should use cross-modal alignment techniques to either retrieve an existing image from a database or synthesize a new one that aligns with the textual context. This relies on robust multimodal representation learning, where embeddings from different modalities (text, image, video) are mapped into a shared latent space, enabling precise cross-modal retrieval or generation.
\item \textbf{Diversity in Multimodal Outputs:} Achieving a balance between diversity, relevance, and quality in multimodal generation requires controlled mechanisms. For instance, in text-to-image generation, models should produce diverse yet faithful representations of textual descriptions. Techniques like conditional sampling can guide models to explore varied latent spaces while adhering to input constraints. 
\end{itemize}

\subsection{Dataset \& Evaluation}
The future direction of datasets and evaluation in MRAG should focus on addressing current gaps and challenges while harnessing the unique capabilities of MLLMs. Below are refined suggestions for advancing datasets and evaluation methodologies in this field.
\begin{itemize}[leftmargin=1em]
\item \textbf{Comprehensive Benchmark Development}: To enhance the evaluation of MLLMs and LLMs in retrieval-augmented generation, it is crucial to develop comprehensive benchmarks that address key limitations in current assessment frameworks. These benchmarks should focus on the following areas:
1) Instruction Following: Create tasks to evaluate the model’s ability to comprehend and execute complex, multi-step instructions across diverse modalities. This includes assessing precision in adhering to nuanced directives and handling ambiguous or incomplete inputs.
2) Multiturn Dialogue: Develop datasets that simulate real-world conversational dynamics, emphasizing the model’s capacity for context retention, coherence, and adaptability over extended interactions. Scenarios should include cross-modal references and long-term memory challenges.
3) Complex Multimodal Reasoning: Design tasks requiring the integration of multiple modalities (e.g., text, images, audio) to solve real-world problems, such as interpreting charts, maps, or combining visual and textual data for decision-making.
4) Creativity Evaluation: Introduce benchmarks to assess generative capabilities in creative tasks, such as composing stories, poems, or designing visual artifacts from multimodal inputs. These tasks should measure originality, relevance, and the ability to synthesize diverse inputs into coherent outputs.
5) Diverse Modalities: Expand evaluation frameworks to include emerging modalities like audio, 3D models, and sensor data, ensuring robustness and versatility in handling a wide range of input types.
\item \textbf{Multimodal Retrieval-Augmented Generation}: The development of robust metrics for evaluating retrieval and generation in multimodal systems requires assessing relevance, precision, diversity, and cross-modal alignment to ensure semantic consistency and contextual appropriateness. Metrics should quantify the system’s ability to filter noise and redundancy, delivering concise and meaningful outputs. For generation quality, coherence, fluency, creativity, and adaptability are essential, alongside factual accuracy and consistency with retrieved data and external knowledge. Effective multimodal integration is crucial to unify diverse inputs into contextually rich outputs. Comprehensive benchmarks must simulate real-world scenarios, incorporating varied queries, multimodal sources, and differing complexity levels to evaluate the end-to-end performance of retrieval-augmented generation (RAG) pipelines.
\end{itemize}

\section{Conclusion}
\label{sec:Conclusion}
In conclusion, this survey comprehensively examines the emerging field of Multimodal Retrieval-Augmented Generation (MRAG), highlighting its potential to enhance the capabilities of large language models (LLMs) by integrating multimodal data such as text, images, and videos. Unlike traditional text-based RAG systems, MRAG addresses the challenges of retrieving and generating information across different modalities, thereby improving the accuracy and relevance of responses while reducing hallucinations. The survey systematically analyzes MRAG from four key perspectives: essential components and technologies, datasets, evaluation methods and metrics, and existing limitations. It identifies current challenges, such as effectively integrating multimodal knowledge and ensuring the reliability of generated outputs, while also proposing future research directions. By providing a structured overview and forward-looking insights, this survey aims to guide researchers in advancing MRAG, ultimately contributing to the development of more robust and versatile Multimodal Retrieval-Augmented Generation.

\bibliographystyle{ACM-Reference-Format}
\bibliography{3-sample-base}

\appendix

\end{document}